\documentclass[12pt]{article}
\usepackage[a4paper,margin=1in]{geometry}
\usepackage[round,authoryear]{natbib}
\usepackage{chngcntr}

\usepackage[T1]{fontenc}
\usepackage[utf8]{inputenc}
\usepackage{lmodern}
\usepackage{amsmath,amssymb,amsthm,mathtools,bm,booktabs}
\usepackage{microtype,graphicx,float,placeins,pdflscape}
\usepackage{xurl}
\usepackage[unicode,hidelinks]{hyperref}
\usepackage{bookmark}
\usepackage{needspace}
\usepackage{algorithm}
\usepackage{algpseudocode}
\allowdisplaybreaks[2]
\setcitestyle{round,semicolon,aysep={,},yysep={;}}

\providecommand{\doi}[1]{\url{https://doi.org/#1}}

\theoremstyle{plain}
\newtheorem{theorem}{Theorem}
\newtheorem{corollary}{Corollary}
\newtheorem{proposition}{Proposition}
\newtheorem{lemma}{Lemma}[section]
\theoremstyle{definition}
\newtheorem{external}{External result}
\newtheorem{assumption}{Assumption}
\theoremstyle{remark}
\newtheorem*{remark}{Remark}

\newcommand{\vct}[1]{\bm{#1}}
\newcommand{\mat}[1]{\bm{\mathrm{#1}}}
\newcommand{\indc}{\mathbf{1}}

\newcommand{\Ex}{\mathbb E_{\xi,n}}
\newcommand{\Op}{O_{\mathbb P}}\newcommand{\tr}{\operatorname{tr}}
\newcommand{\diag}{\operatorname{diag}}\newcommand{\rank}{\operatorname{rank}}
\newcommand{\Cov}{\operatorname{Cov}}\newcommand{\Var}{\operatorname{Var}}
\newcommand{\one}{\vct{1}}\newcommand{\T}{^{\mathsf T}}
\newcommand{\sfac}{s_{f,p}}\newcommand{\alp}{\vct{\alpha}}
\newcommand{\hQ}{\widehat{\mat{Q}}_A}
\newcommand{\hP}{\widehat{\mat{P}}_A}\newcommand{\dP}{\mat{\Delta}_P}
\newcommand{\Em}{\mat{E}}\newcommand{\Fm}{\mat{F}_0}
\newcommand{\Sg}{\mat{\Sigma}_\varepsilon}
\newcommand{\Lnp}{L_{n,p}}\newcommand{\rn}{\rho_n}
\newcommand{\ind}{\mathrel{\perp\!\!\!\perp}}
\newcommand{\Ymat}{\mat{Y}}\newcommand{\Rmat}{\mat{R}}
\newcommand{\Oz}{\mat{O}_z}\newcommand{\Gmat}{\mat{G}}\newcommand{\Xmat}{\mat{X}}
\newcommand{\Jtrue}{\mat{J}_0}\newcommand{\Jhat}{\widehat{\mat{J}}}
\newcommand{\Ptime}{\mat{\Pi}_0}\newcommand{\Phat}{\widehat{\mat{\Pi}}}
\newcommand{\epsT}{a_{T,n}}\newcommand{\hT}{h_{T,n}}
\newcommand{\bvec}{b_{v,n}}\newcommand{\bsum}{b_{\mathrm{sum},n}}\newcommand{\bmax}{b_{\max,n}}
\newcommand{\vdiag}{a_{\Omega,n}}
\newcommand{\mnoise}{m_{\varepsilon,p}}\newcommand{\taunoise}{\tau_{\varepsilon,p}}
\newcommand{\snoise}{s_{\varepsilon,p}}
\newcommand{\Ieff}{\mathcal I_n}\newcommand{\cstar}{c_p}
\newcommand{\Tsum}{T_{\mathrm{sum}}}\newcommand{\Tmax}{T_{\mathrm{max}}}
\newcommand{\TCauchy}{T_{\mathrm{cauchy}}}

\newcommand{\E}{\mathbb E}

\newcommand{\Ops}{O_{\mathbb P^*,\log}}
\newcommand{\Opl}{O_{\mathbb P,\log}}
\newcommand{\op}{o_{\mathbb P}}
\newcommand{\ops}{o_{\mathbb P^*}}
\newcommand{\Prob}{\mathbb P}
\newcommand{\Prb}{\mathbb P}
\newcommand{\norm}[1]{\left\lVert#1\right\rVert}

\newcommand{\FG}{F_G}
\hypersetup{pdftitle={Adaptive Tests of Pricing Restrictions with Observed and Latent Factors},pdfauthor={}}

\begin{document}
\title{Adaptive Tests of Pricing Restrictions with Observed and Latent Factors}
\author{Anlin Liu\textsuperscript{1}, Guowei Yan\textsuperscript{2}, Zhaojun Wang\textsuperscript{1}\\
\small \textsuperscript{1}School of Statistics and Data Science, LEBPS, KLMDASR and LMPC,\\
\small Nankai University, Tianjin 300071, China\\
\small \textsuperscript{2}School of Statistics and Data Science, LEBPS, AAIS, KLMDASR and LMPC,\\
\small Nankai University, Tianjin 300071, China\\
\small Corresponding author: Guowei Yan, \texttt{1120250076@mail.nankai.edu.cn }}
\date{}
\maketitle
\begin{abstract}
This paper develops a high-dimensional test for asset pricing models with observed and latent factors. The null hypothesis allows the intercepts from an observed-factor regression to be nonzero, provided that they lie in the latent factor loading space. We estimate the pricing errors in three steps: regress returns on the observed factors, extract latent factors from the lagged covariance structure of the regression residuals, and regress returns on both observed and estimated latent factors. The intercept from the final regression serves as the estimated pricing error. We construct Sum and Max tests based on the estimated pricing errors to detect diffuse and concentrated alternatives, respectively, and combine them using a Cauchy procedure. Under suitable regularity conditions, we establish the joint null distribution of the component statistics, their asymptotic independence, and their power against alternative hypotheses. We also propose a residual bootstrap and establish its first-order validity. Simulation studies examine the finite-sample performance of the tests under different pricing-error patterns and factor-number choices. An application to S\&P 500 stocks illustrates the use of the proposed framework in evaluating an observed-factor pricing model after latent-factor adjustment.

\end{abstract}

\section{Introduction}\label{sec:introduction}
Linear factor pricing models relate expected asset returns to exposures to systematic risks. The capital asset pricing model and the Fama--French model provide familiar benchmarks \citep{sharpe1964,lintner1965,famaFrench1993}, while arbitrage pricing theory motivates a broader multifactor representation \citep{ross1976}. In the traded-factor benchmark considered here, observed factor means supply the corresponding risk premia. The time-series regression intercept then measures the average return remaining after observed-factor compensation. Testing whether these intercepts are jointly zero is central to evaluating investment performance and the explanatory content of a pricing benchmark \citep{jensen1968,grs1989}.

An observed benchmark may omit common sources of risk. Its regression intercept can then contain compensation for latent exposures, and the residual return dynamics can help identify those exposures. We study whether the mean return remaining after the observed benchmark can be explained by a latent loading space. Write the observed-factor intercept as $\vct{\mu}$ and let $\mat{A}$ have orthonormal columns spanning that space. The decomposition
\[
\vct{\mu}=\mat{A}\vct{\lambda}+\alp,
\qquad \mat{A}\T\alp=\vct{0},
\]
separates the component compatible with latent risk compensation from the remaining pricing error. Our null is $\alp=\vct{0}$, or equivalently $\vct{\mu}\in\operatorname{col}(\mat{A})$. It allows a nonzero observed-factor intercept and evaluates the pricing restriction after latent compensation has been included. The observed traded-factor premia remain fixed at their means.

Several studies provide closely related approaches to inference with latent pricing factors. \citet{giglioLiaoXiu2021} combine observed-factor regressions with residual principal components to estimate individual alphas and control false discoveries across many tests. Their construction already accommodates latent factors alongside observed factors. \citet{li2025} develop a test of latent beta-pricing models, and \citet{massacci2026} propose randomized tests of jointly zero pricing errors, including a latent-factor extension. These contributions establish the relevance of accounting for latent exposures when assessing pricing errors. Our focus is a global test of the restriction above, with the latent space estimated from residual temporal dependence and its mean compensation fitted by a final time-series regression.

The choice of pricing restriction also distinguishes this problem from risk-premium estimation. The two-pass approach of \citet{famaMacBeth1973} estimates premia through cross-sectional regressions on estimated exposures, and \citet{giglioXiu2021} account for omitted factors when estimating observed-factor risk premia. \citet{pesaranSmith2025} study departures of traded-factor premia from factor means. Here the given traded-factor benchmark determines observed compensation, and the additional component is fitted within the latent loading space. Section~\ref{sec:model} specifies this target and explains the role of a freely fitted common intercept in an affine pricing restriction.

High dimensionality creates a further inference problem. The classical test of \citet{grs1989} uses an inverse residual covariance matrix, whereas modern return panels can contain more assets than time observations. \citet{pesaranYamagata2024} address this setting by aggregating squared individual regression statistics. Power also depends on the concentration of pricing errors: sum-type tests accumulate diffuse departures, while max-type tests respond to large coordinatewise departures. \citet{feng2022} develop a maximum-based alpha test and a combination with a sum-type test; \citet{fan2015} enhance power through screening; and \citet{xiaZhang2024} adaptively aggregate ordered coordinate statistics. For dependent observations, \citet{ma2024} construct Sum, Max and Cauchy tests. The asymptotic-independence results of \citet{feng2024} and the Cauchy transformation of \citet{liuXie2020} provide tools for combining complementary evidence. In our setting, their use requires control of the additional estimation error from the latent adjustment.

Our estimator builds on dynamic factor estimation from positive-lag covariances \citep{lamYaoBathia2011,lamyao}. In particular, \citet{cgy} estimate an observed regression first and then recover latent factors from its residuals. We use this regression--LAG construction to obtain factor directions, retain the estimated factor means, and regress returns on both observed and estimated latent factor levels. The resulting intercept estimates $\alp$. The factor-adjusted location tests of \citet{wangLocation2026} are especially close in their use of lag-based extraction and combined Sum and Max evidence. The present pricing problem includes an estimated observed-factor regression and allows nonzero latent compensation under the null. The final regression fits the return exposures associated with that compensation. All three steps use the same observations, so inference must account for their estimation effects jointly.

This paper makes three contributions. First, we develop a three-step LAG procedure for global testing of the stated pricing restriction with a given traded-factor benchmark. The procedure combines observed-factor OLS, residual factor extraction and augmented OLS, and we characterize the correction supplied by the final regression. Under an additional Gaussian condition, we establish a mean-square comparison with direct projection and identify when the final regression gives a strict improvement. Second, we show that the feasible Sum and Max statistics admit first-order null approximations by statistics based on the true factor levels. We derive their asymptotically independent normal and Gumbel limits, Cauchy calibration, and power under diffuse and concentrated alternatives. Third, we establish first-order validity of a residual bootstrap that preserves fitted latent compensation under the null and refits the estimation procedure in each generated panel.

The simulation study examines calibration across observed and latent compensation configurations and power across pricing-error sparsity and factor-number choices. Residual PCA and existing pricing tests provide comparisons. An application to weekly S\&P 500 returns uses the Fama--French three-factor model as the observed benchmark and assesses the remaining mean component after latent adjustment. Residual diagnostics and rolling-window tests show how the pricing assessment changes when latent compensation is included.

Section~\ref{sec:model} introduces the model and pricing target. Section~\ref{sec:method} presents estimation and testing, and Section~\ref{sec:theory} gives the theoretical results. Sections~\ref{sec:simulation} and~\ref{sec:realdata} report simulations and the empirical study, followed by concluding remarks. Appendix~\ref{app:technical} develops preliminary results for the main testing, bootstrap and mean-square proofs in Appendices~\ref{app:mainproofs}--\ref{app:mse}. Appendix~\ref{app:pca} collects the PCA comparison.

\section{Model and pricing target}\label{sec:model}
At times $t=1,\ldots,n$, we observe asset excess returns $\vct{Y}_t\in\mathbb R^p$ and traded-factor returns $\vct{z}_t\in\mathbb R^m$. We model them as
\[
\vct{Y}_t=\vct{\mu}+\mat{B}\vct{z}_t
+\mat{A}\vct{f}_t+\vct{\varepsilon}_t,
\]
where $\mat{B}\in\mathbb R^{p\times m}$ gives exposures to observed factors, $\mat{A}\in\mathbb R^{p\times r}$ gives exposures to $r$ unobserved factors $\vct{f}_t$, and $\vct{\varepsilon}_t\in\mathbb R^p$ is idiosyncratic error. The factor number $r$ and the column space of $\mat{A}$ are unknown. We set $\mathbb E\vct{f}_t=\mathbb E\vct{\varepsilon}_t=0$. The centered observed and latent factors are contemporaneously orthogonal, $\mathbb E\{(\vct{z}_t-\mathbb E\vct{z}_t)\vct{f}_t\T\}=\mat{0}$, so the latent component represents common variation remaining after the observed-factor regression. Time dependence in the factors is allowed; Section~\ref{sec:theory} gives the formal conditions.

The intercept $\vct{\mu}=\mathbb E\vct{Y}_t-\mat{B}\mathbb E\vct{z}_t$ is the mean return left after observed-factor compensation. Normalize $\mat{A}\T\mat{A}=\mat{I}_r$ and set $\mat{P}_A=\mat{A}\mat{A}\T$. Its decomposition is
\[
\vct{\mu}=\mat{A}\vct{\lambda}+\alp,
\qquad \vct{\lambda}=\mat{A}\T\vct{\mu},
\qquad \alp=(\mat{I}_p-\mat{P}_A)\vct{\mu}.
\]
The first component is compatible with compensation for latent-factor exposure; $\alp$ is the pricing error outside that space. We test
\[
H_0:\alp=0\quad\text{against}\quad H_1:\alp\ne0.
\]
In particular, the null allows $\vct{\mu}=\mat{A}\vct{\lambda}\ne0$.

Let $\vct{F}_t=\vct{\lambda}+\vct{f}_t$ denote the latent factor level, so $\mathbb E\vct{F}_t=\vct{\lambda}$. The model is equivalently
\[
\vct{Y}_t=\alp+\mat{B}\vct{z}_t
+\mat{A}\vct{F}_t+\vct{\varepsilon}_t.
\]
Retaining the estimated factor means accounts for latent compensation in the final regression. The pricing target is invariant to the orthonormal basis: replacing $\mat{A}$ by $\mat{A}\mat{H}$ for an orthogonal $\mat{H}$ leaves $\mat{P}_A$ and $\alp$ unchanged.

\noindent\textit{Relation to an affine pricing restriction.} A freely fitted common intercept would enlarge the pricing restriction. With observed-factor compensation held fixed, the affine restriction is
\[
\vct{\mu}=\gamma_0\vct{1}_p+\mat{A}\vct{\gamma},
\qquad (\gamma_0,\vct{\gamma})\in\mathbb R\times\mathbb R^r.
\]
Our null is $\vct{\mu}\in\operatorname{col}(\mat{A})$. Under the affine restriction, $\alp=\gamma_0(\mat{I}_p-\mat{P}_A)\vct{1}_p$: our pricing error includes this component, so the global test also assesses common deviations outside the latent loading space.

\noindent\textbf{Notation.} For a sequence $\{\vct{x}_t\}_{t=1}^n$, write $\bar{\vct{x}}=n^{-1}\sum_{t=1}^n\vct{x}_t$. For a vector $\vct{x}$, $\|\vct{x}\|=(\sum_i x_i^2)^{1/2}$ and $\|\vct{x}\|_\infty=\max_i|x_i|$. For a matrix $\mat{H}$, $\|\mat{H}\|=\sup_{\|\vct{x}\|=1}\|\mat{H}\vct{x}\|$, $\|\mat{H}\|_F=\{\tr(\mat{H}\T\mat{H})\}^{1/2}$, and $\|\mat{H}\|_* =\sum_j\sigma_j(\mat{H})$, where $\sigma_j(\mat{H})$ are its singular values. Constants $c,C>0$ may change between displays.

\Needspace{18\baselineskip}
\section{Method}\label{sec:method}
We first estimate the pricing error, then construct tests from its squared norm and largest standardized coordinate. The procedure also supplies the fitted factors and residuals used for bootstrap calibration. Throughout, the numbers of observed and latent factors are small relative to the time-series length, while the number of assets may exceed it.

\subsection{Estimation in three steps}\label{sec:selectors}
The procedure consists of two least-squares problems separated by a LAG eigenanalysis. The first regression removes the observed-factor component, the LAG step extracts the latent factor directions, and the second regression estimates the associated return exposures and the pricing error.

\paragraph{Step 1: observed-factor regression.}
Regress the asset returns on a constant and the observed factors:
\[
(\widehat{\vct{\mu}},\widehat{\mat{B}})
=\mathop{\arg\min}_{\vct{a}\in\mathbb R^p,\,\mat{D}\in\mathbb R^{p\times m}}
\sum_{t=1}^n\|\vct{Y}_t-\vct{a}-\mat{D}\vct{z}_t\|^2.
\]
Let
$\mat{S}_z=n^{-1}\sum_{t=1}^n(\vct{z}_t-\bar{\vct{z}})
(\vct{z}_t-\bar{\vct{z}})\T$. The explicit OLS solution is
\[
\widehat{\mat{B}}=
\left\{\frac1n\sum_{t=1}^n
(\vct{Y}_t-\bar{\vct{Y}})(\vct{z}_t-\bar{\vct{z}})\T\right\}
\mat{S}_z^{-1},
\qquad
\widehat{\vct{\mu}}=\bar{\vct{Y}}-\widehat{\mat{B}}\bar{\vct{z}}.
\]
The centered residuals and their sample covariance are
\[
\widehat{\vct{\eta}}_t
=\vct{Y}_t-\widehat{\vct{\mu}}-\widehat{\mat{B}}\vct{z}_t,
\qquad
\Rmat=(\widehat{\vct{\eta}}_1,\ldots,\widehat{\vct{\eta}}_n),
\qquad
\widehat{\mat{C}}_\eta=\Rmat\Rmat\T/n.
\]

\paragraph{Step 2: LAG eigenanalysis.}
For a fixed number of positive lags $k_0$, form
\[
\widehat{\mat{\Gamma}}_\eta(h)=
\frac1{n-h}\sum_{t=1}^{n-h}
\widehat{\vct{\eta}}_{t+h}\widehat{\vct{\eta}}_t\T,
\qquad
\widehat{\mat{M}}_A=
\sum_{h=1}^{k_0}\widehat{\mat{\Gamma}}_\eta(h)
\widehat{\mat{\Gamma}}_\eta(h)\T.
\]
Writing $\vct{\eta}_t=\mat{A}\vct{f}_t+\vct{\varepsilon}_t$ and
$\mat{\Gamma}_f(h)=\mathbb E(\vct{f}_{t+h}\vct{f}_t\T)$, independence between the factor and error processes, together with serially uncorrelated errors, gives
\[
\mat{\Gamma}_\eta(h)
:=\operatorname{Cov}(\vct{\eta}_{t+h},\vct{\eta}_t)
=\mat{A}\mat{\Gamma}_f(h)\mat{A}\T,
\qquad h>0.
\]
When the selected positive lags identify all latent directions, the leading eigenspace of $\widehat{\mat{M}}_A$ estimates the loading space \citep{lamYaoBathia2011,cgy}. Let
$\widehat\lambda_1\ge\cdots\ge\widehat\lambda_p$ be its eigenvalues. For a fixed cap $R_*<\min(p,n-m-1)$, use the adjacent eigenvalue-ratio selector \citep{lamyao}:
\begin{equation}\label{eq:rank-selectors}
\widehat r=\min\mathop{\arg\min}_{1\le j\le R_*}
\frac{\widehat\lambda_{j+1}}{\widehat\lambda_j}.
\end{equation}
A ratio with zero denominator is set to one, and the smallest minimizer resolves ties. Let $\widehat{\mat{A}}$ contain the leading $\widehat r$ orthonormal eigenvectors. The estimated latent factor levels are
\[
\widehat{\vct{F}}_t
=\widehat{\mat{A}}\T(\vct{Y}_t-\widehat{\mat{B}}\vct{z}_t)
=\widehat{\mat{A}}\T\widehat{\vct{\mu}}
+\widehat{\mat{A}}\T\widehat{\vct{\eta}}_t.
\]
Their sample mean
$\bar{\widehat{\vct{F}}}=\widehat{\mat{A}}\T\widehat{\vct{\mu}}$
retains the estimated latent risk compensation, while
$\widehat{\vct{f}}_t=\widehat{\vct{F}}_t-\bar{\widehat{\vct{F}}}
=\widehat{\mat{A}}\T\widehat{\vct{\eta}}_t$
is the centered factor score.

\paragraph{Step 3: augmented regression.}
Regress the original returns on a constant, the observed factors and the estimated latent factor levels:
\[
(\widehat{\alp},\widetilde{\mat{B}},\widetilde{\mat{A}})
=\mathop{\arg\min}_{\vct{a}\in\mathbb R^p,\,\mat{D}\in\mathbb R^{p\times m},\,\mat{C}\in\mathbb R^{p\times\widehat r}}
\sum_{t=1}^n
\|\vct{Y}_t-\vct{a}-\mat{D}\vct{z}_t-\mat{C}\widehat{\vct{F}}_t\|^2.
\]
The first-stage normal equations make the centered factor scores sample-orthogonal to the centered observed factors. Hence $\widetilde{\mat{B}}=\widehat{\mat{B}}$, and the remaining OLS solutions are
\begin{equation}\label{eq:post}
\widetilde{\mat{A}}
=\widehat{\mat{C}}_\eta\widehat{\mat{A}}
(\widehat{\mat{A}}\T\widehat{\mat{C}}_\eta\widehat{\mat{A}})^{-1},
\qquad
\widehat{\alp}
=\widehat{\vct{\mu}}-\widetilde{\mat{A}}\bar{\widehat{\vct{F}}}.
\end{equation}
Here $\widehat{\mat{A}}$ gives the LAG extraction directions, whereas $\widetilde{\mat{A}}$ estimates the contemporaneous return exposures to the extracted factors. The final intercept subtracts the fitted latent compensation from the observed-factor intercept and estimates the pricing error $\alp$. An orthogonal change of basis in
$\widehat{\mat{A}}$ preserves the span of the factor regressors, the final
intercept $\widehat{\alp}$, and the test statistics.

\begin{remark}
Direct projection of $\widehat{\vct{\mu}}$ subtracts
$\widehat{\mat{A}}\bar{\widehat{\vct{F}}}$, so the adjustment is confined
to the estimated LAG loading space. The second OLS subtracts
$\widetilde{\mat{A}}\bar{\widehat{\vct{F}}}$, which also accounts for
contemporaneous covariation between the extracted factors and residual
returns orthogonal to that space. Proposition~\ref{prop:mse} in
Section~\ref{sec:apr} gives conditions under which this adjustment
reduces the mean-square error of the pricing-error estimator.
\end{remark}

Algorithm~\ref{alg:lag-alpha} summarizes the procedure. For each asset, the two regressions use $1+m$ and $1+m+\widehat r$ regressors, respectively. Thus ordinary least squares remains feasible when $p>n$, provided the corresponding factor designs have full column rank.

\begin{algorithm}[htbp]
\caption{LAG-Adjusted Alpha Estimation}
\label{alg:lag-alpha}
\begin{algorithmic}[1]
\Procedure{LAGAdjust}{$\{\vct{Y}_t,\vct{z}_t\}_{t=1}^{n},\,k_0,\,R_*$}

    \State \textbf{Step 1.} Fit the observed-factor regression and compute residuals.
    \Statex \hspace{\algorithmicindent}
    $\displaystyle
    (\widehat{\vct{\mu}},\widehat{\mat{B}})
    \gets \operatorname*{arg\,min}_{\vct{a},\,\mat{D}}
    \sum_{t=1}^{n}
    \|\vct{Y}_t-\vct{a}-\mat{D}\vct{z}_t\|^2.$
    \Statex \hspace{\algorithmicindent}
    $\widehat{\vct{\eta}}_t
    \gets \vct{Y}_t-\widehat{\vct{\mu}}
    -\widehat{\mat{B}}\vct{z}_t,\quad t=1,\ldots,n.$

    \State \textbf{Step 2.} Extract the latent factors by LAG eigenanalysis.
    \Statex \hspace{\algorithmicindent}
    $\displaystyle
    \widehat{\mat{\Gamma}}_\eta(h)
    \gets \frac{1}{n-h}\sum_{t=1}^{n-h}
    \widehat{\vct{\eta}}_{t+h}\widehat{\vct{\eta}}_t\T,
    \quad h=1,\ldots,k_0.$
    \Statex \hspace{\algorithmicindent}
    $\displaystyle
    \widehat{\mat{M}}_A
    \gets \sum_{h=1}^{k_0}
    \widehat{\mat{\Gamma}}_\eta(h)
    \widehat{\mat{\Gamma}}_\eta(h)\T.$
    \Statex \hspace{\algorithmicindent}
    Select $\widehat r$ by \eqref{eq:rank-selectors} and form
    $\widehat{\mat{A}}$ from the leading $\widehat r$
    eigenvectors of $\widehat{\mat{M}}_A$.
    \Statex \hspace{\algorithmicindent}
    $\widehat{\vct{F}}_t
    \gets \widehat{\mat{A}}\T
    (\vct{Y}_t-\widehat{\mat{B}}\vct{z}_t),
    \quad t=1,\ldots,n.$

    \State \textbf{Step 3.} Fit the augmented regression.
    \Statex \hspace{\algorithmicindent}
    $\displaystyle
    (\widehat{\alp},\widetilde{\mat{B}},\widetilde{\mat{A}})
    \gets \operatorname*{arg\,min}_{\vct{a},\mat{D},\mat{C}}
    \sum_{t=1}^{n}\|\vct{Y}_t-\vct{a}-\mat{D}\vct{z}_t-\mat{C}\widehat{\vct{F}}_t\|^2.$

    \State \Return $\widehat{\alp}$.
\EndProcedure
\end{algorithmic}
\end{algorithm}

\subsection{Sum, Max, and Cauchy tests}\label{sec:tests}
Given the pricing-error estimator $\widehat{\alp}$, we test $H_0:\alp=\vct{0}$ by measuring its departure from zero. The squared Euclidean norm aggregates evidence across assets and is sensitive to diffuse pricing errors. The maximum of the standardized coordinates targets alternatives with a few large pricing errors.

Calibration requires the null center and variance of $\|\widehat{\alp}\|^2$ and the coordinate variances of $\widehat{\alp}$. To motivate their estimation, let $\widehat{\alp}^{\circ}$ be the intercept from a regression on the true factor levels, and let $\Ieff$ be the squared length of the constant after projection on those factor regressors. Conditional on the factor paths, the error assumptions in Section~\ref{sec:theory} give
\[
\mathbb E_0(\widehat{\alp}^{\circ}\mid\text{factors})=\vct{0},
\qquad
\Cov_0(\widehat{\alp}^{\circ}\mid\text{factors})=\Sg/\Ieff.
\]
Its squared norm therefore has conditional null center $\tr(\Sg)/\Ieff$ and leading variance $2\tr(\Sg^2)/\Ieff^2$. This variance is exact for Gaussian errors; under the stated non-Gaussian conditions, the fourth-cumulant contribution is of smaller order. The feasible calibration estimates these leading quantities using the fitted factors and residuals. The first-order approximation in Section~\ref{sec:theory} accounts for their estimation from the same sample.

Let $\Ymat=(\vct{Y}_1,\ldots,\vct{Y}_n)$, let $\mat{Z}$ be the $n\times m$ matrix with row $\vct{z}_t\T$, and form
\[
\widehat{\mat{D}}_x
=[\mat{Z},(\widehat{\vct{F}}_1,\ldots,\widehat{\vct{F}}_n)\T],
\qquad
\widehat{\mat{M}}_x
=\mat{I}_n-\widehat{\mat{D}}_x
(\widehat{\mat{D}}_x\T\widehat{\mat{D}}_x)^{-1}
\widehat{\mat{D}}_x\T.
\]
The matrix $\widehat{\mat{M}}_x$ removes the factor regressors. The final intercept can be written as $\widehat{\alp}=\Ymat\widehat{\mat{M}}_x\one_n/(\one_n\T\widehat{\mat{M}}_x\one_n)$. For null calibration, form the residual matrix from the same factor regression with the intercept constrained to zero:

\[
\mat{U}=\Ymat\widehat{\mat{M}}_x,\qquad
k=n-m-\widehat r,\qquad
\widehat{\mathcal I}=\one_n\T\widehat{\mat{M}}_x\one_n,\qquad
\widehat{\mat{\Sigma}}_\varepsilon=\frac{\mat{U}\mat{U}\T}{k}.
\]
Here $\widehat{\mathcal I}$ is the squared length of the residualized constant and equals $n$ when the factor regressors have zero sample means. The null residual matrix $\mat{U}$ retains that direction, so its degrees of freedom are $k=n-m-\widehat r$. The unrestricted Step~3 residuals also remove the fitted intercept and will be used for bootstrap resampling. Under $H_0$, $\widehat{\mat{\Sigma}}_\varepsilon$ supplies the covariance functionals needed for calibration.

The following quantities estimate the first-order null center and variance of $\|\widehat{\alp}\|^2$:
\[
\begin{aligned}
\widehat{\mathbb E}_0\|\widehat{\alp}\|^2
&=\frac{\tr(\widehat{\mat{\Sigma}}_\varepsilon)}{\widehat{\mathcal I}},\\
\widehat{\operatorname{Var}}_0(\|\widehat{\alp}\|^2)
&=\frac{2k}{(k+1)\widehat{\mathcal I}^{\,2}}
\left\{\tr(\widehat{\mat{\Sigma}}_\varepsilon^2)
-\frac{\tr(\widehat{\mat{\Sigma}}_\varepsilon)^2}{k}\right\}.
\end{aligned}
\]
The subtraction in the second line corrects the leading high-dimensional bias in estimating $\tr(\Sg^2)$. Standardizing the squared norm with these quantities gives the Sum statistic. For the Max statistic, we use $(\widehat{\mat{\Sigma}}_\varepsilon)_{ii}/\widehat{\mathcal I}$ as the variance scale for coordinate $i$. Thus
\begin{equation}\label{eq:tests}
\begin{aligned}
\Tsum
&=\frac{\widehat{\mathcal I}\|\widehat{\alp}\|^2
-\tr(\widehat{\mat{\Sigma}}_\varepsilon)}
{\left[\frac{2k}{k+1}\left\{
\tr(\widehat{\mat{\Sigma}}_\varepsilon^2)
-\frac{\tr(\widehat{\mat{\Sigma}}_\varepsilon)^2}{k}
\right\}\right]^{1/2}},\\
\Tmax
&=\max_{1\le i\le p}
\frac{\widehat{\mathcal I}\widehat\alpha_i^2}
{(\widehat{\mat{\Sigma}}_\varepsilon)_{ii}}
-2\log p+\log\log p.
\end{aligned}
\end{equation}
The Sum statistic aggregates the squared pricing errors on their original scale. The Max statistic standardizes each coordinate and applies the extreme-value centering $2\log p-\log\log p$. Under the null, their reference distributions are standard normal and Gumbel, respectively, as established in Section~\ref{sec:theory}.

Let $\Phi$ be the standard normal distribution function and $F_G(x)=\exp\{-\pi^{-1/2}e^{-x/2}\}$. Their analytic $p$-values and the Cauchy combination are
\[
p_{\mathrm{sum}}=1-\Phi(\Tsum),\qquad
p_{\mathrm{max}}=1-F_G(\Tmax),\qquad
\TCauchy=\tfrac12\cot(\pi p_{\mathrm{sum}})
+\tfrac12\cot(\pi p_{\mathrm{max}}).
\]
The analytic Cauchy $p$-value is
$p_{\mathrm{cauchy}}=1/2-\arctan(\TCauchy)/\pi$. Write
$F_{\mathrm{sum}}=\Phi$, $F_{\mathrm{max}}=F_G$, and
$F_{\mathrm{cauchy}}(x)=1/2+\arctan(x)/\pi$
for the corresponding reference distribution functions. At significance level $\gamma$, the analytic procedure rejects when the relevant $p$-value is at most $\gamma$. The Cauchy combination is used when the concentration of the pricing errors is unspecified.

\subsection{Bootstrap calibration}\label{sec:bootstrap}
The analytic reference distributions give an asymptotic calibration. For finite samples, we also use a residual bootstrap that repeats the estimation steps on generated panels. Its construction preserves the fitted latent compensation under the null and the empirical cross-sectional dependence of the residuals. First, calculate the centered factor scores and the unrestricted Step~3 residuals
\begin{equation}\label{eq:bootstrap-fit}
\widehat{\vct{f}}_t=\widehat{\vct{F}}_t-\bar{\widehat{\vct{F}}},
\qquad
\widehat{\vct{\varepsilon}}_t
=\widehat{\vct{\eta}}_t-\widetilde{\mat{A}}\widehat{\vct{f}}_t.
\end{equation}
We resample the residuals as whole vectors to preserve their cross-sectional
dependence. To generate temporally dependent factors, standardize the centered
scores $\widehat{\vct{f}}_t/\sqrt p$ using their sample covariance
decomposition:
\[
\frac{1}{np}\sum_{t=1}^n\widehat{\vct{f}}_t\widehat{\vct{f}}_t\T
=\mat{V}_f\diag(d_j)\mat{V}_f\T,\qquad
\widehat{\mat{C}}_f=\mat{V}_f\diag(\sqrt{\widetilde d_j})\mat{V}_f\T.
\]
Here $\widetilde d_j$ is $d_j$ clipped to $[10^{-3},10^3]$. Set
$\widehat{\vct{w}}_t=\widehat{\mat{C}}_f^{-1}\widehat{\vct{f}}_t/\sqrt p$.
The working VAR(1) transition is obtained from the singular value decomposition
of the standardized lag-one covariance:
\begin{equation}\label{eq:bootstrap-transition}
\widehat{\mat{H}}_{f,0}
=\frac{1}{n-1}\sum_{t=1}^{n-1}\widehat{\vct{w}}_{t+1}\widehat{\vct{w}}_t\T
=\mat{U}_H\diag(h_j)\mat{V}_H\T,\qquad
\widehat{\mat{H}}_f=\mat{U}_H\diag(\widetilde h_j)\mat{V}_H\T.
\end{equation}
Clip $h_j$ to $[0.05,0.95]$ to obtain $\widetilde h_j$. If
$\widehat{\mat{H}}_{f,0}$ is singular, set
$\widehat{\mat{H}}_f=0.4\mat{I}_{\widehat r}$. The upper bound keeps the
generator stable, and the lower bound ensures that lag one identifies every
generated factor direction. Section~\ref{sec:theory} establishes the common
first-order null limit of the bootstrap and original statistics.

Take $\widehat{\mat{J}}_f$ to be the lower triangular Cholesky factor of
$\mat{I}_{\widehat r}-\widehat{\mat{H}}_f\widehat{\mat{H}}_f\T$.
For each bootstrap draw, generate a stationary Gaussian path independently
of the residual draws:
\[
\begin{gathered}
\vct{w}_0^*\sim N(\vct{0},\mat{I}_{\widehat r}),\qquad
\vct{w}_t^*=\widehat{\mat{H}}_f\vct{w}_{t-1}^*
+\widehat{\mat{J}}_f\vct{\zeta}_t^*,\qquad
\vct{\zeta}_t^*\overset{\rm iid}{\sim}N(\vct{0},\mat{I}_{\widehat r}),\\
\widehat{\mat{J}}_f\widehat{\mat{J}}_f\T
=\mat{I}_{\widehat r}-\widehat{\mat{H}}_f\widehat{\mat{H}}_f\T,
\qquad
\vct{F}_t^*=\bar{\widehat{\vct{F}}}+\sqrt p\,\widehat{\mat{C}}_f\vct{w}_t^*.
\end{gathered}
\]
Independently draw $I_1,\ldots,I_n\overset{\rm iid}{\sim}\operatorname{Unif}\{1,\ldots,n\}$ and generate returns as
\begin{equation}\label{eq:bootstrap-panel}
\vct{\varepsilon}_t^*=\widehat{\vct{\varepsilon}}_{I_t}
-\bar{\widehat{\vct{\varepsilon}}},\qquad
\vct{Y}_t^*=\widehat{\mat{B}}\vct{z}_t
+\widetilde{\mat{A}}\vct{F}_t^*+\vct{\varepsilon}_t^*.
\end{equation}
Conditional on the data, the observed factor path and the generating coefficients $\widehat{\mat{B}}$ and $\widetilde{\mat{A}}$ are fixed. The generated factors have mean $\bar{\widehat{\vct{F}}}$, and the generated errors have mean zero. Hence
\[
\mathbb E^*(\vct{Y}_t^*\mid\{\vct{z}_s\}_{s=1}^n)
-\widehat{\mat{B}}\vct{z}_t
=\widetilde{\mat{A}}\bar{\widehat{\vct{F}}}.
\]
The bootstrap mean after observed-factor compensation therefore lies in the generating loading space. This implements the composite pricing null while retaining fitted latent compensation, adapting the residual bootstrap of \citet{wangLocation2026}.

For each generated panel, repeat the two regressions and LAG eigenanalysis in Section~\ref{sec:selectors}, and recompute the statistics in Section~\ref{sec:tests}. Hold the factor number fixed at $\widehat r$, while re-estimating the loading directions, regression coefficients and scales. With $B_{\rm boot}$ independent panels, calculate
\begin{equation}\label{eq:bootstrap-pvalue}
p_j^{\rm boot}=\frac{1+\sum_{b=1}^{B_{\rm boot}}
\indc\{T_{j,b}^*\ge T_j\}}{B_{\rm boot}+1},
\qquad j\in\{\mathrm{sum},\mathrm{max},\mathrm{cauchy}\}.
\end{equation}
In each draw, the Cauchy statistic uses the analytic component $p$-values from Section~\ref{sec:tests}. Each statistic is then calibrated against its own bootstrap draws, and the test rejects at level $\gamma$ when $p_j^{\rm boot}\le\gamma$.

\noindent\textit{Implementation.} The simulation study uses $k_0=1$ and $R_*=8$ for LAG, and the weekly-return application uses $k_0=5$ and $R_*=8$. These choices use short positive lags and a rank cap above the anticipated number of latent factors. Both studies use $B_{\rm boot}=499$ and bootstrap calibration for the proposed tests. The factor number is selected once for each original panel and held fixed during its bootstrap refits.

\section{Theoretical results}\label{sec:theory}
We establish a first-order approximation for the LAG-based statistics and derive their null distributions, power and bootstrap validity. A separate condition yields the mean-square comparison of the final OLS intercept with direct projection. Write $\Fm=(\vct{f}_1,\ldots,\vct{f}_n)$ and $\Em=(\vct{\varepsilon}_1,\ldots,\vct{\varepsilon}_n)$. The assumptions below describe the population model, dimensional growth, and factor-number selection.

The dimensions $m,r,k_0$ are fixed positive integers and $p=p_n\to\infty$.
Parameters may form triangular arrays, with constants uniform along them.
Write $\vct{m}_z=\mathbb E\vct{z}_t$ and $\vct{z}_t^0=\vct{z}_t-\vct{m}_z$,
$\sfac=p^{(1-\delta)/2}$,
$\vct{\chi}_t=((\vct{z}_t^0)\T,\sfac^{-1}\vct{f}_t\T)\T$, and
$\mat{\Gamma}_z=\mathbb E(\vct{z}_t^0\vct{z}_t^{0\mathsf T})$.
The scale $\sfac^2=p^{1-\delta}$ describes the order of the factor covariance eigenvalues under $\mat{A}\T\mat{A}=\mat{I}_r$; $\delta=0$ gives order $p$. The conditions below adapt standard factor-process and loading assumptions
\citep{bai2003,lamYaoBathia2011,wangLocation2026}.

\begin{assumption}[Factor process]\label{ass:A}
The centered process $\vct{\chi}_t$ is weakly stationary, uniformly geometrically
$\alpha$-mixing and directionally sub-Gaussian. Also,
\[
\mathbb E(\vct{z}_t^0\vct{f}_t\T)=0,\qquad
c\mat{I}_m\preceq\mat{\Gamma}_z\preceq C\mat{I}_m,\qquad
\|\vct{m}_z\|+\|\vct{\lambda}\|/\sfac\le C.
\]
\end{assumption}
Here uniform geometric mixing means, for
$\mathcal F_a^b=\sigma(\vct{\chi}_t:a\le t\le b)$,
\[
\sup_{a\in\mathbb Z}\sup_{\substack{E_1\in\mathcal F_{-\infty}^a\\
E_2\in\mathcal F_{a+h}^{\infty}}}
|\mathbb P(E_1\cap E_2)-\mathbb P(E_1)\mathbb P(E_2)|\le Ce^{-ch}.
\]
This is the time-origin-uniform definition \citep{bradley2005}.
The directional moment condition is
$\sup_t\sup_{\|\vct{u}\|=1}\|\vct{u}\T\vct{\chi}_t\|_{\psi_2}\le C$,
with $\|X\|_{\psi_2}=\inf\{b>0:\mathbb E\exp(X^2/b^2)\le2\}$.

\begin{assumption}[Factor strength and dispersion]\label{ass:B}
\[
c\sfac^2\mat{I}_r\preceq\mat{\Gamma}_f(0)\preceq C\sfac^2\mat{I}_r,
\qquad \max_i\|\vct{e}_i\T\mat{A}\|^2\le C/p.
\]
\end{assumption}

\begin{assumption}[Idiosyncratic errors]\label{ass:C}
The errors satisfy $\vct{\varepsilon}_t=\Sg^{1/2}\vct{\xi}_t$.
The innovation vectors are i.i.d., independent of the factor process,
and have independent standardized sub-Gaussian coordinates. Also,
$c\mat{I}_p\preceq\Sg\preceq C\mat{I}_p$.
\end{assumption}
The square root is symmetric; standardized coordinates have mean zero and variance one,
with $\sup_{t,i}\|\xi_{ti}\|_{\psi_2}\le C$.

\begin{assumption}[Dimensional growth]\label{ass:D}
For fixed $K<\infty$ and $\eta>0$,
\begin{equation}\label{eq:growth}
0\le\delta<1/4,\qquad p=O(n^K),\qquad
\frac{n}{p^{1-\delta}}=O(n^{-\eta}),\qquad
\frac{p^{1/2+\delta}}n=O(n^{-\eta}).
\end{equation}
\end{assumption}
These conditions balance estimation of the latent space with aggregation across $p$ assets. For $p\asymp n^\kappa$, they require $1/(1-\delta)<\kappa<2/(1+2\delta)$, which becomes $1<\kappa<2$ when $\delta=0$. Appendix~\ref{app:prelim} records the resulting rate bounds.

\begin{assumption}[Lag identification]\label{ass:lag}
\begin{equation}\label{eq:lag-strength}
c\sfac^4\mat{I}_r
\preceq\sum_{h=1}^{k_0}\mat{\Gamma}_f(h)\mat{\Gamma}_f(h)\T
\preceq C\sfac^4\mat{I}_r.
\end{equation}
\end{assumption}
This condition requires the selected positive lags to identify every latent direction. Together with temporal independence of the idiosyncratic errors, it gives the population separation used in Step~2.

\begin{assumption}[Factor-number selection]\label{ass:rank}\label{cond:population-rank}
For a fixed search cap $R_*>r$, $\mathbb P(\widehat r=r)\to1$.
\end{assumption}
Assumption~\ref{ass:rank} supplies the transfer from a known factor number to the selected one. Appendix~\ref{app:lag-rank-status} gives a sufficient eigenvalue-ratio separation condition and establishes consistency under that condition. Regression and scale regularity are established in Appendix~\ref{app:regularity}.

\Needspace{10\baselineskip}
\subsection{Null distributions, power and bootstrap calibration}
\label{sec:common-approximation}
The results below use the selected rank under Assumption~\ref{ass:rank}. For known $r$, Assumptions~\ref{ass:A}--\ref{ass:lag} suffice. Let $\mat{M}_0$ be the residual projection for the true factor design $[\mat{Z},(\vct{\lambda}\one_n\T+\Fm)\T]$. The intercept that would be obtained using the true factor levels and its information are
\[
\widehat{\alp}^{\circ}
=\frac{\Ymat\mat{M}_0\one_n}{\Ieff},
\qquad
\Ieff=\one_n\T\mat{M}_0\one_n.
\]
The corresponding oracle statistics use the same intercept-based construction as Section~\ref{sec:tests}:
\[
\begin{aligned}
T_{\mathrm{sum}}^{\circ}
&=\frac{\Ieff\|\widehat{\alp}^{\circ}\|^2-\tr(\Sg)}
{\sqrt{2\tr(\Sg^2)}},\\
T_{\mathrm{max}}^{\circ}
&=\max_{1\le i\le p}
\frac{\Ieff(\widehat\alpha_i^{\circ})^2}{(\Sg)_{ii}}
-2\log p+\log\log p.
\end{aligned}
\]
For the dense signal shift, write
\[
\Delta_n=\frac{n\|\alp\|^2}{\cstar\sqrt{2\tr(\Sg^2)}},\qquad
\cstar=1+\vct{m}_z\T\mat{\Gamma}_z^{-1}\vct{m}_z
+\vct{\lambda}\T\mat{\Gamma}_f(0)^{-1}\vct{\lambda}.
\]
The factor $\cstar$ records the information adjustment for nonzero factor means. The feasible statistics and their oracle counterparts have the following first-order relationship.
\begin{proposition}[First-order approximation]\label{prop:common-approximation}
Under Assumptions~\ref{ass:A}--\ref{ass:rank}, let $\alp$ be deterministic with
$\mat{A}\T\alp=0$ and $\|\alp\|\le C_\alpha$. Use the selected-rank LAG estimator. Then
\[
\Tsum-T_{\mathrm{sum}}^{\circ}\longrightarrow_{\mathbb P}0.
\]
Under $H_0$, $\Tmax-T_{\mathrm{max}}^{\circ}\longrightarrow_{\mathbb P}0$.
\end{proposition}
The proposition shows that estimating the factors and covariance scales leaves the first-order null calibration unchanged. For Sum, the approximation also holds for the bounded signal sequences used in the power analysis. We can therefore derive the feasible tests from the oracle statistics, whose joint null limit is given next.

\begin{theorem}[Joint calibration]\label{thm:null}
Under Assumptions~\ref{ass:A}--\ref{ass:rank} and $H_0$, use the selected-rank LAG estimator. Then
\[
(\Tmax,\Tsum)\Rightarrow(G_0,Z_0),\qquad G_0\ind Z_0,
\]
where $\mathbb P(G_0\le x)=F_G(x)$ and $Z_0\sim N(0,1)$. Moreover,
\[
\TCauchy\Rightarrow\operatorname{Cauchy}(0,1).
\]
\end{theorem}
For any fixed significance level $\gamma\in(0,1)$, the three analytic tests consequently have null rejection probability converging to $\gamma$. Asymptotic independence justifies the Cauchy reference distribution for the combination of the two component $p$-values.

\noindent\textit{Analytic power against diffuse and concentrated alternatives.}
The signal regimes are specified next.
\begin{assumption}[Signal sequences]\label{ass:signal}
The pricing error is deterministic, $\mat{A}\T\alp=0$, and $\|\alp\|\le C_\alpha$.
\end{assumption}
The local dense and consistent dense regimes are $\Delta_n\to\Delta<\infty$
and $\Delta_n\to\infty$, respectively. A sufficient sparse regime is
$\sqrt n\|\alp\|_\infty/\sqrt{\log(2p)}\to\infty$.
Since $\tr(\Sg^2)\asymp p$ and $\cstar$ is bounded, nondegenerate dense local
signals satisfy $\|\alp\|\asymp p^{1/4}/\sqrt n$.
\begin{theorem}[Power for bounded signals]\label{thm:power}
Under Assumptions~\ref{ass:A}--\ref{ass:rank} and \ref{ass:signal}, use the selected-rank LAG estimator. In the local dense regime,
$\Tsum\Rightarrow N(\Delta,1)$; in the consistent dense regime,
$\Tsum/\Delta_n\to_{\mathbb P}1$.
\end{theorem}
At level $\gamma$, the limiting local power of the Sum test is
\[
1-\Phi\{\Phi^{-1}(1-\gamma)-\Delta\}.
\]
\begin{corollary}[Sparse power]\label{cor:max}
Under the hypotheses of Theorem~\ref{thm:power}, in the sparse regime, $\Tmax\to_{\mathbb P}\infty$.
\end{corollary}
\begin{corollary}[Cauchy power]\label{cor:cauchy}
Under the hypotheses of Theorem~\ref{thm:power}, the Cauchy statistic diverges
to $+\infty$ in either the consistent dense or the sparse regime.
\end{corollary}
The Sum result measures signal through $\|\alp\|^2$, while the sparse result is driven by $\|\alp\|_\infty$. The Cauchy test is consistent in either stated regime, supporting its use when the concentration of the pricing errors is unknown.

\noindent\textit{Bootstrap calibration.}\label{sec:boot-validity}
The next result connects the resampling construction in Section~\ref{sec:bootstrap} to the same joint null limit. It applies to the procedure that re-estimates the loading directions and regression coefficients while holding the outer selected factor number fixed.
\begin{theorem}[Bootstrap validity]\label{thm:column-bootstrap}
Under Assumptions~\ref{ass:A}--\ref{ass:rank} and $H_0$, use the LAG estimator. The procedure of
Section~\ref{sec:bootstrap}, holding the outer selected rank fixed
inside each refit, satisfies
\[
\sup_{x,y}\left|
\mathbb P^*(T_{\mathrm{sum}}^*\le x,T_{\mathrm{max}}^*\le y)-\Phi(x)F_G(y)
\right|\longrightarrow_{\mathbb P}0.
\]
Moreover, $T_{\mathrm{cauchy}}^*\Rightarrow_{\mathbb P^*}\operatorname{Cauchy}(0,1)$
in probability. For each $j\in\{\mathrm{sum},\mathrm{max},\mathrm{cauchy}\}$,
\begin{equation}\label{eq:boot-validity-target}
\sup_x\left|\mathbb P^*(T_j^*\le x)-\mathbb P_0(T_j\le x)\right|
\longrightarrow_{\mathbb P}0.
\end{equation}
\end{theorem}
Thus the conditional bootstrap distributions consistently approximate the null distributions of the three statistics. With the number of bootstrap draws tending to infinity, the corresponding empirical critical values yield asymptotic level $\gamma$ tests.

\subsection{Mean-square comparison of LAG estimators}\label{sec:apr}
We compare two estimators that share the first two steps. Direct projection removes the estimated loading space from the first OLS intercept; our three-step estimator fits the additional OLS slopes before adjusting that intercept. The criterion is the mean-square error of the pricing-error estimate. Define
\[
\widehat{\mat{P}}_A=\widehat{\mat{A}}\widehat{\mat{A}}\T,
\qquad \widehat{\mat{Q}}_A=\mat{I}_p-\widehat{\mat{P}}_A,
\qquad \widehat{\vct{\alpha}}_{\rm dir}
=\widehat{\mat{Q}}_A\widehat{\vct{\mu}}.
\]
The OLS intercept satisfies
\begin{equation}\label{eq:apr-direct}
\begin{aligned}
\widehat{\vct{\alpha}}
&=\bigl(\mat{I}_p-\widetilde{\mat{A}}\widehat{\mat{A}}\T\bigr)
\widehat{\vct{\mu}}\\
&=\widehat{\vct{\alpha}}_{\rm dir}
-\widehat{\mat{Q}}_A\widehat{\mat{C}}_\eta\widehat{\mat{A}}
(\widehat{\mat{A}}\T\widehat{\mat{C}}_\eta\widehat{\mat{A}})^{-1}
\widehat{\mat{A}}\T\widehat{\vct{\mu}}.
\end{aligned}
\end{equation}
The second line identifies the correction supplied by the final regression. It uses the contemporaneous residual covariance to fit return exposures to the extracted factors, and adjusts the fitted mean compensation accordingly. The following result quantifies its effect on mean-square risk under an additional Gaussian condition.
\begin{assumption}[Mean-square comparison]\label{ass:mse}
The factor process $\vct{\chi}_t$ and the errors are Gaussian, and, for every $N$,
\[
c\mat{I}_{N(m+r)}\preceq
\Cov\{(\vct{\chi}_1\T,\ldots,\vct{\chi}_N\T)\T\}
\preceq C\mat{I}_{N(m+r)}.
\]
For the selected-rank comparison, also impose the strengthened selection condition \eqref{eq:mse-rank-condition} in Appendix~\ref{app:mse}.
\end{assumption}
Write $\mat{Q}_A=\mat{I}_p-\mat{P}_A$ and
\[
\mathcal R_{\rm dir}=\mathbb E\|\widehat{\vct{\alpha}}_{\rm dir}-\alp\|^2,
\qquad
\mathcal R_{\rm OLS}=\mathbb E\|\widehat{\vct{\alpha}}-\alp\|^2.
\]
\begin{proposition}[Mean-square efficiency of LAG post-regression]\label{prop:mse}
Under Assumptions~\ref{ass:A}--\ref{ass:lag}, \ref{ass:signal} and \ref{ass:mse}, at the true rank or at the selected rank,
\[
\frac n{\tr(\mat{Q}_A\Sg\mat{Q}_A)}
(\mathcal R_{\rm dir}-\mathcal R_{\rm OLS})
-\frac{\vct{\lambda}\T\mat{\Delta}\vct{\lambda}}{p^{1-\delta}}\longrightarrow0,
\qquad \mat{\Delta}\succeq\mat{0},
\]
where $\mat{\Delta}$ is the covariance of the difference between the
lag-based and contemporaneous factor weights, defined in
Appendix~\ref{app:mse}. The limiting normalized risk difference is nonnegative. If $\liminf_n p^{\delta-1}\vct{\lambda}\T\mat{\Delta}\vct{\lambda}>0$, the three-step estimator has strictly smaller mean-square risk for all sufficiently large $n$.
\end{proposition}
The leading gain depends on the latent compensation $\vct{\lambda}$ and on the difference between the lag-based extraction weights and the contemporaneous regression weights. When $\vct{\lambda}=\vct{0}$, this leading difference vanishes. A positive limiting quadratic form gives a strict improvement, explaining the role of the final regression when the observed-factor intercept contains latent compensation.

\section{Simulation}\label{sec:simulation}
The experiments examine calibration when observed and latent compensation vary, power as pricing errors become more or less concentrated, and the effect of factor-number selection. All experiments use $n=160$, $p=400$, two observed and two latent factors, $\delta=0$, 1,000 Monte Carlo replications and nominal level $5\%$. Methods using bootstrap calibration use 499 draws. The proposed LAG tests use the procedure in Section~\ref{sec:bootstrap}.

For comparison, residual PCA extracts the leading eigenvectors of $\widehat{\mat{C}}_\eta$, with factor number selected by $\mathrm{IC}_{p2}$ \citep{baiNg2002} unless stated otherwise. The estimated factor levels, final regression and tests are then computed as in Section~\ref{sec:method}. For this comparator, the final intercept equals $(\mat{I}_p-\widehat{\mat{A}}_{\rm pc}\widehat{\mat{A}}_{\rm pc}\T)\widehat{\vct{\mu}}$. Appendix~\ref{app:pca} collects its construction, theoretical properties and proofs.

\subsection{Data generation and implementation}\label{sec:sim-design}
We generate
\[
\vct{Y}_t=\mat{B}\vct{z}_t+\mat{A}(\vct{\lambda}_0+\vct{f}_t)+\alp+\vct{\varepsilon}_t,
\qquad \mat{A}^{\top}\mat{A}=\mat{I}_2.
\]
There are two observed and two latent factors. The columns of $\mat{A}$ are obtained by QR orthogonalization of a matrix with independent $U(-1,1)$ entries, and the entries of $\mat{B}$ are independent $U(-2,2)$. The observed factors follow
\[
\vct{z}_t=\vct{m}_z+\vct{z}_t^0,\qquad
\vct{z}_t^0=
\begin{pmatrix}5/8&1/8\\1/8&5/8\end{pmatrix}\vct{z}_{t-1}^0+\vct{u}_{z,t},
\quad \vct{u}_{z,t}\sim N_2(\vct{0},\mat{I}_2).
\]
The centered latent factors satisfy
\[
f_{j,t}=\phi_j f_{j,t-1}+\sqrt{p^{1-\delta}(1-\phi_j^2)}u_{j,t},
\qquad (\phi_1,\phi_2)=(0.8,-0.8),\qquad u_{j,t}\sim N(0,1).
\]
Both processes start in their stationary distributions, so that $\operatorname{Var}(\vct{f}_t)=p^{1-\delta}\mat{I}_2$. Innovations are independent across time and between the observed factors, latent factors, and idiosyncratic errors.

The size experiments use independent standard normal errors, $\vct{\varepsilon}_t\overset{\rm iid}{\sim}N_p(\vct{0},\mat{I}_p)$.

To assess how the proposed and benchmark methods accommodate latent risk compensation, the size experiments set $\alp=\vct{0}$ and use four configurations of our pricing null:
\[
\begin{array}{c|cc}
 & \vct{m}_z & \vct{\lambda}_0\\\hline
\mathrm{M00}&(0,0)^{\top}&(0,0)^{\top}\\
\mathrm{MZ}&(0.4,-0.3)^{\top}&(0,0)^{\top}\\
\mathrm{ML}&(0,0)^{\top}&p^{1/2}(0.6,-0.4)^{\top}\\
\mathrm{MZL}&(0.4,-0.3)^{\top}&p^{1/2}(0.6,-0.4)^{\top}.
\end{array}
\]
All four satisfy $\alp=(\mat{I}_p-\mat{A}\mat{A}^{\top})\vct{\mu}=\vct{0}$.

The proposed LAG method uses \eqref{eq:rank-selectors} with one positive lag and is labelled LAG--ER. The PCA comparator uses $\mathrm{IC}_{p2}$ and is labelled PCA--$\mathrm{IC}_{p2}$. Both searches are capped at eight factors.

The four-configuration size experiment also includes the observed-factor tests of \citet{ma2024} (labelled Ma), the location tests of \citet{wangLocation2026} (labelled WHF) applied to returns and to mean-preserving observed-factor residuals, and the PCA-based randomized test of \citet{massacci2026} (labelled Massacci PC-R). Ma uses the observed-factor regression and bandwidth two, following the formulas in arXiv:2401.14052v1. The two WHF implementations retain their original bootstrap calibration, five-lag loading estimator, rank cap ten, and rejection convention $p<0.05$. Their Cauchy procedure combines the marginal bootstrap $p$-values. Massacci PC-R is applied to $\vct{Y}_t-\widehat{\mat{B}}\vct{z}_t$, retaining its mean, with $\mathrm{IC}_{p2}$ factor selection, $\nu=5$, and the Gumbel critical value. Its premium regression includes a constant, its reported alpha subtracts only the fitted loading premium, and its pooled scale uses centered PCA residuals. Each replication uses one auxiliary Gaussian randomization, independent across replications and paired across the four mean configurations. This observed-residual implementation is an adaptation of the latent-factor procedure.

\subsection{Rejection frequencies under four configurations of the pricing null}
All four configurations satisfy our pricing null, but their implications for the benchmark tests differ. In ML and MZL, $\vct\mu=\mat A\vct\lambda_0\ne0$ generally violates Ma's zero-intercept null. In M00 and MZ the intercept is zero, while the observed-factor regression errors still contain serially dependent latent factors. WHF also differs in its treatment of observed factors and nonzero latent compensation.

Table~\ref{tab:sim-null} reports rejection frequencies under independent standard normal errors. Across the four configurations, PCA and LAG frequencies range from $4.1\%$ to $5.8\%$ and from $4.0\%$ to $6.1\%$, respectively. Both select two factors in every replication. The MZL Cauchy frequencies are $4.1\%$ for PCA and $4.3\%$ for LAG. Massacci PC-R frequencies range from $3.5\%$ to $4.5\%$.

Ma's Cauchy frequencies are $24.6\%$ and $24.0\%$ under M00 and MZ, and $100\%$ under both ML and MZL; the latter two configurations violate its zero-intercept null. WHF's direct and residual implementations give MZL Cauchy frequencies of $91.2\%$ and $98.4\%$, respectively. In the direct implementation, the rank selector retains fewer than the four combined observed and latent directions in $7.1\%$ of replications. These frequencies reflect the respective testing targets and their sensitivity to the residual factor structure. LAG--ER, PCA--$\mathrm{IC}_{p2}$ and Massacci PC-R give frequencies close to the nominal level across the four configurations of the proposed pricing null.

\begin{table}[!htbp]\centering\small
\caption{Rejection frequencies under the composite pricing null with independent standard normal errors (percent).}\label{tab:sim-null}
\setlength{\tabcolsep}{5pt}
\begin{tabular}{llrrrr}\toprule
Method & Test & M00 & MZ & ML & MZL\\\midrule
PCA--$\mathrm{IC}_{p2}$ & Sum & 5.8 & 4.6 & 4.8 & 4.4\\
 & Max & 4.7 & 5.2 & 4.7 & 4.6\\
 & Cauchy & 5.6 & 5.0 & 4.4 & 4.1\\
\addlinespace[2pt]
LAG--ER & Sum & 6.1 & 5.0 & 5.3 & 4.0\\
 & Max & 4.6 & 5.5 & 4.2 & 4.3\\
 & Cauchy & 5.9 & 5.1 & 4.3 & 4.3\\
\addlinespace[2pt]
Ma & Sum & 25.1 & 24.6 & 100.0 & 99.8\\
 & Max & 13.1 & 14.4 & 100.0 & 100.0\\
 & Cauchy & 24.6 & 24.0 & 100.0 & 100.0\\
\addlinespace[2pt]
WHF raw & Sum & 26.6 & 74.7 & 86.0 & 95.8\\
 & Max & 8.0 & 11.6 & 12.4 & 12.4\\
 & Cauchy & 21.6 & 66.7 & 78.2 & 91.2\\
\addlinespace[2pt]
WHF residual & Sum & 37.9 & 83.7 & 94.4 & 99.2\\
 & Max & 7.9 & 22.4 & 6.9 & 12.3\\
 & Cauchy & 33.5 & 81.6 & 91.2 & 98.4\\
\addlinespace[2pt]
Massacci PC-R & Randomized & 3.5 & 3.9 & 3.9 & 4.5\\
\bottomrule\end{tabular}
\par\smallskip\begin{minipage}{\linewidth}\footnotesize $n=160$, $p=400$, $\delta=0$, 1,000 replications, and nominal level $5\%$. Each bootstrap uses 499 draws. All configurations satisfy the proposed pricing null. The text describes the testing targets of the benchmark methods. Massacci entries are frequencies of the one-shot randomized test.\end{minipage}
\end{table}

\subsection{Power across signal sparsity and factor-number choices}\label{sec:sim-power}
Both power experiments use the common generator in Section~\ref{sec:sim-design} under the MZL mean configuration.
We use independent errors $\vct\varepsilon_t=\vct v_t$ and spike errors
\[
\vct\varepsilon_t=\vct v_t+\sqrt{12}\,\vct b\zeta_t,
\qquad \vct v_t\sim N_p(\vct0,\mat I_p),\quad
\zeta_t\sim N(0,1),\quad \|\vct b\|_2=1,\quad \mat A^\top\vct b=0.
\]
All innovations are independent across time and between blocks. The direction
$\vct b$ is obtained by projecting independent $U(-1,1)$ coordinates onto
the orthogonal complement of $\mat A$ and normalizing; it is fixed together
with the loadings. Thus the spike covariance is $\mat I_p+12\vct b\vct b^\top$.

\subsubsection{Statistics across signal sparsity}
Under independent errors, we fix the factor count at its true value $r=2$
for LAG and PCA while estimating the loading directions. The signal norm
is $\kappa\in\{0.6,0.8\}$ and its support size varies over
\[
s\in\{5,10,20,40,80,160,240,320,400\}.
\]
Each replication independently draws nested supports and Rademacher signs,
shared across implementations and strengths. Let $\mat E_s$ select the
support coordinates and $\mat A_s=\mat E_s^\top\mat A$. With a Rademacher
vector $\vct\xi_s$, define
\[
\begin{gathered}
\widetilde{\vct q}_s=\mat E_s
\{\mat I_s-\mat A_s(\mat A_s^\top\mat A_s)^\dagger\mat A_s^\top\}\vct\xi_s,
\\
\vct q_s=\widetilde{\vct q}_s/\|\widetilde{\vct q}_s\|_2,
\qquad \alp_s=\kappa\vct q_s.
\end{gathered}
\]
Here $\dagger$ denotes the Moore--Penrose inverse. Projecting within the
support ensures that the tested pricing error itself has $s$ nonzero
coordinates, $\mat A^\top\alp_s=0$ and $\|\alp_s\|_2=\kappa$;
the projected signal generally has unequal nonzero magnitudes.

\begin{figure}[!htbp]\centering
\includegraphics[width=\linewidth]{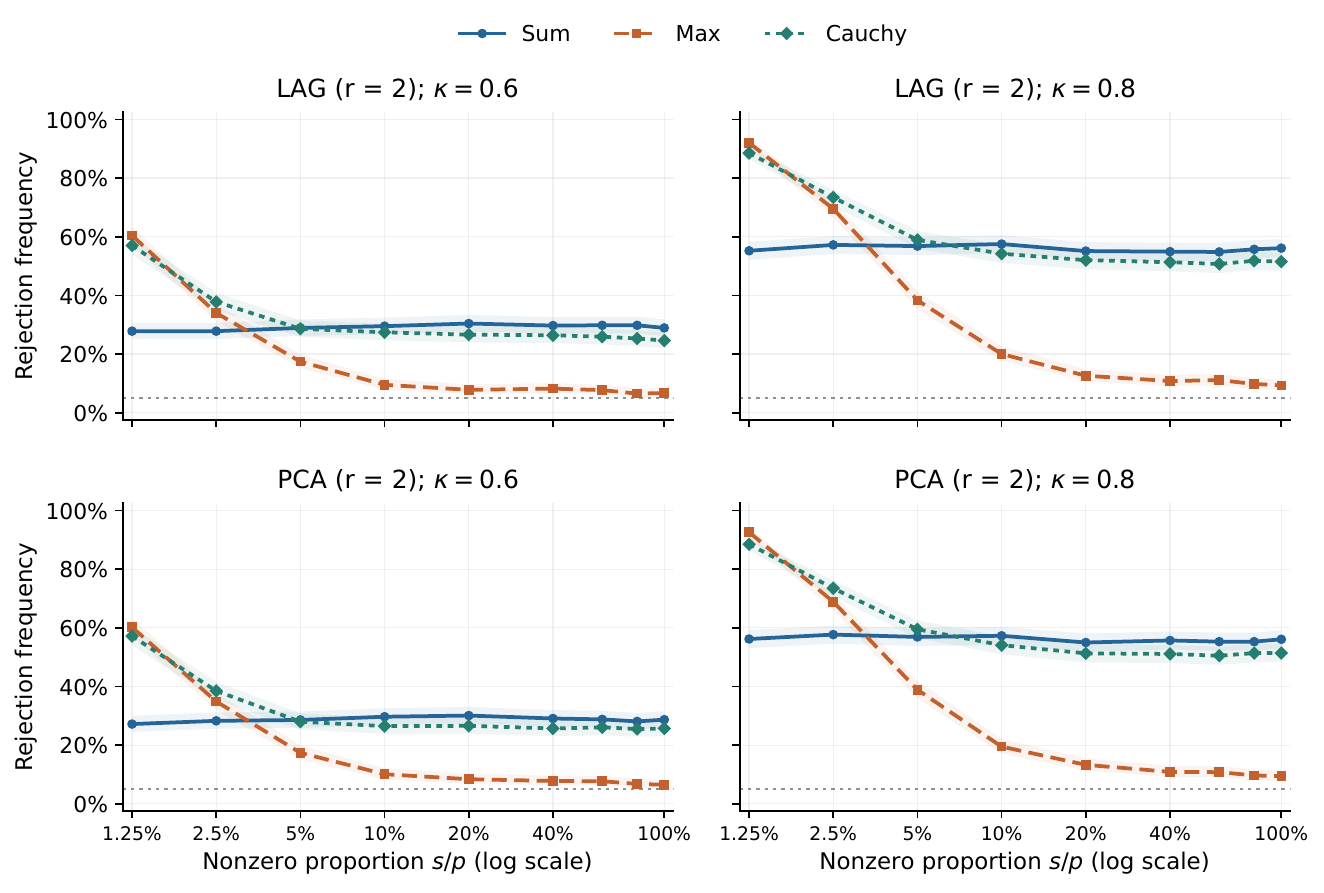}
\caption{Power across sparsity at fixed total signal norm.}\label{fig:sim-power-sparsity}
\par\smallskip\begin{minipage}{\linewidth}\footnotesize Independent errors,
$n=160$, $p=400$, $\delta=0$, 1,000 replications and 499 bootstrap draws.
Rows use LAG and PCA with known $r=2$; columns use $\kappa=0.6,0.8$.
The horizontal axis shows $s/p$ on a logarithmic scale.
Bands are pointwise $95\%$ Wilson Monte Carlo intervals.
\end{minipage}
\end{figure}

Figure~\ref{fig:sim-power-sparsity} shows that Sum power is comparatively
stable as the signal disperses, whereas Max power declines with reduced coordinate concentration.
Cauchy responds to coordinate concentration while retaining sensitivity
to diffuse signals, with a combination cost relative to the better
component. LAG and PCA exhibit the same main pattern.

\subsubsection{Loading implementation and factor-number choice}
The second experiment uses the dense signal $\alp(\kappa)=\kappa\vct b$
for $\kappa=0,0.2,\ldots,1.6$, with covariance and signal direction
held fixed; $\kappa=0$ is the MZL null. This alignment has an economic
interpretation: assets affected by common shocks or similar mechanisms
exhibit similar pricing errors, causing the overall signal direction to
overlap with the disturbance direction.
We compare LAG--ER and LAG with known $r=2$ with PCA--$\mathrm{IC}_{p2}$,
PCA with known $r=2$, and Massacci PC-R. LAG--ER applies
\eqref{eq:rank-selectors} to the lag-one matrix. Massacci PC-R uses the
implementation described in Section~\ref{sec:sim-design}, on the same
panels and signal strengths.

\begin{figure}[!htbp]\centering
\includegraphics[width=\linewidth]{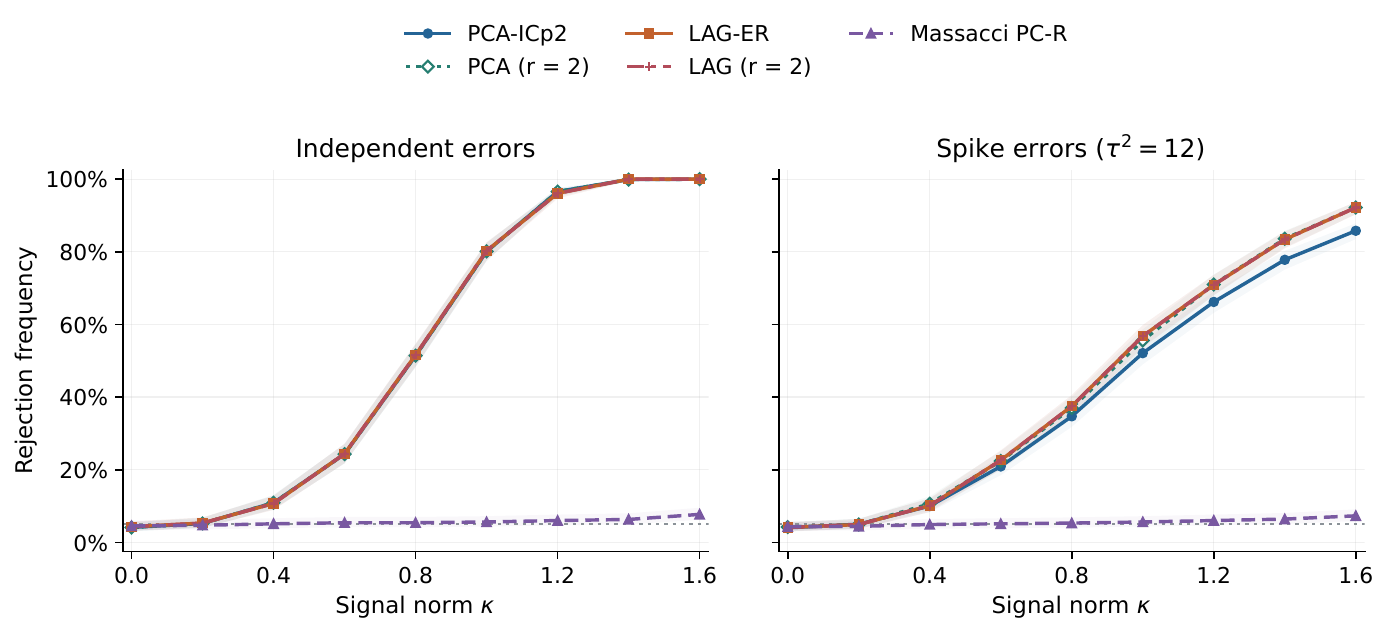}
\caption{Power comparison for a fixed dense direction.}\label{fig:sim-power-selection}
\par\smallskip\begin{minipage}{\linewidth}\footnotesize Left: independent
errors; right: spike errors with $\tau^2=12$. The signal is
$\alp=\kappa\vct b$; zero gives MZL empirical size. Other dimensions and
Monte Carlo conventions match Figure~\ref{fig:sim-power-sparsity}.
LAG and PCA use the Cauchy test; Massacci PC-R uses its randomized test.
Bands are pointwise $95\%$ Wilson intervals. Legend entries identify each implementation, including coincident curves.
\end{minipage}
\end{figure}

Figure~\ref{fig:sim-power-selection} compares the Cauchy tests for LAG and
PCA with the randomized Massacci PC-R test. Under independent errors,
the four LAG and PCA implementations have comparable power. Under spike
errors, PCA--$\mathrm{IC}_{p2}$ has lower power than LAG and known-rank PCA.
Massacci PC-R has substantially lower power over this dense-signal grid
under both error designs.

\begin{table}[!htbp]\centering\small
\caption{Factor-number selection frequencies in the power design (percent).}\label{tab:power-selection}
\begin{tabular}{lrrrrrr}\toprule
&\multicolumn{3}{c}{Independent errors}&\multicolumn{3}{c}{Spike errors}\\
\cmidrule(lr){2-4}\cmidrule(lr){5-7}
Method&$\widehat r<2$&$\widehat r=2$&$\widehat r>2$&$\widehat r<2$&$\widehat r=2$&$\widehat r>2$\\\midrule
LAG--ER & 0.0 & 100.0 & 0.0 & 0.0 & 100.0 & 0.0\\
PCA--$\mathrm{IC}_{p2}$ & 0.0 & 100.0 & 0.0 & 0.0 & 92.2 & 7.8\\
\bottomrule\end{tabular}
\par\smallskip\begin{minipage}{\linewidth}\footnotesize The table reports the data-driven selectors; known-rank implementations use two factors in every replication. Massacci PC-R uses the same residual PCA--$\mathrm{IC}_{p2}$
selection. Frequencies use 1,000 replications per error design.
\end{minipage}
\end{table}

Table~\ref{tab:power-selection} helps explain the power difference in
Figure~\ref{fig:sim-power-selection}. With independent errors, both
selectors recover the true factor number in every replication. With spike
errors, PCA--$\mathrm{IC}_{p2}$ selects more than two factors in $7.8\%$
of replications, while LAG--ER always selects two factors. The selected-
and known-rank LAG curves coincide in both designs, as do the two PCA
curves under independent errors. Each selector uses centered data, so its
choice within a replication is the same across signal strengths.

The spike contributes contemporaneous variance but no population
positive-lag covariance. After removing the observed-factor component,
the contemporaneous covariance has eigenvalues
$1+p,1+p,13,1,\ldots,1$.
Because the signal is aligned with $\vct b$, selecting an additional
noise direction can also remove pricing signal, as quantified by the
identities in Appendix~\ref{app:pca-method}.

\begin{table}[!htbp]\centering\small
\caption{MZL null rejection frequencies of LAG and PCA tests in the power design (percent).}\label{tab:power-null}
\begin{tabular}{lrrrrrr}\toprule
&\multicolumn{3}{c}{Independent errors}&\multicolumn{3}{c}{Spike errors}\\
\cmidrule(lr){2-4}\cmidrule(lr){5-7}
Method&Sum&Max&Cauchy&Sum&Max&Cauchy\\\midrule
LAG--ER & 4.0 & 4.3 & 4.3 & 4.0 & 4.6 & 4.1\\
LAG, $r=2$ & 4.0 & 4.3 & 4.3 & 4.0 & 4.6 & 4.1\\
PCA--$\mathrm{IC}_{p2}$ & 4.4 & 4.6 & 4.1 & 4.2 & 4.1 & 4.0\\
PCA, $r=2$ & 4.4 & 4.6 & 4.1 & 4.5 & 4.2 & 4.2\\
\bottomrule\end{tabular}
\par\smallskip\begin{minipage}{\linewidth}\footnotesize The $\kappa=0$
cases of the selection experiment; nominal level $5\%$, 1,000 replications
and 499 bootstrap draws. All tests use their nominal 5\% critical values.
\end{minipage}
\end{table}

\Needspace{5\baselineskip}
The independent-error MZL entries in Table~\ref{tab:power-null} coincide with those in Table~\ref{tab:sim-null}: the two tables use identical outer panels and bootstrap streams for the corresponding implementations. Table~\ref{tab:power-null} reports Cauchy null frequencies between $4.0\%$ and $4.3\%$. Massacci PC-R has null rejection frequencies of $4.5\%$ and $4.3\%$ under independent and spike errors, respectively. All these observed frequencies are below the nominal level in this finite sample. 

\section{Real data}\label{sec:realdata}
\subsection{Data and diagnostics}
We use weekly excess returns for 378 stocks from January 14, 2005 to December 26, 2025, giving 1,094 observations per stock. The balanced sample comprises end-of-period S\&P 500 constituents with complete adjusted-price histories. Weekly returns are aligned with Kenneth French's factor file and net of the weekly risk-free return. The observed factors are market excess return, SMB, and HML \citep{FrenchData}. We use the complete aligned weekly series and retain the original return observations.

We move a 260-week window forward one week at a time, obtaining 835 windows with $p=378>n=260$. The table below displays the first and last rolling windows and three interior snapshots ending in 2014, 2019 and 2023, spanning different parts of the sample. Figure~\ref{fig:sp500-components} shows all 835 windows. Within each window, $\vct{\mu}$ is the FF3 intercept from Section~\ref{sec:model}. Ma tests the FF3 restriction $\vct{\mu}=0$, whereas LAG tests the latent-adjusted restriction $\mat{Q}_A\vct{\mu}=0$. The comparison asks whether rejection of FF3 pricing persists after allowing latent compensation.

\Needspace{3\baselineskip}
We implement LAG as in Section~\ref{sec:method}, using the calibration in Section~\ref{sec:bootstrap} with 499 bootstrap draws. Ma uses the analytical calibration with lag bandwidth $b_{\rm Ma}=3$. All methods use identical data.

\noindent\textit{Residual serial-correlation diagnostics.}
We compare FF3 residuals with the final residuals after LAG adjustment using stock-level Ljung--Box tests at three lags and Benjamini--Hochberg (BH) 5\% discovery proportions \citep{LjungBox1978,BH1995}.

\begin{table}[!htbp]\centering\small
\caption{Residual serial correlation: BH 5\% discovery proportions.}
\label{tab:residual-lb3}
\begin{tabular*}{\linewidth}{@{\extracolsep{\fill}}lrr@{}}\toprule
Window end & FF3 & FF3+LAG \\ \midrule
2010-01-01 & 10.05 & 2.38 \\
2014-11-21 & 0.00 & 0.00 \\
2019-11-22 & 0.00 & 0.00 \\
2023-11-24 & 4.23 & 1.06 \\
2025-12-26 & 0.00 & 0.00 \\
\bottomrule\end{tabular*}
\vspace{4pt}\begin{minipage}{\linewidth}\footnotesize
\textit{Notes:} Entries are percentages of 378 stocks. Stock-level Ljung--Box tests use three lags and the $\chi^2_3$ reference; BH adjustment is applied within each window and residual series. All windows contain 260 weeks.
\end{minipage}\end{table}

After LAG adjustment, the discovery proportion is at most 2.38\% in the five windows (Table~\ref{tab:residual-lb3}). The largest reductions occur in the windows ending in 2010 and 2023. These diagnostics indicate limited residual serial correlation at the examined lags and provide an empirical check on the temporal specification used for calibration.

\subsection{Empirical results and interpretation}\label{sec:sp500-results}
Table~\ref{tab:sp500-main} reports Sum, Max, and Cauchy $p$-values for the five selected windows.

\begin{table}[!htbp]\centering\small
\caption{Five-window comparison: Ma and LAG $p$-values.}
\label{tab:sp500-main}
\begin{tabular*}{\linewidth}{@{\extracolsep{\fill}}llrrrr@{}}\toprule
Method & Calibration & $\widehat r$ & Sum & Max & Cauchy \\
\midrule
\multicolumn{6}{@{}l}{\textit{2005-01-14 to 2010-01-01}} \\[2pt]
Ma & Analytical & -- & $0.0982$ & $0.0727$ & $0.0836$ \\
LAG & Bootstrap (499) & 4 & $0.3840$ & $0.9560$ & $0.9240$ \\
\midrule
\multicolumn{6}{@{}l}{\textit{2009-12-04 to 2014-11-21}} \\[2pt]
Ma & Analytical & -- & $\mathbf{1.081\times10^{-4}}$ & $0.0589$ & $\mathbf{2.158\times10^{-4}}$ \\
LAG & Bootstrap (499) & 2 & $0.7740$ & $0.8140$ & $0.7980$ \\
\midrule
\multicolumn{6}{@{}l}{\textit{2014-12-05 to 2019-11-22}} \\[2pt]
Ma & Analytical & -- & $\mathbf{0.0379}$ & $0.2011$ & $0.0645$ \\
LAG & Bootstrap (499) & 2 & $\mathbf{0.0500}$ & $0.6880$ & $0.0740$ \\
\midrule
\multicolumn{6}{@{}l}{\textit{2018-12-07 to 2023-11-24}} \\[2pt]
Ma & Analytical & -- & $0.7337$ & $\mathbf{7.319\times10^{-7}}$ & $\mathbf{1.464\times10^{-6}}$ \\
LAG & Bootstrap (499) & 1 & $0.9020$ & $0.7900$ & $0.8380$ \\
\midrule
\multicolumn{6}{@{}l}{\textit{2021-01-08 to 2025-12-26}} \\[2pt]
Ma & Analytical & -- & $0.2607$ & $0.1473$ & $0.1902$ \\
LAG & Bootstrap (499) & 1 & $0.5140$ & $0.3240$ & $0.4240$ \\
\bottomrule\end{tabular*}
\vspace{4pt}\begin{minipage}{\linewidth}\footnotesize
\textit{Notes:} Bold entries indicate $p\leq0.05$. These windows are exact subsets of the rolling calculations in Figure~\ref{fig:sp500-components}.
\end{minipage}\end{table}
\FloatBarrier
Ma's Cauchy test rejects in the windows ending in 2014 and 2023. The corresponding LAG Cauchy $p$-values are $0.798$ and $0.838$, respectively. In 2019, the Sum $p$-values are $0.0379$ for Ma and $0.050$ for LAG. The latter is exactly at the prescribed 5\% rejection boundary.

\begin{figure}[!htbp]\centering
\includegraphics[width=\linewidth]{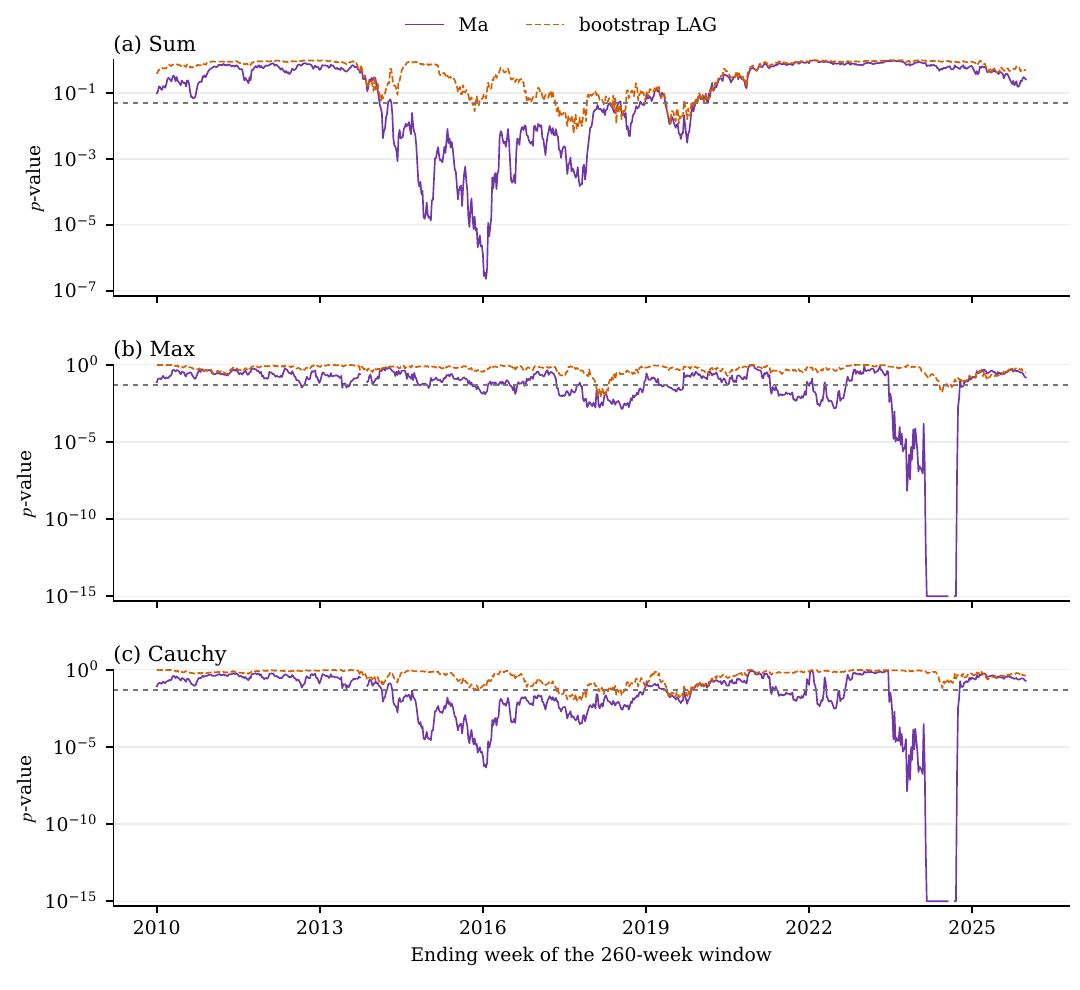}
\caption{Rolling $p$-values for Sum, Max, and Cauchy tests. Each window contains 260 weeks and 378 stocks. Dates indicate window endpoints; the dashed horizontal line marks 0.05. Panel-specific vertical axes are logarithmic. Values below $10^{-15}$ are placed at $10^{-15}$ for display. Unavailable Ma values remain gaps.}
\label{fig:sp500-components}\end{figure}

Figure~\ref{fig:sp500-components} places these snapshots in the full sequence of 835 rolling windows. LAG $p$-values are available in every window. In 11 windows, Ma's coordinate long-run variance estimate is nonpositive for at least one stock, leaving its Max statistic and the dependent Cauchy combination undefined. These values appear as gaps. The sequence describes overlapping windows and their changing pricing assessments.

The windows ending in 2014 and 2023 illustrate the pricing question addressed by latent adjustment. Ma's Cauchy test rejects zero FF3 intercepts in both windows, while the LAG results are compatible with $\vct{\mu}=\mat{A}\vct{\lambda}$ after selecting two and one latent factors, respectively. Other rolling windows retain Cauchy rejections after adjustment, indicating evidence of a mean component outside the estimated latent loading space. These outcomes describe how the assessment of the FF3 intercept changes when latent compensation is allowed. Their economic interpretation concerns compatibility with that pricing restriction, with the factor space estimated from return dynamics.

\FloatBarrier

\section{Conclusion}\label{sec:conclusion}
We develop tests of a pricing restriction that allows latent compensation in addition to an observed traded-factor benchmark. The target is the component of the benchmark intercept orthogonal to the latent loading space. A three-step procedure combines observed-factor OLS, residual LAG eigenanalysis and augmented OLS to estimate this pricing error. Sum and Max statistics measure diffuse and concentrated departures, and their Cauchy combination accommodates uncertainty about the signal pattern.

The theory establishes first-order oracle approximations, joint null limits and power under the stated signal conditions. A residual bootstrap provides a further calibration with first-order validity, and a mean-square comparison explains the contribution of the final regression. The simulations illustrate the complementary power of Sum and Max and the effect of factor-number selection on the comparisons with residual PCA. In the S\&P 500 application, the tests identify windows in which FF3-adjusted mean returns are compatible with latent compensation and windows in which evidence of pricing errors persists after adjustment.

The framework uses a fixed loading space over each observation window and identifies latent factors through positive-lag dependence. Extensions with time-varying or characteristic-driven loadings could adapt the estimation step while retaining the same decomposition of expected returns.

\urlstyle{same}
\bibliographystyle{plainnat}
\bibliography{reference}

\clearpage
\appendix
\counterwithin{table}{section}
\renewcommand{\thetable}{\Alph{section}.\arabic{table}}
\renewcommand{\thelemma}{\Alph{section}.\arabic{lemma}}
\renewcommand{\thesubsection}{\Alph{section}.\arabic{subsection}}
\counterwithin{proposition}{section}
\counterwithin{theorem}{section}
\renewcommand{\theHproposition}{appendix.\Alph{section}.\arabic{proposition}}
\renewcommand{\theHtheorem}{appendix.\Alph{section}.\arabic{theorem}}
\renewcommand{\theHtable}{appendix.\Alph{section}.\arabic{table}}

\section{Preliminary results for the testing theory}\label{app:technical}
We introduce common notation and the loading condition, then establish
moment bounds, LAG loading accuracy, regression geometry, covariance
scales, statistical approximation and the oracle joint limit.
These results supply the proofs in Appendices~\ref{app:mainproofs}
and~\ref{app:bootstrap-proof}, and the PCA analysis in Appendix~\ref{app:pca}.

\subsection{Notation for the proofs}\label{app:proof-notation}

All fitted quantities below are evaluated at the true rank unless stated
otherwise. In distributional statements, a statistic is set to zero if a
required inverse or positive scale is unavailable. Proposition~\ref{prop:regularity}
and the bootstrap regularity argument show that these extensions occur
with probability tending to zero. In numerical calculations, undefined statistics are recorded as unavailable.
Write $\mat{Z}_c=\mat{Z}-\one_n\bar{\vct{z}}\T$ and
\[
\mat{M}_1=\mat{I}_n-\one_n\one_n\T/n,\qquad
\mat{M}_c=\mat{M}_1-\mat{Z}_c(\mat{Z}_c\T\mat{Z}_c)^{-1}\mat{Z}_c\T.
\]
Then $\Rmat=\Ymat\mat{M}_c$. Put
\[
\hP=\widehat{\mat{A}}\widehat{\mat{A}}\T,\quad
\hQ=\mat{I}_p-\hP,\quad \mat{\Delta}_P=\hP-\mat{P}_A,
\]
\[
\widehat{\vct{\ell}}=\widehat{\mat{A}}\T\widehat{\vct{\mu}},\quad
\widehat{\mat{K}}_g=\widehat{\mat{A}}\T\widehat{\mat{C}}_\eta\widehat{\mat{A}},\quad
c_{z,n}=1+\bar{\vct{z}}\T\mat{S}_z^{-1}\bar{\vct{z}},\quad
\widehat{\mathcal I}=\one_n\T\widehat{\mat{M}}_x\one_n.
\]

The relation $X_n\asymp_{\mathbb P} a_n$ means that $X_n/a_n$ and its reciprocal are $O_\mathbb P(1)$; $O_\mathbb P(a_n)$ denotes tightness after division by $a_n$. The coordinate vector is $\vct{e}_i$.
Put $\Lnp=\log(2np)$ and $\rn=p^{\delta/2}/\sqrt n$. Define
\[
\begin{gathered}
\epsT=\sfac^{-1}+\frac{\rn p^{\delta/2}}{\sqrt n},
\qquad \bvec=\sqrt n\,\epsT,\\
\hT=\epsT^2+(\sqrt n\sfac)^{-1}
+\frac{\rn p^{\delta/2}}{\sqrt n},\\
\vdiag=\sqrt{\Lnp/n}+\Lnp/n+\epsT,\\
\bsum=\bvec+\sqrt p\,\rn^2+n^{-1/2},
\quad\bmax=\bvec\sqrt{\Lnp}+\vdiag\Lnp.
\end{gathered}
\]

For the proofs, write
\[
\widehat c=n/\widehat{\mathcal I},\qquad
\mnoise=\tr(\Sg),\qquad
\taunoise=\tr(\Sg^2),\qquad
\snoise=\sqrt{2\taunoise}.
\]
Also put
\[
\vct{\beta}=\mat{M}_0\one_n/\sqrt{\Ieff},\qquad
\vct{\xi}_n^\circ=\Em\vct{\beta}.
\]
\[
Q_n^\circ=\frac{\|\vct{\xi}_n^\circ\|^2-\mnoise}{\snoise},\qquad
M_n^\circ=\max_i\frac{(\xi_{n,i}^\circ)^2}{(\Sg)_{ii}}
-2\log p+\log\log p.
\]
Under $H_0$, these equal $T_{\mathrm{sum}}^\circ$ and
$T_{\mathrm{max}}^\circ$, respectively. For estimating $\tr(\Sg^2)$, set
\[
\widehat v_\varepsilon=
\frac{k}{k+1}\left\{\tr(\widehat{\mat{\Sigma}}_\varepsilon^2)
-\frac{\tr(\widehat{\mat{\Sigma}}_\varepsilon)^2}{k}\right\}.
\]
Starred fitted quantities, including $\widehat v_\varepsilon^*$, are defined by the same formulas on each bootstrap sample.
Retain $\mat{M}_0$, $\Ieff$, $\Delta_n$ and $\cstar$ from
Section~\ref{sec:common-approximation}.

\subsection{Loading condition for the common bounds}\label{app:loading-condition}
For the common estimation bounds, consider a true-rank loading estimator
with orthonormal columns, measurable with respect to the centered residual
matrix and invariant to an intercept shift. The required loading-space
rate is
\begin{equation}\label{eq:admission}
\|\widehat{\mat{P}}_A^{[r]}-\mat{P}_A\|=\Op(\rn),
\quad \rn=p^{\delta/2}/\sqrt n,\quad \sqrt p\,\rn^2\to0.
\end{equation}
The rate relations below follow from Assumption~\ref{ass:D}.
The LAG argument verifies the loading condition under
Assumptions~\ref{ass:A}--\ref{ass:lag}; the PCA verification is given
with its construction in Appendix~\ref{app:pca}.

\subsection{Supporting identities and rate bounds}\label{app:prelim}
\subsubsection{Rate comparisons}
For $p\asymp n^\kappa$, the admissible window is
$1/(1-\delta)<\kappa<2/(1+2\delta)$; it becomes $1<\kappa<2$ when
$\delta=0$. The fixed positive margin absorbs logarithmic factors in the resampling bounds.
The loading rate specified in \eqref{eq:admission} is
$\rho_n=p^{\delta/2}/\sqrt n$; the two growth restrictions control
$n/\sfac^2$ and $\sqrt p\,\rho_n^2$, respectively. Multiplying the two vanishing ratios in \eqref{eq:growth} gives $p^{2\delta-1/2}$, yielding the strict bound $\delta<1/4$. At the boundary, the power-law window collapses to $\kappa=4/3$.

\begin{lemma}[Rate relations]\label{lem:rates}
Under \eqref{eq:growth}, with $\rn=p^{\delta/2}/\sqrt n$,
\[
p/n\to\infty,\qquad \rn,\epsT,\bsum,\bmax\to0,
\]
\[
\hT=O\{(\sqrt n\sfac)^{-1}+\rn p^{\delta/2}/\sqrt n\},
\qquad \rn^2p/n\to0.
\]
\end{lemma}
\begin{proof}
Put $u_n=n/p^{1-\delta}$ and $v_n=p^{1/2+\delta}/n$.
Assumption~\ref{ass:D} gives $u_n+v_n=O(n^{-\eta})$ and $\Lnp=O(\log n)$.
The definition of $\rn$ gives $\epsT=\sfac^{-1}+p^\delta/n$. Thus
\[
\frac np\le u_n,\quad \rn^2=\frac{v_n}{\sqrt p},\quad
\bvec^2\le2u_n+2v_np^{\delta-1/2},\quad
\sqrt p\,\rn^2=v_n,\quad \frac{\rn^2p}{n}=v_n^2p^{-\delta}\le v_n^2.
\]
Consequently $\bvec\sqrt{\Lnp}\to0$, $\bsum\to0$, and
\[
\vdiag\Lnp=\frac{\Lnp^{3/2}}{\sqrt n}+\frac{\Lnp^2}{n}
+\frac{\bvec\Lnp}{\sqrt n}\to0.
\]
For $b_n=\rn p^{\delta/2}/\sqrt n\to0$, $\sfac/\sqrt n\to\infty$ gives
$\epsT^2\le2\sfac^{-2}+2b_n^2
=O\{(\sqrt n\sfac)^{-1}+b_n\}$, proving the bound for $\hT$.

\end{proof}

\subsubsection{The post-regression identities}

For any nonsingular $r\times r$ matrix $\mat{H}$ and
$\mat{K}=\widehat{\mat{A}}\T\widehat{\mat{C}}_\eta\widehat{\mat{A}}$,
\[
\widehat{\mat{A}}\mat{H}(\mat{H}\T\mat{K}\mat{H})^{-1}
\mat{H}\T\widehat{\mat{A}}\T
=\widehat{\mat{A}}\mat{K}^{-1}\widehat{\mat{A}}\T.
\]
Thus changing the loading basis preserves the factor time span, intercept,
information factor, and test statistics.

\begin{lemma}[Post-regression identities]\label{lem:fwl}
Whenever the required inverses exist,
\[
\widehat{\vct{\alpha}}=\frac{\Ymat\widehat{\mat{M}}_x\one_n}{\widehat{\mathcal I}},
\qquad \widehat c=\frac{n}{\widehat{\mathcal I}}.
\]
\end{lemma}
\begin{proof}
Let $\widehat{\mat{G}}_c=\widehat{\mat{A}}\T\Rmat$ and
\[
\vct{h}_*=n^{-1}(\one_n-\mat{Z}_c\mat{S}_z^{-1}\bar{\vct{z}})
-n^{-1}\widehat{\mat{G}}_c\T
\widehat{\mat{K}}_g^{-1}\widehat{\vct{\ell}}.
\]
Centering and the normal equations give
\[
\begin{gathered}
\widehat{\mat{G}}_c\one_n=0,\quad
\widehat{\mat{G}}_c\mat{Z}_c=0,\quad
\widehat{\mat{G}}_c\widehat{\mat{G}}_c\T=n\widehat{\mat{K}}_g,\\
\one_n\T\vct{h}_*=1,\quad\widehat{\mat{D}}_x\T\vct{h}_*=0,\\
\Ymat\vct{h}_*=\widehat{\vct{\mu}}-
\widehat{\mat{C}}_\eta\widehat{\mat{A}}\widehat{\mat{K}}_g^{-1}
\widehat{\vct{\ell}}=\widehat{\vct{\alpha}},\qquad
n\|\vct{h}_*\|^2=c_{z,n}+\widehat{\vct{\ell}}\T
\widehat{\mat{K}}_g^{-1}\widehat{\vct{\ell}}=\widehat c.
\end{gathered}
\]
Substituting $\widetilde{\mat{A}}$ and
$\bar{\widehat{\vct{F}}}=\widehat{\mat{A}}\T\widehat{\vct{\mu}}$
into \eqref{eq:post} gives \eqref{eq:apr-direct}.

Since $\vct{h}_*\in\mathcal M([\one_n,\widehat{\mat{D}}_x])$ and
$\widehat{\mat{D}}_x\T\vct{h}_*=0$, normalization gives
\[
\vct{h}_*=\frac{\widehat{\mat{M}}_x\one_n}{\widehat{\mathcal I}},
\qquad \|\vct{h}_*\|^2=\widehat{\mathcal I}^{-1},
\]

\end{proof}

\subsection{Moment bounds and consistency of the loading estimators}\label{app:moments}
\subsubsection{Conditional localization and independent moments}
The notation $\Ex$ integrates only over the innovations with the factor realization fixed. For $a>0$, deterministic $b_n>0$, and factor-measurable $W_n\ge0$ with $W_n=\Op(1)$ and $\Ex|V_n|^a\le b_n^aW_n$,
\begin{equation}\label{eq:local}
\mathbb P(|V_n|>Kb_n)\le \mathbb P(W_n>M)+M/K^a.
\end{equation}
This follows by integrating the conditional Markov bound on $\{W_n\le M\}$.

For a centered innovation coordinate $X$ and an independent copy $X'$,
sub-Gaussian tails give $\mathbb E|X-X'|^{2j}\le C^j j!$.
Symmetrization and the exponential series therefore imply
\[
\mathbb E e^{tX}\le\mathbb E\cosh\{t(X-X')\}
\le\sum_{j\ge0}(Ct^2)^j/j!=e^{Ct^2}.
\]
Multiplication over independent coordinates, Chernoff's bound, and tail
integration give, for deterministic $\vct{b}$ and fixed $a>0$,
\[
\mathbb E e^{t\vct{b}\T\vct{\xi}}\le e^{Ct^2\|\vct{b}\|^2},\qquad
\mathbb P(|\vct{b}\T\vct{\xi}|>x)\le2e^{-cx^2/\|\vct{b}\|^2},\qquad
\mathbb E|\vct{b}\T\vct{\xi}|^a\le C_a\|\vct{b}\|^a.
\]
For symmetric $\mat{H}$,
\[
\vct{\xi}\T \mat{H}\vct{\xi}-\tr \mat{H}
=\sum_j\mat{H}_{jj}(\xi_j^2-1)+2\sum_{j<l}\mat{H}_{jl}\xi_j\xi_l.
\]
Cross-products containing an innovation index that occurs only once have zero expectation. Thus
\begin{equation}\label{eq:quadmoment}
\Var(\vct{\xi}\T \mat{H}\vct{\xi})=2\tr(\mat{H}^2)+\sum_j\kappa_j\mat{H}_{jj}^2
\le C\|\mat{H}\|_F^2,\qquad\kappa_j=\mathbb E\xi_{1j}^4-3.
\end{equation}

\begin{external}[Hanson--Wright]\label{ext:hw}
For independent centered inputs with uniformly bounded $\psi_2$ norms and a deterministic symmetric matrix, \citet[Theorem 1.1]{hw} give
\[
\mathbb P(|\vct{\xi}\T \mat{H}\vct{\xi}-\mathbb E\vct{\xi}\T \mat{H}\vct{\xi}|>x)
\le2\exp\{-c\min(x^2/\|\mat{H}\|_F^2,x/\|\mat{H}\|)\}.
\]
\end{external}

\subsubsection{The time-domain noise matrix}
\begin{lemma}[Matrix and directional moment bounds]\label{lem:noise}
Write $\mat{K}_\varepsilon=\Em\T\Em-\mnoise \mat{I}_n$. For a factor-measurable unit vector $\vct{a}$ and a factor-measurable orthogonal projection $\mat{M}$ of rank $k\asymp n$,
\[
\|\mat{K}_\varepsilon\|=\Op(\sqrt{np}+n),\quad
\|\mat{K}_\varepsilon \vct{a}\|=\Op(\sqrt{np}),\quad
\|\mat{A}\T\Em \vct{a}\|=\Op(1),
\]
\[
\|\mat{A}\T\Em\|=\Op(\sqrt n),\quad\|\Em\|=\Op(\sqrt n+\sqrt p),
\quad\max_i\|\vct{e}_i\T\Em\|=\Op(\sqrt{n+\Lnp}),
\]
\[
\tr(\mat{M}\mat{K}_\varepsilon)=\Op(\sqrt{np}).
\]
\[
\max_i\left|\frac{\vct{e}_i\T\Em \mat{M}\Em\T \vct{e}_i}{k}-(\Sg)_{ii}\right|
=\Op\{\sqrt{\Lnp/n}+\Lnp/n\}.
\]
\end{lemma}
\begin{proof}
For fixed $\|\vct{u}\|=1$, the stacked-innovation coefficient matrix of
$\vct{u}\T\mat{K}_\varepsilon\vct{u}$ is
$(\vct{u}\vct{u}\T)\otimes\Sg$. Hanson--Wright gives
$\mathbb P(|\vct{u}\T\mat{K}_\varepsilon\vct{u}|>x)
\le2e^{-c\min(x^2/p,x)}$.
A $1/4$-net $\mathcal N$ of the unit sphere has $|\mathcal N|\le9^n$
by volume packing, and
$\|\mat{K}\|\le2\max_{\vct{u}\in\mathcal N}|\vct{u}\T\mat{K}\vct{u}|$
for symmetric $\mat{K}$. Hence, for sufficiently large $A$,
\[
\mathbb P\{\|\mat{K}_\varepsilon\|>2A(\sqrt{np}+n)\}
\le2e^{n\log9-cAn},\quad
\|\Em\|^2\le\mnoise+\|\mat{K}_\varepsilon\|=\Op(p+n).
\]
Conditional time-index pairing and \eqref{eq:quadmoment} yield
\[
\begin{gathered}
\Ex|(\mat{K}_\varepsilon\vct{a})_t|^2
=\taunoise\sum_{s\ne t}a_s^2+a_t^2\Var(\|\vct{\varepsilon}_t\|^2)
\le Cp,\\
\Ex\|\mat{A}\T\Em\vct{a}\|^2=\tr(\mat{A}\T\Sg\mat{A})\le Cr,
\qquad \mathbb E\|\mat{A}\T\Em\|_F^2\le Cnr,\\
\Var_\xi\{\tr(\mat{M}\mat{K}_\varepsilon)\}\le Ck\taunoise.
\end{gathered}
\]
Markov's inequality proves the directional, projected-noise and trace bounds.
Each error row has independent sub-Gaussian time coordinates. Thus
\[
\mathbb P_\xi\{ |\|\vct{e}_i\T\Em\mat{M}\|^2-k(\Sg)_{ii}|>x\}
\le2e^{-c\min(x^2/k,x)}.
\]
Set $x=C\{\sqrt{k\Lnp}+\Lnp\}$ and take a union bound over $i$.
Taking $\mat{M}=\mat{I}_n$ also gives the row-norm bound.
\end{proof}

\subsubsection{Mixing and sample moments}

\begin{lemma}[Finite-window moments]\label{lem:weak-window}
Suppose $V_t$ is measurable with respect to
$\sigma(\vct{\chi}_u,\vct{\xi}_u:t\le u\le t+\ell_0)$,
where $\ell_0$ is fixed, and $\sup_t \mathbb E|V_t|^4\le C_0$. Then, uniformly in $a\in\mathbb Z$ and $N\ge1$,
\[
\mathbb E\left|\frac1N\sum_{t=a+1}^{a+N}(V_t-\mathbb E V_t)\right|^2\le\frac{C}{N}.
\]
\end{lemma}
\begin{proof}
For past and future events separated by $h$, let $a_1,a_2$ be their
conditional probabilities given the factors. Then
\[
|\mathbb P(E_1\cap E_2)-\mathbb P(E_1)\mathbb P(E_2)|
=|\Cov(a_1,a_2)|\le\alpha_{\rm mix}(h),
\]
by integrating $\indc\{a_j>u\}$ over $u\in[0,1]$.
Let $\phi_M(v)=\max(-M,\min(v,M))$. Uniform fourth moments and
Cauchy--Schwarz give
\[
\|V_t-\phi_M(V_t)\|_2\le C/M,\qquad
|\Cov(V_s,V_t)|\le CM^{-1}+4M^2\alpha_{\rm mix}(|t-s|-\ell_0)
\]
when $|t-s|>\ell_0$. The covariance of the truncated variables is bounded
by integrating their level-set indicators over $[-M,M]^2$.
Taking $M=e^{c(|t-s|-\ell_0)/3}$ gives
$|\Cov(V_s,V_t)|\le Ce^{-c(|t-s|-\ell_0)/3}$; for the remaining pairs
Cauchy--Schwarz gives a constant bound. Therefore
\[
\frac1{N^2}\sum_{s,t=a+1}^{a+N}|\Cov(V_s,V_t)|
\le\frac{C}{N}\left(2\ell_0+1+2\sum_{h\ge1}e^{-ch/3}\right)
\le C/N.
\]
\end{proof}

\begin{lemma}[Stable second-order sample moments]\label{lem:weak-moments}
Under Assumption~\ref{ass:A}, for $n>2k_0$,
\[
\mathbb E\left\|\frac1n\sum_{t=1}^n\vct{\chi}_t\right\|^2
+\max_{0\le h\le k_0}\mathbb E\left\|
\frac1{n-h}\sum_{t=1}^{n-h}
\{\vct{\chi}_{t+h}\vct{\chi}_t\T-\mathbb E(\vct{\chi}_{t+h}\vct{\chi}_t\T)\}
\right\|_F^2
\le C/n.
\]
\end{lemma}
\begin{proof}
Apply Lemma~\ref{lem:weak-window} to $\chi_{t,j}$ and
$\chi_{t+h,i}\chi_{t,j}$, whose fourth moments are uniformly bounded.
Their expectations are $0$ and $\mathbb E(\chi_{t+h,i}\chi_{t,j})$.
Sum the bounds $C/n$ and $C/(n-h)\le2C/n$ over the fixed coordinates.
\end{proof}

\subsubsection{Common regression and residual covariance}
\begin{lemma}[Centered regression and covariance bounds]\label{lem:regression}
Let $\vct{\eta}_t=\mat{A}\vct{f}_t+\vct{\varepsilon}_t$, $\mat{C}_\eta=\mat{A}\mat{\Gamma}_f(0)\mat{A}\T+\Sg$, and $\mat{E}_B=\widehat{\mat{B}}-\mat{B}$. Under Assumptions~\ref{ass:A}--\ref{ass:C},
\[
\|\mat{E}_B\|_F=\Op(\sqrt{p/n}),\qquad
\|\widehat{\mat{C}}_\eta-\mat{C}_\eta\|_F=\Op(p/\sqrt n).
\]
\end{lemma}
\begin{proof}
Loading dispersion gives $\|\vct{e}_i\T \mat{A}\|\sfac\le C$. Thus $\sup_{t,i}\|\eta_{t,i}\|_{\psi_2}\le C$. Lemmas~\ref{lem:weak-moments} and~\ref{lem:weak-window} imply
\[
\|\mat{S}_z-\mat{\Gamma}_z\|=\Op(n^{-1/2}),\quad
\|\bar{\vct{z}}-\vct{m}_z\|=\Op(n^{-1/2}),\quad
\|\bar{\vct{\eta}}\|=\Op(\sqrt{p/n}).
\]
The product fourth moments are bounded, and $\mathbb E\vct{\eta}_t\vct{z}_t^{0\mathsf T}=0$ at every time,
\[
\mathbb E\left\|n^{-1}\sum_t\vct{\eta}_t\vct{z}_t^{0\mathsf T}\right\|_F^2\le Cp/n,
\qquad
\mathbb E\left\|n^{-1}\sum_t(\vct{\eta}_t\vct{\eta}_t\T-\mat{C}_\eta)\right\|_F^2\le Cp^2/n.
\]
Set $\mat{V}_{z\vct{\eta}}=n^{-1}\sum_t(\vct{\eta}_t-\bar{\vct{\eta}})(\vct{z}_t-\bar{\vct{z}})\T$. On $\lambda_{\min}(\mat{S}_z)\ge c/2$,
\[
\mat{E}_B=\mat{V}_{z\vct{\eta}}\mat{S}_z^{-1},\quad
\mat{V}_{z\vct{\eta}}=n^{-1}\sum_t\vct{\eta}_t\vct{z}_t^{0\mathsf T}-\bar{\vct{\eta}}\bar{\vct{z}}^{0\mathsf T},
\quad\|\mat{V}_{z\vct{\eta}}\|_F=\Op(\sqrt{p/n}).
\]
The centered normal equations give the exact covariance identity
\[
\widehat{\mat{C}}_\eta=n^{-1}\sum_t(\vct{\eta}_t-\bar{\vct{\eta}})(\vct{\eta}_t-\bar{\vct{\eta}})\T
-\mat{V}_{z\vct{\eta}}\mat{S}_z^{-1}\mat{V}_{z\vct{\eta}}\T.
\]
Consequently,
\[
\begin{aligned}
\|\widehat{\mat{C}}_\eta-\mat{C}_\eta\|_F
&\le\left\|n^{-1}\sum_t(\vct{\eta}_t\vct{\eta}_t\T-\mat{C}_\eta)\right\|_F\\
&\quad+\|\bar{\vct{\eta}}\|^2+\|\mat{V}_{z\vct{\eta}}\|_F^2\|\mat{S}_z^{-1}\|\\
&=\Op(p/\sqrt n+p/n).
\end{aligned}
\]
\end{proof}

\subsubsection{Spectral inequalities}
The variational characterizations of eigenvalues and singular values give
\[
|\lambda_j(\mat{T}+\mat{E})-\lambda_j(\mat{T})|\le\|\mat{E}\|,
\qquad
\min_{\rank\mat{D}\le j}\|\mat{X}-\mat{D}\|=\sigma_{j+1}(\mat{X}),
\]
for symmetric $\mat{T},\mat{E}$, and
\[
\min_{\rank\mat{D}\le j}\|\mat{X}-\mat{D}\|_F^2
=\sum_{\ell>j}\sigma_\ell(\mat{X})^2.
\]
The lower bounds follow by restricting to the complement of a rank-$j$
space; truncating the singular-value expansion attains them.
If $c_j$ are the singular values of $\mat{A}\T\widehat{\mat{A}}$ and
$s_j=\sqrt{1-c_j^2}$, the difference of the two projectors has eigenvalues
$\pm s_j$ on the corresponding principal planes. Thus
\begin{equation}\label{eq:projection-geometry}
\|\widehat{\mat{P}}_A-\mat{P}_A\|=\|\mat{Q}_A\widehat{\mat{A}}\|
=\max_j s_j,\qquad
\sigma_{\min}(\widehat{\mat{A}}\T\mat{A})
=\sqrt{1-\|\widehat{\mat{P}}_A-\mat{P}_A\|^2}.
\end{equation}

\subsubsection{Residual factor and noise blocks}\label{app:refined-blocks}
\begin{lemma}[Refined residual moment bounds]\label{lem:refined-blocks}
Under Assumptions~\ref{ass:A}--\ref{ass:D}, put
$\mat{F}_c=\Fm\mat{M}_c$, $\mat{W}_c=\Em\mat{M}_c$ and
$\rn=p^{\delta/2}/\sqrt n$. Let $\mat{S}_{h,n}$ have entries
$(S_{h,n})_{t+h,t}=1$ for $1\le t\le n-h$, with all other entries zero,
and let $\mat{S}_{0,n}=\mat{I}_n$. For fixed $0\le h\le k_0$,
\[
\begin{gathered}
\|\mat{F}_c\|=\Op(\sfac\sqrt n),\qquad
\|\mat{W}_c\|=\Op(\sqrt p),\\
\left\|\frac{\mat{F}_c\mat{S}_{h,n}\mat{F}_c\T}{n-h}
-\mat{\Gamma}_f(h)\right\|=\Op(\sfac^2n^{-1/2}),\\
\left\|\frac{\mat{W}_c\mat{S}_{h,n}\mat{F}_c\T}{n-h}\right\|
+\left\|\frac{\mat{F}_c\mat{S}_{h,n}\mat{W}_c\T}{n-h}\right\|
=\Op\{\sfac\sqrt{p/n}\},\qquad
\left\|\frac{\mat{W}_c\mat{S}_{h,n}\mat{W}_c\T}{n-h}\right\|
=\Op(p/n).
\end{gathered}
\]
Consequently, with $\widehat{\mat{\Gamma}}_\eta(0)=\widehat{\mat{C}}_\eta$,
\begin{equation}\label{eq:refined-residual-moment}
\max_{0\le h\le k_0}
\|\widehat{\mat{\Gamma}}_\eta(h)-\mat{A}\mat{\Gamma}_f(h)\mat{A}\T\|
=\Op\{\sfac^2n^{-1/2}+\sfac\sqrt{p/n}+p/n\}
=\Op(\sfac^2\rn).
\end{equation}
\end{lemma}
\begin{proof}
Lemma~\ref{lem:weak-moments} gives $\|\Fm\|=\Op(\sfac\sqrt n)$.
Lemma~\ref{lem:noise} and $p/n\to\infty$ give
$\|\Em\|=\Op(\sqrt p)$. The projection $\mat{M}_c$ is a contraction,
so the first two bounds follow. Write $\mat{P}_c=\mat{I}_n-\mat{M}_c$.
By factor centering, contemporaneous factor orthogonality and the
fixed-dimensional moment bounds,
\[
\begin{gathered}
\|\Fm\one_n/\sqrt n\|=\Op(\sfac),\qquad
\|\Fm\mat{Z}_c\|_F=\Op(\sfac\sqrt n),\qquad
\lambda_{\min}(\mat{Z}_c\T\mat{Z}_c)\asymp_{\mathbb P}n,\\
\|\Fm\mat{P}_c\|_F^2
=\|\Fm\one_n/\sqrt n\|^2
+\|\Fm\mat{Z}_c(\mat{Z}_c\T\mat{Z}_c)^{-1/2}\|_F^2
=\Op(\sfac^2).
\end{gathered}
\]
Since $\|\mat{S}_{h,n}\|\le1$, expansion of
$\mat{M}_c\mat{S}_{h,n}\mat{M}_c-\mat{S}_{h,n}$ gives
\[
\left\|\frac{\mat{F}_c\mat{S}_{h,n}\mat{F}_c\T
-\Fm\mat{S}_{h,n}\Fm\T}{n-h}\right\|
\le\frac{2\|\Fm\mat{P}_c\|\|\Fm\|
+\|\Fm\mat{P}_c\|^2}{n-h}
=\Op(\sfac^2n^{-1/2}).
\]
Lemma~\ref{lem:weak-window}, applied to the fixed-dimensional normalized
factor products, gives
\[
\mathbb E\left\|\frac{\Fm\mat{S}_{h,n}\Fm\T}{n-h}
-\mat{\Gamma}_f(h)\right\|_F^2\le C\sfac^4/n.
\]
This proves the factor-block bound. The two cross blocks and the noise
block satisfy the deterministic inequalities
\[
\begin{gathered}
\frac{\|\mat{W}_c\mat{S}_{h,n}\mat{F}_c\T\|
+\|\mat{F}_c\mat{S}_{h,n}\mat{W}_c\T\|}{n-h}
\le\frac{2\|\mat{W}_c\|\|\mat{F}_c\|}{n-h},\\
\frac{\|\mat{W}_c\mat{S}_{h,n}\mat{W}_c\T\|}{n-h}
\le\frac{\|\mat{W}_c\|^2}{n-h}.
\end{gathered}
\]
Use $\Rmat=\mat{A}\mat{F}_c+\mat{W}_c$ and
$\widehat{\mat{\Gamma}}_\eta(h)=\Rmat\mat{S}_{h,n}\Rmat\T/(n-h)$.
Dividing the resulting perturbation by $\sfac^2$ gives
\[
n^{-1/2}+\frac{\sqrt p}{\sfac\sqrt n}
+\frac p{n\sfac^2}
=n^{-1/2}+\rn+\rn^2=O(\rn).
\]
Here $\rn\to0$ by Lemma~\ref{lem:rates}. The maximum is over a fixed
number of lags, proving \eqref{eq:refined-residual-moment}.
\end{proof}

\subsubsection{Residual lag covariances}
Put $\mat{K}_f=\sum_{h=1}^{k_0}\mat{\Gamma}_f(h)\mat{\Gamma}_f(h)\T$
and $\mat{M}_A=\mat{A}\mat{K}_f\mat{A}\T$.
\begin{lemma}[Refined LAG loading rate]\label{lem:loading}
Under Assumptions~\ref{ass:A}--\ref{ass:lag}, the true-rank estimator satisfies
\[
\|\widehat{\mat{M}}_A-\mat{M}_A\|=\Op(\sfac^4\rn),\qquad
\|\widehat{\mat{P}}_A-\mat{P}_A\|=\Op(\rn),\qquad
\rn=p^{\delta/2}/\sqrt n.
\]
\end{lemma}
\begin{proof}
Set $\mat{\Delta}_h=\widehat{\mat{\Gamma}}_\eta(h)
-\mat{A}\mat{\Gamma}_f(h)\mat{A}\T$ and
$\mat{\Gamma}_\eta(h)=\mat{A}\mat{\Gamma}_f(h)\mat{A}\T$.
Lemma~\ref{lem:refined-blocks} and fixed-dimensional factor moment bounds give
\[
\max_h\|\mat{\Delta}_h\|=\Op(\sfac^2\rn),\qquad
\max_h\|\mat{\Gamma}_\eta(h)\|\le C\sfac^2.
\]
Thus
\[
\begin{aligned}
\widehat{\mat{M}}_A-\mat{M}_A
&=\sum_h\{\mat{\Gamma}_\eta(h)\mat{\Delta}_h\T
+\mat{\Delta}_h\mat{\Gamma}_\eta(h)\T
+\mat{\Delta}_h\mat{\Delta}_h\T\},\\
\|\widehat{\mat{M}}_A-\mat{M}_A\|
&\le\sum_h\{2\|\mat{\Gamma}_\eta(h)\|\|\mat{\Delta}_h\|
+\|\mat{\Delta}_h\|^2\}
=\Op\{\sfac^4(\rn+\rn^2)\}.
\end{aligned}
\]
By Assumption~\ref{ass:lag}, $\lambda_r(\mat{M}_A)\ge c\sfac^4$.
Let $\widehat{\mat{\Lambda}}$ collect the leading $r$ sample eigenvalues.
The eigenvector equation and \eqref{eq:projection-geometry} give
\[
\mat{Q}_A\widehat{\mat{A}}\widehat{\mat{\Lambda}}
=\mat{Q}_A(\widehat{\mat{M}}_A-\mat{M}_A)\widehat{\mat{A}},
\]
\[
\|\widehat{\mat{P}}_A-\mat{P}_A\|
\le\frac{\|\widehat{\mat{M}}_A-\mat{M}_A\|}
{\lambda_r(\mat{M}_A)-\|\widehat{\mat{M}}_A-\mat{M}_A\|}
=\Op(\rn)
\]
on an event of probability tending to one. The spectral map is
centered-residual measurable and shift invariant, and
$\sqrt p\,\rn^2=p^{1/2+\delta}/n\to0$.
\end{proof}

For the fixed-rank LAG estimator, the preceding proof verifies \eqref{eq:admission}.

\subsubsection{Factor-number transfer}
\label{app:rank-transfer}
\begin{proposition}[Same-sample rank transfer]\label{prop:number-transfer}
Let $\vct{T}_n^{[j]}$ be the rank-$j$ vector of all three statistics. Then
\begin{equation}\label{eq:number-tv}
\sup_{D\in\mathcal B(\mathbb R^3)}|\mathbb P\{\vct{T}_n^{[\widehat r]}\in D\}-\mathbb P\{\vct{T}_n^{[r]}\in D\}|\le\pi_n.
\end{equation}
For any common target $Z_n$, $b_n>0$, and $M>0$,
\begin{equation}\label{eq:number-rate}
\mathbb P\{|T_n^{[\widehat r]}-Z_n|>Mb_n\}\le\pi_n+\mathbb P\{|T_n^{[r]}-Z_n|>Mb_n\}.
\end{equation}
Here $\pi_n=\mathbb P(\widehat r\ne r)$.
\end{proposition}

\begin{proof}[Proof of Proposition~\ref{prop:number-transfer}]
On $\mathcal A_n=\{\widehat r=r\}$, the selected-rank and true-rank maps
are identical. Consequently,
\[
\left|\indc\{\vct{T}_n^{[\widehat r]}\in D\}
-\indc\{\vct{T}_n^{[r]}\in D\}\right|\le\indc_{\mathcal A_n^c},
\]
\[
\{|T_n^{[\widehat r]}-Z_n|>Mb_n\}
\subseteq\mathcal A_n^c\cup\{|T_n^{[r]}-Z_n|>Mb_n\}.
\]
Taking probabilities proves both inequalities. Intercept-shift invariance
follows from $(\Ymat+\vct{a}\one_n\T)\mat{M}_c=\Rmat$.
\end{proof}

\subsubsection{LAG rank consistency}
\label{app:lag-rank-status}
A sufficient separation condition for the ratio selector is, for some
deterministic $a_{{\rm sel},n}\downarrow0$, where
$q_{j,n}=\widehat\lambda_{j+1}/\widehat\lambda_j$ uses the ordered
eigenvalues of $\widehat{\mat M}_A$,
\begin{equation}\label{eq:lag-rank-separation}
\frac{p^{\delta}}n=o(a_{{\rm sel},n}),\qquad
\mathbb P\left\{\min_{r<j\le R_*}q_{j,n}>a_{{\rm sel},n}\right\}\to1.
\end{equation}
Together with Assumptions~\ref{ass:A}--\ref{ass:lag}, this gives rank consistency
by Lemma~\ref{lem:lag-rank-original}, following
\citet[Assumption 4.3]{wangLocation2026}.

\begin{lemma}[LAG rank selection]\label{lem:lag-rank-original}
Under Assumptions~\ref{ass:A}--\ref{ass:lag}, and
\eqref{eq:lag-rank-separation}, the selector in \eqref{eq:rank-selectors} satisfies
$\mathbb P(\widehat r=r)\to1$.
\end{lemma}
\begin{proof}
Let $\mat{H}=[\mat{\Gamma}_\eta(1),\ldots,\mat{\Gamma}_\eta(k_0)]$ and $\widehat{\mat{H}}=[\widehat{\mat{\Gamma}}_\eta(1),\ldots,\widehat{\mat{\Gamma}}_\eta(k_0)]$.
The lag-identification and loading bounds give
$\rank(\mat{H})=r$, $\sigma_j(\mat{H})\asymp\sfac^2$ for $j\le r$,
and $\|\widehat{\mat{H}}-\mat{H}\|=\Op(\sfac^2\rn)$.
Singular-value perturbation therefore yields
\[
\widehat\lambda_j\asymp_{\mathbb P}\sfac^4\ (j\le r),
\quad \widehat\lambda_{r+1}=\Op(\sfac^4\rn^2)=\Op(p^{2-\delta}/n),\quad
q_{r,n}=\Op(p^{\delta}/n).
\]
For a sufficiently small $c_0>0$, with probability tending to one,
\[
\min_{j<r}q_{j,n}>c_0,\qquad
q_{r,n}<a_{{\rm sel},n},\qquad
\min_{r<j\le R_*}q_{j,n}>a_{{\rm sel},n}.
\]
The last two inequalities follow from \eqref{eq:lag-rank-separation}. Since
$a_{{\rm sel},n}<c_0$ eventually, $r$ is the unique minimizer.
\end{proof}
\subsection{Conditional design geometry and estimated time spaces}\label{app:geometry}
\subsubsection{The true time-domain design}
Define
\[
\mat{M}_Z=\mat{I}_n-\mat{Z}(\mat{Z}\T \mat{Z})^{-1}\mat{Z}\T,\quad d=n-m,
\quad \Oz\T\Oz=\mat{I}_d,\quad\Oz\Oz\T=\mat{M}_Z,
\]
\[
\omega_z =\|\Oz\T\one_n\|,\quad \vct{u}_z=\Oz\T\one_n/\omega_z ,
\quad\Xmat=\Ymat\Oz,
\quad\Gmat=(\vct{\lambda}\one_n\T+\Fm)\Oz,
\]
\[
\begin{gathered}
\Ptime=\Gmat\T(\Gmat\Gmat\T)^{-1}\Gmat,\qquad\Jtrue=\mat{I}_d-\Ptime,\\
 \vct{v}=\Jtrue \vct{u}_z,\quad\eta_0=\|\vct{v}\|^2,\quad
\Ieff=\omega_z ^2\eta_0,\quad\vct{\beta}=\Oz \vct{v}/\sqrt{\eta_0}.
\end{gathered}
\]

\begin{lemma}[Information and conditional weights]\label{lem:geometry}
Let $\mat{\Gamma}_f(0)=\mathbb E\vct{f}_t\vct{f}_t\T$. Then
\[
c\sfac^2 \mat{I}_r\preceq\mat{\Gamma}_f(0)\preceq C\sfac^2 \mat{I}_r,
\quad \sigma_{\min}(\Gmat)\asymp_{\mathbb P}\|\Gmat\|\asymp_{\mathbb P}\sqrt n\sfac,
\]
\[
\frac{\Ieff}{n}=\cstar^{-1}+\Op(n^{-1/2}),\quad
\cstar=1+\vct{m}_z\T\mat{\Gamma}_z^{-1}\vct{m}_z+\vct{\lambda}\T\mat{\Gamma}_f(0)^{-1}\vct{\lambda},
\]
\[
\mathbb P(\eta_0\ge c)\to1,\quad \omega_z \asymp_{\mathbb P}\sqrt n,
\quad\sum_t\beta_t^2=1,\quad\max_t|\beta_t|=\Op(\sqrt{\Lnp/n}).
\]
\end{lemma}
\begin{proof}
Assumption~\ref{ass:B} bounds $\mat{\Gamma}_f(0)$.
Set $\vct{x}_t=(\vct{z}_t\T,(\vct{\lambda}+\vct{f}_t)\T/\sfac)\T$ and $\vct{\mu}_x=\mathbb E\vct{x}_t$. Then
\[
\mat{\Gamma}_x=\Cov(\vct{x}_t)=\diag(\mat{\Gamma}_z,\mat{\Gamma}_f(0)/\sfac^2),
\quad c\mat{I}\preceq\mat{\Gamma}_x\preceq C\mat{I},
\quad\|\vct{\mu}_x\|\le C.
\]
Define $\bar{\vct{x}}=n^{-1}\sum_t\vct{x}_t$ and $\mat{S}_x=n^{-1}\sum_t(\vct{x}_t-\bar{\vct{x}})(\vct{x}_t-\bar{\vct{x}})\T$. By Lemma~\ref{lem:weak-moments}, the identity
\[
\mat{S}_x-\mat{\Gamma}_x
=\frac1n\sum_t\{(\vct{x}_t-\vct{\mu}_x)(\vct{x}_t-\vct{\mu}_x)\T-\mat{\Gamma}_x\}
-(\bar{\vct{x}}-\vct{\mu}_x)(\bar{\vct{x}}-\vct{\mu}_x)\T
\]
give
\[
\|\mat{S}_x-\mat{\Gamma}_x\|+\|\bar{\vct{x}}-\vct{\mu}_x\|=\Op(n^{-1/2}).
\]
The uncentered Gram matrix $n^{-1}\sum_t\vct{x}_t\vct{x}_t\T=\mat{S}_x+\bar{\vct{x}}\bar{\vct{x}}\T$ is uniformly positive definite on a high-probability event. Hence
\[
\frac{\vct{u}\T\Gmat\Gmat\T \vct{u}}{n\sfac^2}
=\min_{\vct{b}\in\mathbb R^m}\frac1n\sum_t\{\vct{u}\T(\vct{\lambda}+\vct{f}_t)/\sfac-\vct{b}\T \vct{z}_t\}^2
\ge c\|\vct{u}\|^2,
\]
Taking $\vct{b}=0$ gives the upper bound. Let $\mat{M}_0$ be the orthogonal residual projection for $[\mat{Z},(\vct{\lambda}\one_n\T+\Fm)\T]$. Then
\[
\mat{M}_0=\Oz\Jtrue\Oz\T,\quad \rank \mat{M}_0=k,
\quad\Ieff=\one_n\T \mat{M}_0\one_n
=\frac{n}{1+\bar{\vct{x}}\T \mat{S}_x^{-1}\bar{\vct{x}}}.
\]
The last identity follows by substituting
\[
(\mat{S}_x+\bar{\vct{x}}\bar{\vct{x}}\T)^{-1}\bar{\vct{x}}
=\frac{\mat{S}_x^{-1}\bar{\vct{x}}}{1+\bar{\vct{x}}\T \mat{S}_x^{-1}\bar{\vct{x}}}
\]
into $\one_n\T \mat{M}_0\one_n$. On $\lambda_{\min}(\mat{S}_x)\ge c/2$, the inverse-matrix identity gives
\[
\|\mat{S}_x^{-1}-\mat{\Gamma}_x^{-1}\|
\le\|\mat{S}_x^{-1}\|\|\mat{S}_x-\mat{\Gamma}_x\|\|\mat{\Gamma}_x^{-1}\|
=\Op(n^{-1/2}).
\]
Consequently,
\[
\bar{\vct{x}}\T \mat{S}_x^{-1}\bar{\vct{x}}-\vct{\mu}_x\T\mat{\Gamma}_x^{-1}\vct{\mu}_x=\Op(n^{-1/2}),
\qquad 1+\vct{\mu}_x\T\mat{\Gamma}_x^{-1}\vct{\mu}_x=\cstar.
\]
Similarly, $\omega_z ^2=n/c_{z,n}\asymp_{\mathbb P} n$, so $\eta_0$ is bounded away from zero with probability tending to one. Moreover,
\[
\vct{\beta}=\frac{\mat{M}_0\one_n}{\sqrt{\Ieff}},\quad
(\mat{M}_0\one_n)_t=1-\vct{x}_t\T(n^{-1}\mat{X}_{\mathrm{des}}\T\mat{X}_{\mathrm{des}})^{-1}\bar{\vct{x}},
\]
Here $\mat{X}_{\mathrm{des}}$ has rows $\vct{x}_t\T$. Choose $c_x>0$ such that $\mat{\Gamma}_x\succeq c_x\mat{I}$ and define
\[
\mat{V}_{x,n}=n^{-1}\sum_t\{(\vct{x}_t-\vct{\mu}_x)(\vct{x}_t-\vct{\mu}_x)\T-\mat{\Gamma}_x\},
\quad \vct{u}_{x,n}=\bar{\vct{x}}-\vct{\mu}_x,
\]
\[
\mathcal E_{x,n}=
\{\|\mat{V}_{x,n}\|\le c_x/4,\ \|\vct{u}_{x,n}\|^2\le\min(c_x/4,1)\}.
\]
Since $\mathbb E\|\mat{V}_{x,n}\|_F^2+\mathbb E\|\vct{u}_{x,n}\|^2\le C/n$, Markov's inequality gives
\[
\begin{gathered}
\mathbb P(\mathcal E_{x,n}^c)\le C/n,\\
\mat{S}_x=\mat{\Gamma}_x+\mat{V}_{x,n}-\vct{u}_{x,n}\vct{u}_{x,n}\T\succeq c_x\mat{I}/2,\\
\|\bar{\vct{x}}\|\le C\quad\text{on }\mathcal E_{x,n}.
\end{gathered}
\]
\[
\mathbb P\{\max_t\|\vct{x}_t\|>M\sqrt{\Lnp}\}\le Cn\exp(-cM^2\Lnp).
\]
For sufficiently large fixed $M$, define $\mathcal E_{\beta,n}=\mathcal E_{x,n}\cap\{\max_t\|\vct{x}_t\|\le M\sqrt{\Lnp}\}$. Then
\begin{equation}\label{eq:design-event}
\mathbb P(\mathcal E_{\beta,n}^c)\le C/n+C(np)^{-2},\qquad
\max_t|\beta_t|\le C_\beta\sqrt{\Lnp/n}\quad\text{on }\mathcal E_{\beta,n}.
\end{equation}
Here $C_\beta$ is deterministic.
\end{proof}

\subsubsection{From loading accuracy to time-space accuracy}
The observed design $\widehat{\mat{D}}_x$ and $[\mat{Z},\Ymat\T\widehat{\mat{A}}]$ span the same space. In the $\Oz$ coordinates, define
\[
\widehat{\mat{G}}=\widehat{\mat{A}}\T\Xmat,\quad
\Phat=\widehat{\mat{G}}\T(\widehat{\mat{G}}\widehat{\mat{G}}\T)^{-1}\widehat{\mat{G}},
\quad\Jhat=\mat{I}_d-\Phat,
\quad\widehat\eta=\vct{u}_z\T\Jhat \vct{u}_z.
\]
\[
\widehat{\mat{M}}_x=\Oz\Jhat\Oz\T,
\quad\sqrt{\widehat{\mathcal I}}\,\widehat{\alp}=\Xmat\Jhat \vct{u}_z/\sqrt{\widehat\eta},
\quad\widehat{\mathcal I}=\omega_z ^2\widehat\eta.
\]

\begin{lemma}[Directional bounds for the estimated time space]\label{lem:time}
Under \eqref{eq:admission}, for deterministic $\mat{A}\T\alp=0$ and $\|\alp\|\le C_\alpha$,
\[
\|\Phat-\Ptime\|=\Op(\epsT),\quad
\|\Ptime\Jhat \vct{u}_z\|=\Op(\hT),\quad
|\widehat\eta-\eta_0|=\Op(\epsT).
\]
\end{lemma}
\begin{proof}
On $\|\mat{\Delta}_P\|<1/2$, define the aligned left inverse
\[
\mat{H}_A=(\widehat{\mat{A}}\T \mat{A})^{-1}\widehat{\mat{A}}\T=\mat{A}\T+\mat{T}_A,\qquad \mat{T}_A\mat{A}=0,
\]
\[
\|\mat{T}_A\|\le\frac{\|\mat{\Delta}_P\|}{\sqrt{1-\|\mat{\Delta}_P\|^2}}=\Op(\rn).
\]
The matrices $\mat{H}_A\Xmat$ and $\widehat{\mat{G}}$ have the same row space. With $\Xmat=\mat{A}\Gmat+\alp \omega_z \vct{u}_z\T+\mat{W}$ and $\mat{W}=\Em\Oz$,
\[
\mat{H}_A\Xmat=\Gmat+\mat{N}_A,
\quad \mat{N}_A=\mat{A}\T \mat{W}+\mat{T}_A\mat{W}+\mat{T}_A\alp \omega_z \vct{u}_z\T.
\]
By \eqref{eq:admission} and Lemmas~\ref{lem:noise} and~\ref{lem:geometry},
\[
\begin{split}
\|\mat{N}_A\|&\le\|\mat{A}\T \mat{W}\|+\|\mat{T}_A\|\|\mat{W}\|+
\|\mat{T}_A\|\|\alp\|\omega_z 
=\Op(\sqrt n+\rn\sqrt p),\\
\|\mat{N}_A\vct{v}\|&\le\|\mat{A}\T \mat{W}\vct{v}\|+\|\mat{T}_A\|\|\mat{W}\vct{v}\|
+\|\mat{T}_A\|\|\alp\|\omega_z |\vct{u}_z\T \vct{v}|
=\Op(1+\rn\sqrt p).
\end{split}
\]
Since $\vct{v}$ is factor-measurable,
\[
\Ex\|\mat{A}\T \mat{W}\vct{v}\|^2\le C\|\vct{v}\|^2\le C,\qquad
\Ex\|\mat{W}\vct{v}\|^2=\mnoise\|\vct{v}\|^2\le Cp.
\]
Put $g_T=\sigma_{\min}(\Gmat)$. On $\|\mat{N}_A\|\le g_T/2$, take
\[
\widehat{\mat{V}}=(\Gmat+\mat{N}_A)\T\{(\Gmat+\mat{N}_A)(\Gmat+\mat{N}_A)\T\}^{-1/2}.
\]
Since $\Jtrue\Gmat\T=0$,
\[
\sigma_{\min}(\Gmat+\mat{N}_A)\ge g_T-\|\mat{N}_A\|>0,
\qquad
\|\Jtrue\widehat{\mat{V}}\|\le\frac{\|\mat{N}_A\|}{g_T-\|\mat{N}_A\|}.
\]
Equation~\eqref{eq:projection-geometry} gives
\[
\|\Phat-\Ptime\|
=\Op\!\left(\frac{\sqrt n+\rn\sqrt p}{\sqrt n\sfac}\right)
=\Op(\sfac^{-1}+\rn p^{\delta/2}/\sqrt n)=\Op(\epsT).
\]
Since $\Gmat \vct{v}=0$,
\[
\Phat \vct{v}=(\Gmat+\mat{N}_A)\T\{(\Gmat+\mat{N}_A)(\Gmat+\mat{N}_A)\T\}^{-1}\mat{N}_A\vct{v},
\]
\[
\|\Phat \vct{v}\|\le\frac{\|\mat{N}_A\vct{v}\|}{g_T-\|\mat{N}_A\|}
=\Op\{(\sqrt n\sfac)^{-1}+\rn p^{\delta/2}/\sqrt n\}.
\]
\[
\|\Ptime\Jhat\Ptime\|\le\|\Phat-\Ptime\|^2,
\qquad
\Ptime\Jhat \vct{u}_z=\Ptime\Jhat\Ptime \vct{u}_z-\Ptime\Phat \vct{v}.
\]
Hence
\[
\|\Ptime\Jhat \vct{u}_z\|
\le\left(\frac{\|\mat{N}_A\|}{g_T-\|\mat{N}_A\|}\right)^2
+\frac{\|\mat{N}_A\vct{v}\|}{g_T-\|\mat{N}_A\|}.
\]
This is $\Op(\hT)$, and
$|\widehat\eta-\eta_0|\le\|\Phat-\Ptime\|=\Op(\epsT)$.
For the centered scores, let $\mat{F}_c=\Fm \mat{M}_c$, $\mat{S}_f=n^{-1}\Fm \mat{M}_1\Fm\T$, and $\mat{S}_{fz}=n^{-1}\Fm \mat{Z}_c$. Exactly,
\[
n^{-1}\mat{F}_c\mat{F}_c\T=\mat{S}_f-\mat{S}_{fz}\mat{S}_z^{-1}\mat{S}_{fz}\T.
\]
By $\mathbb E(\vct{z}_t^0\vct{f}_t\T)=0$, fixed-dimensional sample moment bounds give
\[
\|\mat{S}_f-\mat{\Gamma}_f(0)\|=\Op(\sfac^2n^{-1/2}),\qquad
\|\mat{S}_{fz}\|=\Op(\sfac n^{-1/2}),
\]
\[
\left\|\frac{\mat{F}_c\mat{F}_c\T}{n\sfac^2}-\frac{\mat{\Gamma}_f(0)}{\sfac^2}\right\|
\le\frac{\|\mat{S}_f-\mat{\Gamma}_f(0)\|+\|\mat{S}_{fz}\|^2\|\mat{S}_z^{-1}\|}{\sfac^2}
=\Op(n^{-1/2}+n^{-1}).
\]
Consequently, $\sigma_{\min}(\mat{F}_c)\asymp_{\mathbb P}\sqrt n\sfac$.
\[
\begin{gathered}
\mat{H}_A\Rmat=\mat{F}_c+\mat{H}_A\Em \mat{M}_c,\qquad
\frac{\|\mat{H}_A\Em \mat{M}_c\|}{\sqrt n\sfac}=\Op(\epsT),\\
\sigma_{\min}(\widehat{\mat{A}}\T \mat{A})\ge\sqrt{1-\|\dP\|^2}.
\end{gathered}
\]
Since $\widehat{\mat{G}}_c=(\widehat{\mat{A}}\T \mat{A})\mat{H}_A\Rmat$, with probability tending to one,
\[
c\sfac^2\mat{I}_r\preceq\widehat{\mat{K}}_g=n^{-1}\widehat{\mat{G}}_c\widehat{\mat{G}}_c\T\preceq C\sfac^2\mat{I}_r.
\]
Together with $\mat{S}_z\succ0$ and Lemma~\ref{lem:fwl}, this proves invertibility of $\widehat{\mat{D}}_x\T\widehat{\mat{D}}_x$ and positivity of $\widehat{\mathcal I}$. 
\end{proof}

\subsection{Centered quadratic forms and noise scales}\label{app:quadratic}
\subsubsection{A quadratic-form identity}
Let $\mat{K}_0\succeq0$ have support projection $\mat{\Pi}$, let
$\mat{D}_0=\mat{D}_0\T$, and put $\mat{J}=\mat{I}-\mat{\Pi}$ and
$\mat{K}=\mat{K}_0+\mat{D}_0$. For $\|\vct{u}_z\|=1$ and an equal-rank orthogonal projection $\widehat{\mat{\Pi}}$, set $\widehat{\mat{J}}=\mat{I}-\widehat{\mat{\Pi}}$, $\vct{v}=\mat{J}\vct{u}_z$, and $\vct{d}_u=\widehat{\mat{J}}\vct{u}_z-\vct{v}$.

Since $\mat{K}_0\vct{v}=0$, expansion gives
\[
\vct{u}_z\T\widehat{\mat{J}} \mat{K}\widehat{\mat{J}}\vct{u}_z-\vct{v}\T \mat{D}_0\vct{v}
=2\vct{v}\T \mat{D}_0\vct{d}_u+\vct{d}_u\T \mat{K}_0\vct{d}_u+\vct{d}_u\T \mat{D}_0\vct{d}_u.
\]
Consequently, if $\|\vct{d}_u\|\le e$ and $\|\mat{\Pi}\vct{d}_u\|\le h$, then
\begin{equation}\label{eq:directionbound}
|\vct{u}_z\T\widehat{\mat{J}} \mat{K}\widehat{\mat{J}}\vct{u}_z-\vct{v}\T \mat{D}_0\vct{v}|
\le2\|\mat{D}_0\vct{v}\|e+\|\mat{K}_0\|h^2+\|\mat{D}_0\|e^2.
\end{equation}

\subsubsection{Directional scales and finite-rank residuals}
Let $\mat{H}=\Xmat\T\Xmat$, $\mat{K}=\mat{H}-\mnoise \mat{I}_d$, and $\mat{K}_0=\Gmat\T\Gmat$. Since $\mat{A}\T\alp=0$,
\begin{equation}\label{eq:perturbation}
\begin{split}
\mat{D}_0={}&\Gmat\T \mat{A}\T \mat{W}+\mat{W}\T \mat{A}\Gmat+\Oz\T \mat{K}_\varepsilon\Oz\\
&+\omega_z ^2\|\alp\|^2\vct{u}_z\vct{u}_z\T+\omega_z \vct{u}_z\alp\T \mat{W}+\omega_z \mat{W}\T\alp \vct{u}_z\T.
\end{split}
\end{equation}
\begin{lemma}[Directional remainders and finite-rank residuals]\label{lem:residual}
For a true-rank loading estimator satisfying \eqref{eq:admission} and a bounded deterministic signal,
\[
\|\mat{D}_0\|=\Op(n\sfac),\quad\|\mat{D}_0\vct{v}\|=\Op(\sqrt{np}),\quad\|\mat{K}_0\|=\Op(n\sfac^2).
\]
Moreover,
\[
\mat{B}_{\mathrm{err}}=\Jhat \mat{K}\Jhat-\Jtrue \mat{D}_0\Jtrue,\quad
\rank(\mat{B}_{\mathrm{err}})\le5r,\quad\|\mat{B}_{\mathrm{err}}\|+\|\mat{B}_{\mathrm{err}}\|_F=\Op(n),
\]
\[
\frac{|\vct{u}_z\T\Jhat \mat{K}\Jhat \vct{u}_z-\vct{v}\T \mat{D}_0\vct{v}|}{\sqrt p}
=\Op(\bvec+\sqrt p\,\rn^2+p^{-1/2}).
\]
\end{lemma}
\begin{proof}
For bounded deterministic $\alp$, $\Ex\|\Em\T\alp\|^2\le Cn$. Equation~\eqref{eq:perturbation} and the noise and design bounds give
\[
\|\mat{D}_0\|=\Op(n\sfac+\sqrt{np}+n)=\Op(n\sfac),
\quad \sqrt{np}/(n\sfac)=\sqrt{p^\delta/n}\to0.
\]
Since $\Gmat \vct{v}=0$,
\[
\|\Gmat\T \mat{A}\T \mat{W}\vct{v}\|=\Op(\sqrt n\sfac),\quad
\|\Oz\T \mat{K}_\varepsilon\Oz \vct{v}\|=\Op(\sqrt{np}),
\]
\[
\begin{gathered}
\|\omega_z ^2\|\alp\|^2\vct{u}_z(\vct{u}_z\T \vct{v})\|=O_\mathbb P(n),\quad
\|\omega_z \vct{u}_z\alp\T \mat{W}\vct{v}\|=\Op(\sqrt n),\\
\|\omega_z \mat{W}\T\alp(\vct{u}_z\T \vct{v})\|=\Op(n).
\end{gathered}
\]
Because $p>n$ and $\sfac\le\sqrt p$, these imply the directional bound.
The exact relation $(\Gmat+\mat{N}_A)\Jhat=0$ implies
\[
\|\Gmat\Jhat\|^2\le\|\mat{N}_A\|^2=\Op(n+\rn^2p).
\]
Further,
\[
\mat{B}_{\mathrm{err}}=\Jhat \mat{K}_0\Jhat
+(\Jhat-\Jtrue)\mat{D}_0\Jhat+\Jtrue \mat{D}_0(\Jhat-\Jtrue),
\qquad \rank(\mat{B}_{\mathrm{err}})\le r+2r+2r.
\]
Thus
\[
\|\mat{B}_{\mathrm{err}}\|\le\|\Gmat\Jhat\|^2+2\|\mat{D}_0\|\|\Jhat-\Jtrue\|
=\Op(n+\rn^2p+\rn\sqrt{np})=\Op(n),
\]
where $\rn^2p/n\to0$ by Lemma~\ref{lem:rates}. Fixed rank yields the Frobenius bound.
For the actual score direction, equation~\eqref{eq:directionbound} gives
\[
\frac{|\vct{u}_z\T\Jhat \mat{K}\Jhat \vct{u}_z-\vct{v}\T \mat{D}_0\vct{v}|}{\sqrt p}
=\Op\left\{\sqrt n\epsT+\frac{n\sfac^2}{\sqrt p}\hT^2
+\frac{n\sfac}{\sqrt p}\epsT^2\right\}.
\]
The deterministic comparisons are
\[
\frac{n\sfac^2}{\sqrt p}\hT^2
\le C\{p^{-1/2}+\sqrt p\,\rn^2\},
\]
\[
\frac{n\sfac}{\sqrt p}\epsT^2
\le C\{n/(\sfac\sqrt p)+\rn^2p^{\delta/2}\}
\le C\{\bvec\sqrt{n/p}+\sqrt p\,\rn^2\}.
\]

\end{proof}

\subsubsection{Exact noise-Gram moments}
\begin{lemma}[Covariance with factor-measurable coefficients]\label{lem:grammom}
Let $\mathfrak k_{\varepsilon,p}=\sum_{j=1}^p\kappa_j(\Sg)_{jj}^2$. For factor-measurable symmetric matrices $\mat{T}_1,\mat{T}_2$,
\[
\Cov_\xi\{\tr(\mat{T}_1\mat{K}_\varepsilon),\tr(\mat{T}_2\mat{K}_\varepsilon)\}
=2\taunoise\tr(\mat{T}_1\mat{T}_2)+\mathfrak k_{\varepsilon,p}\sum_t(\mat{T}_1)_{tt}(\mat{T}_2)_{tt}.
\]
Also $|\mathfrak k_{\varepsilon,p}|\le C\taunoise$.
\end{lemma}
\begin{proof}
Independence over time and coordinate pairing give $\Ex(\mat{K}_\varepsilon)_{ab}=0$ and
\[
\begin{split}
\Ex\{(\mat{K}_\varepsilon)_{ab}(\mat{K}_\varepsilon)_{cd}\}
={}&\taunoise\{\indc_{a=c,b=d}+\indc_{a=d,b=c}\}\\
&+\mathfrak k_{\varepsilon,p}\indc_{a=b=c=d}.
\end{split}
\]
For $a\ne b$, independence and $\mathbb E\vct{\varepsilon}_a=0$ give
\[
\Ex(\vct{\varepsilon}_a\T\vct{\varepsilon}_b)^2
=\Ex\{\vct{\varepsilon}_a\T\Sg\vct{\varepsilon}_a\}=\tr(\Sg^2),
\]
\[
\Ex\{(\|\vct{\varepsilon}_a\|^2-\mnoise)(\|\vct{\varepsilon}_b\|^2-\mnoise)\}=0,
\qquad
\Var_\xi(\|\vct{\varepsilon}_a\|^2)=2\taunoise+\mathfrak k_{\varepsilon,p}.
\]
The last identity is \eqref{eq:quadmoment} with $\mat{H}=\Sg$. A time index occurring only once has zero expectation after conditioning on all other times.
\[
\tr(\mat{T}_1\mat{K}_\varepsilon)=\sum_{a,b}(\mat{T}_1)_{ab}(\mat{K}_\varepsilon)_{ab},
\qquad
\sum_{a,b}(\mat{T}_1)_{ab}(\mat{T}_2)_{ab}=\tr(\mat{T}_1\mat{T}_2).
\]
Multiply and sum to obtain the covariance identity. Since $\sup_j|\kappa_j|\le C$,
\[
|\mathfrak k_{\varepsilon,p}|\le C\sum_j(\Sg)_{jj}^2\le C\taunoise.
\]
Let $\mat{M}$ be a factor-measurable orthogonal projection, $1\le k_M=\rank(\mat{M})\le n$, and $d_M=\sum_t\mat{M}_{tt}^2\le k_M$. Since $\mat{M}^2=\mat{M}$,
\[
\begin{split}
\Ex\|\mat{M}\mat{K}_\varepsilon \mat{M}\|_F^2
&=\Ex\tr(\mat{M}\mat{K}_\varepsilon \mat{M}\mat{K}_\varepsilon)\\
&=\sum_{a,b,c,d}\mat{M}_{ab}\mat{M}_{cd}\Ex\{(\mat{K}_\varepsilon)_{bc}(\mat{K}_\varepsilon)_{da}\}\\
&=\taunoise\sum_{a,b}\mat{M}_{ab}^2
+\taunoise\sum_{a,c}\mat{M}_{aa}\mat{M}_{cc}
+\mathfrak k_{\varepsilon,p}\sum_a \mat{M}_{aa}^2\\
&=k_M(k_M+1)\taunoise+d_M\mathfrak k_{\varepsilon,p}.
\end{split}
\]
\[
\Var_\xi\{\tr(\mat{M}\mat{K}_\varepsilon)\}
=2k_M\taunoise+d_M\mathfrak k_{\varepsilon,p}.
\]
Define the center and centered Gram matrix under the true time projection by
\[
m_M^\circ=\frac1{k_M}\tr(\Em \mat{M}\Em\T)
=\mnoise+\frac{\tr(\mat{M}\mat{K}_\varepsilon)}{k_M},
\]
\[
\mat{R}_M^\circ=\mat{M}\mat{K}_\varepsilon \mat{M}-\frac{\tr(\mat{M}\mat{K}_\varepsilon)}{k_M}\mat{M}.
\]
The identity $\mat{M}^2=\mat{M}$ yields
\[
\|\mat{R}_M^\circ\|_F^2
=\|\mat{M}\mat{K}_\varepsilon \mat{M}\|_F^2-\frac{\{\tr(\mat{M}\mat{K}_\varepsilon)\}^2}{k_M}.
\]
Consequently,
\[
\Ex\|\mat{R}_M^\circ\|_F^2
=\{k_M(k_M+1)-2\}\taunoise
+(1-k_M^{-1})d_M\mathfrak k_{\varepsilon,p}.
\]
For $\tau_M^\circ=\|\mat{R}_M^\circ\|_F^2/\{k_M(k_M+1)\}$,
\[
\Ex\tau_M^\circ-\taunoise
=-\frac{2\taunoise}{k_M(k_M+1)}
+\frac{(1-k_M^{-1})d_M\mathfrak k_{\varepsilon,p}}{k_M(k_M+1)}.
\]
Thus $|\Ex\tau_M^\circ/\taunoise-1|\le C/k_M$.
For a factor-measurable unit vector $\vct{a}$ with $\mat{M}\vct{a}=\vct{a}$, put $Z_a=\|\Em \vct{a}\|^2$.
\[
\Cov_\xi(Z_a,m_M^\circ)
=\frac{2\taunoise+\mathfrak k_{\varepsilon,p}\sum_ta_t^2\mat{M}_{tt}}{k_M}.
\]
With $\mat{T}_\circ=\vct{a}\vct{a}\T-\mat{M}/k_M$,
\[
Z_a-m_M^\circ=\tr(\mat{T}_\circ \mat{K}_\varepsilon),\qquad
\tr \mat{T}_\circ=0,\qquad \tr(\mat{T}_\circ^2)=1-k_M^{-1},
\]
\[
\Var_\xi(Z_a-m_M^\circ)
=2\taunoise(1-k_M^{-1})
+\mathfrak k_{\varepsilon,p}\sum_t(a_t^2-\mat{M}_{tt}/k_M)^2.
\]

\end{proof}

\subsubsection{Same-sample covariance functionals}
\begin{lemma}[Accuracy of the feasible covariance functionals]\label{lem:scales}
For $\mat{A}\T\alp=0$ and $\|\alp\|\le C_\alpha$,
\[
\tr(\widehat{\mat{\Sigma}}_\varepsilon)-\mnoise=\Op(\sqrt{p/n}+1),\qquad
\widehat v_\varepsilon/\taunoise-1=\Op(n^{-1}).
\]
\[
\max_i\left|(\widehat{\mat{\Sigma}}_\varepsilon)_{ii}-(\Sg)_{ii}-\frac{\Ieff}{k}\alpha_{i}^2\right|
=\Op(\vdiag).
\]
\end{lemma}
\begin{proof}
For $\mat{M}_0=\Oz\Jtrue\Oz\T$, Lemmas~\ref{lem:grammom}
and~\ref{lem:residual} give
\[
k(\tr(\widehat{\mat{\Sigma}}_\varepsilon)-\mnoise)
=\tr(\mat{M}_0\mat{K}_\varepsilon)+\Ieff\|\alp\|^2
+2\sqrt{\Ieff}\,\alp\T\Em\vct{\beta}+\tr\mat{B}_{\mathrm{err}}.
\]
The four terms are respectively $\Op(\sqrt{np})$, $O_\mathbb P(n)$, $\Op(\sqrt n)$, and $\Op(n)$. Divide by $k\asymp n$ for the trace bound.
To analyze the bias-corrected estimator of $\tr(\Sg^2)$, define
\[
U_n^\circ=\frac1{n(n-1)}\sum_{s\ne t}(\vct{\varepsilon}_s\T\vct{\varepsilon}_t)^2,
\quad \mathbb E U_n^\circ=\taunoise,
\]
\[
h_1(\vct{x})=\vct{x}\T\Sg\vct{x}-\taunoise,
\quad h_2(\vct{x},\vct{y})=(\vct{x}\T\vct{y})^2-\taunoise-h_1(\vct{x})-h_1(\vct{y}).
\]
For fixed $\vct{x}$, $\mathbb E(\vct{x}\T\vct{\varepsilon}_2)^2=\vct{x}\T\Sg\vct{x}$ and $\mathbb E h_1(\vct{\varepsilon}_2)=0$, so
\[
\mathbb E\{h_2(\vct{x},\vct{\varepsilon}_2)\}
=\vct{x}\T\Sg\vct{x}-\taunoise-(\vct{x}\T\Sg\vct{x}-\taunoise)=0.
\]
For distinct $a,b,c$, conditional independence gives
\[
\begin{split}
\mathbb E\{h_2(\vct{\varepsilon}_a,\vct{\varepsilon}_b)h_2(\vct{\varepsilon}_a,\vct{\varepsilon}_c)\}
&=\mathbb E\big[\mathbb E\{h_2(\vct{\varepsilon}_a,\vct{\varepsilon}_b)\mid\vct{\varepsilon}_a\}
\mathbb E\{h_2(\vct{\varepsilon}_a,\vct{\varepsilon}_c)\mid\vct{\varepsilon}_a\}\big]=0,\\
\mathbb E\{h_1(\vct{\varepsilon}_a)h_2(\vct{\varepsilon}_a,\vct{\varepsilon}_b)\}
&=\mathbb E\big[h_1(\vct{\varepsilon}_a)\mathbb E\{h_2(\vct{\varepsilon}_a,\vct{\varepsilon}_b)\mid\vct{\varepsilon}_a\}\big]=0.
\end{split}
\]
Disjoint index sets are independent. Expanding the pair sum therefore gives
\[
U_n^\circ-\taunoise=\frac2n\sum_t h_1(\vct{\varepsilon}_t)
+\frac2{n(n-1)}\sum_{s<t}h_2(\vct{\varepsilon}_s,\vct{\varepsilon}_t),
\]
\[
\begin{gathered}
\mathbb E\{h_1(\vct{\varepsilon}_a)h_2(\vct{\varepsilon}_s,\vct{\varepsilon}_t)\}=0\quad(s\ne t),\\
\mathbb E\{h_2(\vct{\varepsilon}_a,\vct{\varepsilon}_b)h_2(\vct{\varepsilon}_s,\vct{\varepsilon}_t)\}=0
\quad\text{if }a\ne b,\ s\ne t,\ \{a,b\}\ne\{s,t\},\\
\Var(U_n^\circ)=\frac4n \mathbb E h_1^2+\frac2{n(n-1)}\mathbb E h_2^2,\\
h_1(\vct{\varepsilon}_t)=\vct{\xi}_t\T\Sg^2\vct{\xi}_t-\tr(\Sg^2),\quad
\mathbb E h_1^2\le C\tr(\Sg^4)\le Cp,\\
\mathbb E h_2^2\le4\{\mathbb E(\vct{\varepsilon}_1\T\vct{\varepsilon}_2)^4+\taunoise^2+2\mathbb E h_1^2\}\le Cp^2.
\end{gathered}
\]
The conditional linear-form moment bound gives
\[
\mathbb E[(\vct{\varepsilon}_1\T\vct{\varepsilon}_2)^4\mid\vct{\varepsilon}_1]
\le C(\vct{\varepsilon}_1\T\Sg\vct{\varepsilon}_1)^2,
\quad \mathbb E(\vct{\varepsilon}_1\T\Sg\vct{\varepsilon}_1)^2\le Cp+\taunoise^2\le Cp^2.
\]
Therefore,
\[
U_n^\circ/\taunoise-1=\Op((np)^{-1/2}+n^{-1}).
\]
The exact diagonal decomposition is
\[
\|\mat{K}_\varepsilon\|_F^2
=n(n-1)U_n^\circ+\sum_t(\|\vct{\varepsilon}_t\|^2-\mnoise)^2,
\quad \mathbb E\sum_t(\|\vct{\varepsilon}_t\|^2-\mnoise)^2\le Cnp.
\]
It follows that
\[
\frac{\|\mat{K}_\varepsilon\|_F^2}{n^2\taunoise}-1
=\Op((np)^{-1/2}+n^{-1}),\qquad
\|\mat{K}_\varepsilon\|_F=\Op(n\sqrt p).
\]
With $\mat{P}_0=\mat{I}_n-\mat{M}_0$,
\[
\mat{K}_\varepsilon=\mat{P}_0\mat{K}_\varepsilon \mat{P}_0+\mat{P}_0\mat{K}_\varepsilon \mat{M}_0
+\mat{M}_0\mat{K}_\varepsilon \mat{P}_0+\mat{M}_0\mat{K}_\varepsilon \mat{M}_0.
\]
The four blocks are pairwise orthogonal in the Frobenius inner product. Since $\mat{K}_\varepsilon=\mat{K}_\varepsilon\T$,
\[
\begin{split}
0\le \|\mat{K}_\varepsilon\|_F^2-\|\mat{M}_0\mat{K}_\varepsilon \mat{M}_0\|_F^2
&=\|\mat{P}_0\mat{K}_\varepsilon \mat{P}_0\|_F^2+2\|\mat{P}_0\mat{K}_\varepsilon \mat{M}_0\|_F^2\\
&\le2\|\mat{P}_0\mat{K}_\varepsilon\|_F^2
\le2(m+r)\|\mat{K}_\varepsilon\|^2.
\end{split}
\]
As $p>n$ eventually, Lemma~\ref{lem:noise} bounds the right-hand side by $\Op(np)$. Set
\[
\mat{W}_0=\Jtrue\Oz\T \mat{K}_\varepsilon\Oz\Jtrue,
\qquad \|\mat{W}_0\|_F=\|\mat{M}_0\mat{K}_\varepsilon \mat{M}_0\|_F.
\]
Then
\[
\begin{gathered}
\frac{\|\mat{W}_0\|_F^2}{n^2\taunoise}-1
=\Op\{(np)^{-1/2}+n^{-1}\},\\
\|\mat{W}_0\|=\Op(\sqrt{np}),\qquad \tr \mat{W}_0=\Op(\sqrt{np}).
\end{gathered}
\]
Define
\[
\widehat{\mat{R}}_0=\Jhat(\mat{H}-\mnoise \mat{I}_d)\Jhat=\mat{W}_0+\mat{V}_{\mathrm{err},n},
\qquad t_0=\tr(\widehat{\mat{R}}_0)=k(\tr(\widehat{\mat{\Sigma}}_\varepsilon)-\mnoise).
\]
Write $\vct{a}_\alpha=\Jtrue \mat{W}\T\alp$. Equation~\eqref{eq:perturbation} and Lemma~\ref{lem:residual} give
\[
\mat{V}_{\mathrm{err},n}=\mat{B}_{\mathrm{err}}+\omega_z ^2\|\alp\|^2\vct{v}\vct{v}\T
+\omega_z (\vct{v}\vct{a}_\alpha\T+\vct{a}_\alpha \vct{v}\T).
\]
The last three terms have column space contained in $\operatorname{span}(\vct{v},\vct{a}_\alpha)$. Therefore,
\[
\rank(\mat{V}_{\mathrm{err},n})\le5r+2,\qquad
\|\vct{a}_\alpha\|\le\|\mat{W}\T\alp\|=\Op(\sqrt n),
\]
\[
\begin{gathered}
\|\mat{V}_{\mathrm{err},n}\|=\Op(n),\qquad
\|\mat{V}_{\mathrm{err},n}\|_*\le(5r+2)\|\mat{V}_{\mathrm{err},n}\|=\Op(n),\\
\|\mat{V}_{\mathrm{err},n}\|_F^2=\Op(n^2),\qquad t_0=\Op(\sqrt{np}+n).
\end{gathered}
\]
Since $\Jhat\widehat{\mat{R}}_0=\widehat{\mat{R}}_0$ and $\tr\Jhat=k$,
\[
\widehat{\mat{R}}_c:=\Jhat(\mat{H}-\tr(\widehat{\mat{\Sigma}}_\varepsilon) \mat{I}_d)\Jhat
=\widehat{\mat{R}}_0-\frac{t_0}{k}\Jhat,
\]
\[
\widehat v_\varepsilon=\frac{\|\widehat{\mat{R}}_c\|_F^2}{k(k+1)},\qquad
\|\widehat{\mat{R}}_c\|_F^2=\|\widehat{\mat{R}}_0\|_F^2-\frac{t_0^2}{k}.
\]
Hence
\[
\|\widehat{\mat{R}}_c\|_F^2-\|\mat{W}_0\|_F^2
=2\tr(\mat{W}_0\mat{V}_{\mathrm{err},n})+\|\mat{V}_{\mathrm{err},n}\|_F^2-\frac{t_0^2}{k},
\]
\[
\left|\|\widehat{\mat{R}}_c\|_F^2-\|\mat{W}_0\|_F^2\right|
\le2\|\mat{W}_0\|\|\mat{V}_{\mathrm{err},n}\|_*+\|\mat{V}_{\mathrm{err},n}\|_F^2+\frac{t_0^2}{k}.
\]
Substitution gives
\[
\frac{|\tr(\mat{W}_0\mat{V}_{\mathrm{err},n})|}{n^2\taunoise}=\Op((np)^{-1/2}),
\qquad
\frac{\|\mat{V}_{\mathrm{err},n}\|_F^2}{n^2\taunoise}=\Op(p^{-1}),
\]
\[
\frac{t_0^2}{k n^2\taunoise}=\Op\{n^{-2}+(np)^{-1}\}.
\]
Finally, $n^2/\{k(k+1)\}=1+O(n^{-1})$ yields
\[
\frac{\widehat v_\varepsilon}{\taunoise}-1
=\Op\{(np)^{-1/2}+n^{-1}+p^{-1}\}=\Op(n^{-1}).
\]
For the coordinate variances, put $\mat{X}_0=\mat{W}\Jtrue+\alp \omega_z \vct{u}_z\T\Jtrue$. Then
\[
\Xmat\Jhat-\mat{X}_0
=\mat{A}\Gmat(\Jhat-\Jtrue)+(\mat{W}+\alp \omega_z \vct{u}_z\T)(\Jhat-\Jtrue).
\]
Loading dispersion and Lemmas~\ref{lem:noise} and~\ref{lem:time} give
\[
\begin{split}
\max_i\frac{\|\vct{e}_i\T(\Xmat\Jhat-\mat{X}_0)\|}{\sqrt k}
&\le\left\{\frac{C\|\Gmat\|}{\sqrt{pk}}
+\frac{\max_i\|\vct{e}_i\T \mat{W}\|}{\sqrt k}
+\frac{\omega_z \|\alp\|_\infty}{\sqrt k}\right\}\|\Jhat-\Jtrue\|\\
&=\Op\{(\sfac/\sqrt p+1+\|\alp\|_\infty)\epsT\}
=\Op(\epsT).
\end{split}
\]
Also,
\[
\frac{\|\vct{e}_i\T \mat{X}_0\|^2}{k}
=\frac{\vct{e}_i\T\Em \mat{M}_0\Em\T \vct{e}_i}{k}
+\frac{\Ieff}{k}\alpha_{i}^2
+\frac{2\sqrt{\Ieff}}{k}\alpha_{i}(\Em\vct{\beta})_i.
\]
Set $\vct{\delta}_i=\vct{e}_i\T(\Xmat\Jhat-\mat{X}_0)$. Then
\[
\max_i\frac{\|\vct{e}_i\T \mat{X}_0\|}{\sqrt k}=O_\mathbb P(1),
\]
\[
\begin{split}
\max_i\frac{\big|\|\vct{e}_i\T \mat{X}_0+\vct{\delta}_i\|^2-\|\vct{e}_i\T \mat{X}_0\|^2\big|}{k}
&\le\max_i\left\{\frac{2\|\vct{e}_i\T \mat{X}_0\|\|\vct{\delta}_i\|}{k}+\frac{\|\vct{\delta}_i\|^2}{k}\right\}\\
&=O_\mathbb P(\epsT+\epsT^2).
\end{split}
\]
\[
\max_i\frac{2\sqrt{\Ieff}}{ k}|\alpha_{i}(\Em\vct{\beta})_i|
=O_\mathbb P(\sqrt{\Lnp/n}),\qquad \epsT\to0.
\]
Combining these bounds with Lemma~\ref{lem:noise} gives the assertion.
\end{proof}

\subsubsection{Regularity of the original fitted statistics}\label{app:regularity}
\begin{proposition}[Original-sample regularity]\label{prop:regularity}
Under Assumptions~\ref{ass:A}--\ref{ass:lag} and a bounded pricing error, the original OLS fits, information scale and Sum and Max denominators are defined with probability tending to one at the true rank. Under Assumption~\ref{ass:rank}, the selected-rank LAG procedure inherits this regularity.
\end{proposition}
\begin{proof}
At rank $r$, Lemmas~\ref{lem:geometry} and~\ref{lem:time} give, with
probability tending to one,
\[
\mat{S}_z\succeq c\mat{I}_m,\qquad
\widehat{\mat{K}}_g\succeq c\sfac^2\mat{I}_r,\qquad
\widehat{\mathcal I}\ge cn.
\]
The centered design $[\one_n,\mat{Z}_c,\widehat{\mat{G}}_c\T]$ has
Gram matrix $n\diag(1,\mat{S}_z,\widehat{\mat{K}}_g)$, so both
regressions have full rank. Lemma~\ref{lem:scales} gives
\[
\widehat v_\varepsilon/\tr(\Sg^2)=1+\Op(n^{-1}),\qquad
\max_i\left|(\widehat{\mat{\Sigma}}_\varepsilon)_{ii}-(\Sg)_{ii}
-\Ieff\alpha_{i}^2/k\right|=\Op(\vdiag).
\]
Since $\tr(\Sg^2)\ge c^2p$, $(\Sg)_{ii}\ge c$, and $\vdiag\to0$,
the scales are positive with probability tending to one.
Rank consistency and Proposition~\ref{prop:number-transfer}
transfer these statements to the selected rank.
\end{proof}

\subsection{Approximation of the feasible statistics}\label{app:replacement}
For coordinate standardization, write $\mat{D}_\varepsilon=\diag(\Sg)$ and $\widehat{\mat{D}}_\varepsilon=\diag(\widehat{\mat{\Sigma}}_\varepsilon)$.
Conditional on the factors, $\sum_t\beta_t\vct{\xi}_t$ has independent standardized sub-Gaussian coordinates. Equation~\eqref{eq:quadmoment} and the linear tail bound give
\[
\Ex(Q_n^\circ)^2\le C,\qquad Q_n^\circ=\Op(1),\qquad
\|\vct{\xi}_n^\circ\|_\infty=\Op(\sqrt{\Lnp}).
\]

\begin{lemma}[Full-sample approximation with nonzero factor means]\label{lem:replacement}
Under Assumptions~\ref{ass:A}--\ref{ass:D}, let a true-rank loading estimator satisfy
\eqref{eq:admission}. For deterministic $\alp$ with
$\|\alp\|\le C_\alpha$ and $\mat{A}\T\alp=0$, we have
\begin{equation}\label{eq:sumreplacement}
\Tsum=Q_n^\circ+\Delta_n+\Op(\bsum).
\end{equation}
Under the null,
\begin{equation}\label{eq:maxreplacement}
\Tmax=M_n^\circ+\Op(\bmax+\bmax^2).
\end{equation}
\end{lemma}
\begin{proof}
Let $B_v=\vct{v}\T \mat{D}_0\vct{v}$. Exactly,
\[
\frac{B_v}{\eta_0}
=\|\vct{\xi}_n^\circ+\sqrt{\Ieff}\,\alp\|^2-\mnoise,
\]
Since $\sqrt{\widehat{\mathcal I}}\,\widehat{\alp}=\Xmat\Jhat \vct{u}_z/\sqrt{\widehat\eta}$ and $\Jhat^2=\Jhat$,
\[
\vct{u}_z\T\Jhat^2\vct{u}_z=\widehat\eta,\qquad
\frac{\vct{u}_z\T\Jhat(\mnoise \mat{I}_d)\Jhat \vct{u}_z}{\widehat\eta}=\mnoise.
\]
Therefore
\begin{equation}\label{eq:centeringidentity}
\begin{split}
\|\sqrt{\widehat{\mathcal I}}\,\widehat{\alp}\|^2-\tr(\widehat{\mat{\Sigma}}_\varepsilon)
&=\frac{\vct{u}_z\T\Jhat \mat{H}\Jhat \vct{u}_z}{\widehat\eta}-\tr(\widehat{\mat{\Sigma}}_\varepsilon)\\
&=\frac{\vct{u}_z\T\Jhat(\mat{H}-\mnoise \mat{I}_d)\Jhat \vct{u}_z}{\widehat\eta}
-(\tr(\widehat{\mat{\Sigma}}_\varepsilon)-\mnoise).
\end{split}
\end{equation}
By Lemma~\ref{lem:residual}, the first term differs from
$B_v/\widehat\eta$ by $\sqrt p$ times
\[
\Op\!\left(\bvec+\sqrt p\,\rn^2+p^{-1/2}\right).
\]
Conditional on the factors, $\vct{\xi}_n^\circ$ is centered with covariance $\Sg$, so
\[
\Var_\xi\!\left(\frac{2\sqrt{\Ieff}\,\alp\T\vct{\xi}_n^\circ}{\snoise}\right)
=\frac{2\Ieff}{\taunoise}\alp\T\Sg\alp
\le Cn\|\alp\|^2/p.
\]
Together with Lemma~\ref{lem:geometry}, this gives
\[
\frac{\Ieff\|\alp\|^2}{\snoise}-\Delta_n=\Op(\sqrt{n/p}),
\]
\[
\frac{2\sqrt{\Ieff}\,\alp\T \vct{\xi}_n^\circ}{\snoise}
=\Op(\sqrt{n/p}),\qquad
\Delta_n\le Cn/\sqrt p.
\]
On $|\widehat\eta/\eta_0-1|\le1/2$ and
$|\widehat v_\varepsilon/\taunoise-1|\le1/2$, normalization adds
\[
\Op\{(\epsT+n^{-1})(1+\Delta_n)\}.
\]
Term by term,
\[
\epsT\Delta_n
\le C\left(\frac{n}{\sfac\sqrt p}+\rn p^{\delta/2}\sqrt{n/p}\right)
\le C\bvec,
\]
\[
n^{-1}(1+\Delta_n)\le C(n^{-1}+p^{-1/2}),
\qquad\frac{|\tr(\widehat{\mat{\Sigma}}_\varepsilon)-\mnoise|}{\sqrt p}=\Op(n^{-1/2}+p^{-1/2}).
\]
These bounds prove \eqref{eq:sumreplacement}.
For the coordinatewise score difference, let $\vct{d}_u=\Jhat \vct{u}_z-\vct{v}$. Then
\[
\Xmat\Jhat \vct{u}_z-\Em\Oz \vct{v}-\alp \omega_z \widehat\eta
=\mat{A}\Gmat\Ptime\Jhat \vct{u}_z+\mat{W}\vct{d}_u.
\]
Consequently,
\[
\|\mat{A}\Gmat\Ptime\Jhat \vct{u}_z\|_\infty
\le\frac{C}{\sqrt p}\|\Gmat\|\|\Ptime\Jhat \vct{u}_z\|
=\Op(p^{-1/2}+\rn),
\]
\[
\|\mat{W}\vct{d}_u\|_\infty
\le\max_i\|\vct{e}_i\T\Em\|\|\vct{d}_u\|
=\Op(\sqrt n\epsT)=\Op(\bvec).
\]
On $\widehat\eta,\eta_0\ge c$, the exact identity is
\[
\begin{split}
\sqrt{\widehat{\mathcal I}}\,\widehat{\alp}-\vct{\xi}_n^\circ-\sqrt{\Ieff}\,\alp
={}&\left(\sqrt{\eta_0/\widehat\eta}-1\right)\vct{\xi}_n^\circ
+\omega_z (\sqrt{\widehat\eta}-\sqrt{\eta_0})\alp\\
&+\frac{\mat{A}\Gmat\Ptime\Jhat \vct{u}_z+\mat{W}\vct{d}_u}{\sqrt{\widehat\eta}}.
\end{split}
\]
The first two terms are $\Op(\epsT\sqrt{\Lnp})$ and
$\Op(\sqrt n\epsT)$ in sup norm. Since $\Lnp=O(\log n)$,
\begin{equation}\label{eq:vectorreplacement}
\|\sqrt{\widehat{\mathcal I}}\,\widehat{\alp}-\vct{\xi}_n^\circ-\sqrt{\Ieff}\,\alp\|_\infty=\Op(\bvec).
\end{equation}
Under the null, Lemma~\ref{lem:scales} gives
\[
\|\widehat{\mat{D}}_\varepsilon^{-1/2}\sqrt{\widehat{\mathcal I}}\,\widehat{\alp}-\mat{D}_\varepsilon^{-1/2}\vct{\xi}_n^\circ\|_\infty
=\Op(\bvec+\vdiag\sqrt{\Lnp}).
\]
The squared-difference identity and the oracle maximum bound $O_\mathbb P(\sqrt{\Lnp})$ prove \eqref{eq:maxreplacement}.
\end{proof}

\subsection{Joint limiting distributions}\label{app:joint}
\subsubsection{A common coupling for each conditional design}
\begin{external}[Sakhanenko's scalar strong approximation]\label{ext:sakha}
For independent centered $X_t$ satisfying $\mathbb E|X_t|^3e^{|X_t|/a}\le a\mathbb E X_t^2$, Theorem~1 in \citet{zaitsev} gives a coupling with independent centered Gaussian $G_t$, $\mathbb E G_t^2=\mathbb E X_t^2$, such that
\[
\mathbb E\exp\left\{\frac{c}{a}\max_{l\le n}\left|\sum_{t\le l}(X_t-G_t)\right|\right\}
\le1+\frac{(\sum_t\mathbb E X_t^2)^{1/2}}{a}.
\]
\end{external}
\begin{lemma}[Joint comparison for each conditional design]\label{lem:coupling}
Fix any factor realization in $\mathcal E_{\beta,n}$. On its coupling space one can construct $\vct{\xi}^\dagger,\vct{G}_\varepsilon$ with marginals equal to the conditional law of $\vct{\xi}_n^\circ$ and $N_p(0,\Sg)$, respectively, such that
\[
\mathbb P_{\rm c}\{\|\mat{D}_\varepsilon^{-1/2}(\vct{\xi}^\dagger-\vct{G}_\varepsilon)\|_\infty>C\epsilon_n\Lnp\}
\le C(np)^{-2},\qquad \epsilon_n=\log(2n)\sqrt{\Lnp/n},
\]
\[
\mathbb E_{\rm c}\left|\frac{\|\vct{\xi}^\dagger\|^2-\|\vct{G}_\varepsilon\|^2}{\snoise}\right|^2
\le C\epsilon_n^2.
\]
Here $\mathbb P_{\rm c},\mathbb E_{\rm c}$ denote the coupling law;
$C$ is uniform over $\mathcal E_{\beta,n}$.
\end{lemma}
\begin{proof}
Equation~\eqref{eq:design-event} gives $\max_t|\beta_t|\le C_\beta\sqrt{\Lnp/n}$. Choose $a=C\max_t|\beta_t|$ with a sufficiently large fixed $C$. Then
\[
\mathbb E|\beta_t\xi_{tj}|^3e^{|\beta_t\xi_{tj}|/a}
\le C|\beta_t|^3\le a\beta_t^2=a\mathbb E(\beta_t\xi_{tj})^2.
\]
The normalization $\sum_t\beta_t^2=1$ also implies $\max_t|\beta_t|\ge n^{-1/2}$. External result~\ref{ext:sakha} gives, for each innovation coordinate, a pair $u_j,g_j$ satisfying
\[
u_j\overset d=\sum_t\beta_t\xi_{tj},\qquad g_j\sim N(0,1),\qquad
\mathbb E_{\rm c}e^{c|u_j-g_j|/a}\le1+1/a.
\]
For a sufficiently large $C_1$, Jensen's inequality reduces the exponent:
\[
\mathbb E_{\rm c}\exp\left\{\frac{|u_j-g_j|}{C_1a\log(2+1/a)}\right\}\le2,
\quad\|u_j-g_j\|_{\psi_1}\le C\epsilon_n.
\]
Here $\|X\|_{\psi_1}:=\inf\{b>0:\mathbb E e^{|X|/b}\le2\}$.
\[
a\log(2+1/a)\le C\sqrt{\Lnp/n}\log(2n)=C\epsilon_n.
\]
Take independent coupling pairs across $j$. With $d_j=u_j-g_j$,
\[
\mathbb E_{\rm c}d_j=0,\qquad \mathbb E_{\rm c}|d_j|^b\le b!(C\epsilon_n)^b\quad(b\ge2).
\]
For $\sum_jb_j^2=1$ and $|t|\le c/\epsilon_n$, expansion of the exponential gives
\[
\begin{split}
\mathbb E_{\rm c}e^{tb_jd_j}
&\le1+\sum_{a=2}^{\infty}(|tb_j|C\epsilon_n)^a\\
&\le1+C t^2b_j^2\epsilon_n^2\le e^{Ct^2b_j^2\epsilon_n^2}.
\end{split}
\]
Multiply over $j$, choose $t=\min\{x/(4C\epsilon_n^2),c/(2\epsilon_n)\}$, and apply the same calculation to the opposite tail:
\[
\mathbb P_{\rm c}(|\sum_jb_jd_j|>x)
\le2\exp\{-c\min(x^2/\epsilon_n^2,x/\epsilon_n)\}.
\]
Set $\vct{\xi}^\dagger=\Sg^{1/2}\vct{u}$ and $\vct{G}_\varepsilon=\Sg^{1/2}\vct{g}$. For $\mat{B}=\mat{D}_\varepsilon^{-1/2}\Sg^{1/2}$,
\[
\|\vct{e}_i\T \mat{B}\|^2=1,\qquad
\mathbb P_{\rm c}(\|\mat{B}(\vct{u}-\vct{g})\|_\infty>C_2\epsilon_n\Lnp)
\le2p e^{-cC_2\Lnp}\le C(np)^{-2}
\]
for sufficiently large $C_2$. For the quadratic form define
\[
a_j=u_j^2-g_j^2,\qquad b_{j\ell}=u_ju_\ell-g_jg_\ell,\qquad
e_j=\mathbb E_{\rm c}(u_j-g_j)^2.
\]
Moment matching and independence across coupling pairs give
\[
\mathbb E_{\rm c}a_j=0,\quad
\mathbb E_{\rm c}a_j^2\le\{\mathbb E_{\rm c}(u_j-g_j)^4\}^{1/2}\{\mathbb E_{\rm c}(u_j+g_j)^4\}^{1/2}
\le C\epsilon_n^2,
\]
\[
e_j=2-2\mathbb E_{\rm c}(u_jg_j),\qquad
\mathbb E_{\rm c}b_{j\ell}^2=e_j+e_\ell-e_je_\ell/2\le C\epsilon_n^2.
\]
Centering the singly occurring coordinates gives
\[
\mathbb E_{\rm c}\{(u_ju_\ell-g_jg_\ell)(u_ju_a-g_jg_a)\}=0\quad(j,\ell,a\text{ distinct}).
\]
Likewise $\mathbb E_{\rm c}(a_jb_{ab})=0$. Hence
\[
\begin{split}
\mathbb E_{\rm c}(\|\vct{\xi}^\dagger\|^2-\|\vct{G}_\varepsilon\|^2)^2
={}&\sum_j(\Sg)_{jj}^2\mathbb E_{\rm c}a_j^2
+4\sum_{j<\ell}(\Sg)_{j\ell}^2\mathbb E_{\rm c}b_{j\ell}^2\\
\le{}&C\epsilon_n^2\taunoise.
\end{split}
\]

\end{proof}

\subsubsection{Gaussian maxima and an unstandardized quadratic form}
\begin{lemma}[Gaussian joint limit]\label{lem:gaussian}
If $\vct{G}_\varepsilon\sim N_p(0,\Sg)$, then
\[
\left(\max_i\frac{G_{\varepsilon,i}^2}{(\Sg)_{ii}}-2\log p+\log\log p,
\frac{\|\vct{G}_\varepsilon\|^2-\mnoise}{\snoise}\right)
\Rightarrow(G_0,Z_0),\quad G_0\ind Z_0,
\]
where $\mathbb P(G_0\le x)=F_G(x)$ and $Z_0\sim N(0,1)$.
\end{lemma}
\begin{proof}
Let $\mat{R}=\mat{D}_\varepsilon^{-1/2}\Sg\mat{D}_\varepsilon^{-1/2}$. Its bounded spectrum and unit diagonal imply
\[
|R_{ij}|\le1-c'=:\rho_*<1\quad(i\ne j),\qquad \sum_{i,j}R_{ij}^2\le Cp.
\]
For the Gaussian density $\phi_s$ with covariance
$\mat{R}(s)=\mat{I}+s(\mat{R}-\mat{I})$, differentiation gives
$\partial_s\phi_s=\sum_{i<j}R_{ij}\partial_{ij}\phi_s$.
Integrating over $[-a,a]^p$ reduces each mixed derivative to four boundary
integrals, each bounded by its bivariate marginal density. Therefore
\[
\begin{aligned}
|F_1(a)-F_0(a)|
&\le\sum_{i<j}|R_{ij}|\int_0^1
\frac{2e^{-a^2/(1+|sR_{ij}|)}}{\pi\sqrt{1-s^2R_{ij}^2}}\,ds\\
&\le C\sum_{i<j}|R_{ij}|e^{-a^2/(1+|R_{ij}|)},
\end{aligned}
\]
where $F_s(a)=\int_{[-a,a]^p}\phi_s$.
Take $a^2=2\log p-\log\log p+x$ and split the sum at
$\epsilon_p=(\log p)^{-2}$. Cauchy--Schwarz for the small correlations
and $|R_{ij}|\le\epsilon_p^{-1}R_{ij}^2$ for the others give
\[
|F_1(a)-F_0(a)|
\le C_x\{p^{-1/2}\log p+
 p^{-(1-\rho_*)/(1+\rho_*)}(\log p)^{2+1/(1+\rho_*)}\}\to0.
\]
The normal tail bound
$a\phi(a)/(1+a^2)\le1-\Phi(a)\le\phi(a)/a$ yields
$2p\{1-\Phi(a)\}\to\pi^{-1/2}e^{-x/2}$ and
$F_0(a)=[1-2\{1-\Phi(a)\}]^p\to F_G(x)$. Hence
\[
M_G:=\max_i\frac{G_{\varepsilon,i}^2}{(\Sg)_{ii}}
-2\log p+\log\log p\Rightarrow G_0.
\]
Diagonalize $\Sg$ with eigenvalues $\lambda_j\in[c,C]$. For
$Q_G=(\|\vct{G}_\varepsilon\|^2-\mnoise)/\snoise$,
\[
Q_G\overset d=\snoise^{-1}\sum_j\lambda_j(g_j^2-1),\qquad
\snoise^2=2\sum_j\lambda_j^2,\quad g_j\stackrel{\rm iid}{\sim}N(0,1).
\]
For every fixed real $t$, the Gaussian moment generating function gives
\[
\begin{aligned}
\log\mathbb E e^{tQ_G}
&=\sum_j\left\{-t\lambda_j/\snoise
 -\tfrac12\log(1-2t\lambda_j/\snoise)\right\}\\
&=t^2/2+O_t\left(\frac{\sum_j\lambda_j^3}{\snoise^3}\right)
=t^2/2+O_t(p^{-1/2}).
\end{aligned}
\]
Taking $t=iu$ gives $Q_G\Rightarrow N(0,1)$ by the continuity theorem
\citep{billingsley1999}.

To prove independence, exponentially tilt by
$d\mathbb P_t/d\mathbb P=e^{tQ_G}/\mathbb E e^{tQ_G}$.
The tilted vector is Gaussian with covariance
$\Sg(\mat{I}-2t\Sg/\snoise)^{-1}$.
Couple it as $\vct{G}_t=(\mat{I}-2t\Sg/\snoise)^{-1/2}\vct{G}_\varepsilon$.
The spectral bound and a Gaussian union bound imply
\[
\begin{gathered}
\|(\mat{I}-2t\Sg/\snoise)^{-1/2}-\mat{I}\|\le C_t p^{-1/2},\\
\max_i|G_{t,i}-G_{\varepsilon,i}|
=\Op\!\left(\sqrt{\log p/p}\right),\qquad
|M_t-M_G|=\Op(\log p/\sqrt p),
\end{gathered}
\]
where $M_t$ uses $\vct{G}_t$ with the same original coordinate denominators.
It follows that $\mathbb P_t(M_G\le x)\to F_G(x)$ and hence
\[
\mathbb E[e^{tQ_G}\indc\{M_G\le x\}]
=\mathbb E e^{tQ_G}\mathbb P_t(M_G\le x)
\longrightarrow e^{t^2/2}F_G(x).
\]
For $\mu_{p,x}(B)=\mathbb P(M_G\le x,Q_G\in B)$,
$\mu_{p,x}(\mathbb R)\to F_G(x)>0$. For every fixed $A>0$,
\[
\int e^{A|y|}\,\mu_{p,x}(dy)
\le\mathbb E e^{AQ_G}+\mathbb E e^{-AQ_G}\le C_A.
\]
The normalized measures are tight; the bound with $2A$ gives uniform
integrability. Every subsequential limit has Laplace transform $e^{t^2/2}$,
so $\mathbb P(M_G\le x,Q_G\le y)\to F_G(x)\Phi(y)$.
\end{proof}

\begin{lemma}[Oracle joint limit]\label{lem:oracle}
We have $(M_n^\circ,Q_n^\circ)\Rightarrow(G_0,Z_0)$ with $G_0\ind Z_0$.
\end{lemma}
\begin{proof}
On $\mathcal E_{\beta,n}$, use the common coupling in
Lemma~\ref{lem:coupling}. A Gaussian union bound gives, for a sufficiently
large fixed $C$,
\[
\mathbb P\!\left\{\max_i
\frac{|G_{\varepsilon,i}|}{\sqrt{(\Sg)_{ii}}}>C\sqrt{\Lnp}\right\}
\le2p e^{-C^2\Lnp/2}\le C(np)^{-2}.
\]
The squared-difference identity therefore gives,
with $d_{n,p}=C(\epsilon_n\Lnp^{3/2}+\epsilon_n^2\Lnp^2)\to0$,
\[
\mathbb P_{\rm c}(|M^\dagger-M_G|>d_{n,p})\le C(np)^{-2},\qquad
\mathbb P_{\rm c}(|Q^\dagger-Q_G|>h)\le C\epsilon_n^2/h^2.
\]
Here $(M^\dagger,Q^\dagger)$ has the conditional oracle law and
$(M_G,Q_G)$ has the law in Lemma~\ref{lem:gaussian}.
For fixed $h>0$ and large $n$, put
$F_{G,p}(x,y)=\mathbb P(M_G\le x,Q_G\le y)$.
Integrating the conditional sandwich over the factor law yields
\[
\begin{aligned}
F_{G,p}(x-h,y-h)-r_{n,p}(h)
&\le\mathbb P(M_n^\circ\le x,Q_n^\circ\le y)\\
&\le F_{G,p}(x+h,y+h)+r_{n,p}(h),
\end{aligned}
\]
\[
r_{n,p}(h)\le C\{n^{-1}+(np)^{-2}+\epsilon_n^2/h^2\}\to0
\]
by \eqref{eq:design-event}. Apply Lemma~\ref{lem:gaussian} and let
$h\downarrow0$.
\end{proof}

\clearpage
\section{Proofs of the main testing results}\label{app:mainproofs}
This appendix proves the approximation, calibration and power results in their order of appearance in the main text. Appendix~\ref{app:bootstrap-proof} treats bootstrap calibration, and Appendix~\ref{app:mse} proves the mean-square comparison. The supporting identities and probabilistic bounds have been established in Appendix~\ref{app:technical}.

\subsection{First-order approximation}
\begin{proof}[Proof of Proposition~\ref{prop:common-approximation}]
Lemma~\ref{lem:loading} establishes \eqref{eq:admission}.
By Lemmas~\ref{lem:fwl}, \ref{lem:geometry} and~\ref{lem:time},
\[
\frac{\widehat{\mathcal I}}{\Ieff}-1=\Op(\epsT),\quad
\frac{\Ieff}{n}-\cstar^{-1}=\Op(n^{-1/2}),\qquad
\widehat c-\cstar=\Op(\epsT+n^{-1/2}).
\]
Lemmas~\ref{lem:replacement} and~\ref{lem:rates} give
\[
T_{\mathrm{sum}}-Q_n^\circ-\Delta_n=\Op(\bsum),\qquad
T_{\mathrm{max}}-M_n^\circ=\Op(\bmax+\bmax^2)\quad(H_0).
\]
The true-factor regression satisfies
$\sqrt{\Ieff}\,\widehat{\alp}^{\circ}
=\sqrt{\Ieff}\,\alp+\vct{\xi}_n^\circ$. Hence
\[
T_{\mathrm{sum}}^\circ
=Q_n^\circ+\frac{\Ieff\|\alp\|^2
+2\sqrt{\Ieff}\,\alp\T\vct{\xi}_n^\circ}{\snoise}
=Q_n^\circ+\Delta_n+o_\mathbb P(1),
\]
where the last equality uses the bounds in the proof of Lemma~\ref{lem:replacement}. Under $H_0$, $T_{\mathrm{max}}^\circ=M_n^\circ$. The assertion follows since $\epsT,\bsum,\bmax\to0$. Under Assumption~\ref{ass:rank},
the selected-rank quantities agree with their true-rank counterparts on
$\{\widehat r=r\}$, whose probability tends to one.
\end{proof}

\subsection{Joint null distribution}\label{app:null-proof}
\begin{proof}[Proof of Theorem~\ref{thm:null}]
Under $H_0$, $(T_{\mathrm{max}}^\circ,T_{\mathrm{sum}}^\circ)=(M_n^\circ,Q_n^\circ)$. Proposition~\ref{prop:common-approximation} and Lemma~\ref{lem:oracle} give
\[
(T_{\mathrm{max}},T_{\mathrm{sum}})=(M_n^\circ,Q_n^\circ)+o_\mathbb P(1)
\Rightarrow(G_0,Z_0),\qquad G_0\ind Z_0.
\]
Thus $(p_{\mathrm{max}},p_{\mathrm{sum}})\Rightarrow(U_1,U_2)$ with independent
$U_j\sim\operatorname{Unif}(0,1)$. For $C_j=\cot(\pi U_j)$,
\[
\mathbb E e^{it(C_1+C_2)/2}
=e^{-|t|/2}e^{-|t|/2}=e^{-|t|}.
\]
Continuity of the cotangent map on $(0,1)^2$ gives the Cauchy limit.
Assumption~\ref{ass:rank} and Proposition~\ref{prop:number-transfer}
transfer the conclusions to the selected-rank LAG statistics.
\end{proof}

\subsection{Power of the Sum test}
\begin{proof}[Proof of Theorem~\ref{thm:power}]
Lemmas~\ref{lem:replacement} and~\ref{lem:oracle} give
$T_{\mathrm{sum}}=Q_n^\circ+\Delta_n+\Op(\bsum)$ and $Q_n^\circ\Rightarrow N(0,1)$.
Since $\bsum\to0$, Slutsky's theorem yields the local limit and
its stated rejection probability. When $\Delta_n\to\infty$,
\[
T_{\mathrm{sum}}/\Delta_n-1=\Op\{(1+\bsum)/\Delta_n\}\to_{\mathbb P}0.
\]
Assumption~\ref{ass:rank} and Proposition~\ref{prop:number-transfer} give the
selected-rank assertions.
\end{proof}

\subsection{Power of the Max test}
\begin{proof}[Proof of Corollary~\ref{cor:max}]
Choose deterministic $i_0$ with $|\alpha_{i_0}|=\|\alp\|_\infty$. Equation~\eqref{eq:vectorreplacement} gives
\[
\sqrt{\widehat{\mathcal I}}\,\widehat\alpha_{i_0}=\sqrt{\Ieff}\alpha_{i_0}+\xi_{n,i_0}^\circ+O_\mathbb P(\bvec),
\quad \xi_{n,i_0}^\circ=O_\mathbb P(1).
\]
Lemma~\ref{lem:scales} gives $c\le(\widehat{\mat{\Sigma}}_\varepsilon)_{i_0i_0}\le C$ on a high-probability event. Hence
\[
\frac{|\sqrt{\widehat{\mathcal I}}\,\widehat\alpha_{i_0}|}{\sqrt{(\widehat{\mat{\Sigma}}_\varepsilon)_{i_0i_0}}\sqrt{\log(2p)}}
\ge c\frac{\sqrt n\|\alp\|_\infty}{\sqrt{\log(2p)}}
-\frac{O_\mathbb P(1)+O_\mathbb P(\bvec)}{\sqrt{\log(2p)}}\to_{\mathbb P}\infty.
\]
Squaring and subtracting the extreme-value center proves consistency. 
\end{proof}

\subsection{Power of the Cauchy test}\label{app:cauchy-power-proof}
\begin{proof}[Proof of Corollary~\ref{cor:cauchy}]
The Sum approximation and $\Delta_n\ge0$ imply that the negative part of $T_{\mathrm{sum}}$ is tight. For the Max statistic define the deterministic set
\[
\mathcal J_n=\{i:|\alpha_{i}|\le(n\Lnp)^{-1/2}\},\qquad
p-|\mathcal J_n|\le C_\alpha^2 n\Lnp=o(p).
\]
On this set, $\sqrt{\Ieff}|\alpha_{i}|\le\Lnp^{-1/2}$ and Lemma~\ref{lem:scales} gives
\[
\max_{i\in\mathcal J_n}|(\widehat{\mat{\Sigma}}_\varepsilon)_{ii}-(\Sg)_{ii}|
=\Op(\vdiag+(n\Lnp)^{-1}).
\]
The score bound \eqref{eq:vectorreplacement} therefore yields
\[
\max_{i\in\mathcal J_n}
\left|\frac{\widehat{\mathcal I}\,\widehat\alpha_i^2}{(\widehat{\mat{\Sigma}}_\varepsilon)_{ii}}
-\frac{(\xi_{n,i}^\circ)^2}{(\Sg)_{ii}}\right|
=\Op\{1+\bvec\sqrt{\Lnp}+\vdiag\Lnp\}=\Op(1).
\]
Put $p_J=|\mathcal J_n|$. The principal submatrix on $\mathcal J_n$ has spectrum in $[c,C]$, and
\[
\begin{gathered}
0\le1-p_J/p\le Cn\Lnp/p\to0,\\
|2\log(p_J/p)-\log(\log p_J/\log p)|=O(n\Lnp/p).
\end{gathered}
\]
Lemmas~\ref{lem:coupling} and~\ref{lem:gaussian}, restricted to these coordinates, therefore give
\[
\max_{i\in\mathcal J_n}\frac{(\xi_{n,i}^\circ)^2}{(\Sg)_{ii}}
-2\log p+\log\log p\Rightarrow G_0.
\]
Consequently the negative part of $T_{\mathrm{max}}$ is tight. For $j\in\{\mathrm{sum},\mathrm{max}\}$, continuity and monotonicity of $F_j$ imply
\[
\{\cot(\pi p_j)\}^{-}=O_\mathbb P(1),\qquad x^-:=\max(-x,0).
\]
For a consistent component $j$ and the other component $j'$, Theorem~\ref{thm:power} or Corollary~\ref{cor:max} gives
\[
p_j\to_{\mathbb P}0,\quad\cot(\pi p_j)\to_{\mathbb P}\infty,\quad
2T_{\mathrm{cauchy}}\ge\cot(\pi p_j)-\{\cot(\pi p_{j'})\}^{-}\to_{\mathbb P}\infty.
\]
\end{proof}

\clearpage
\section{Proof of bootstrap validity}\label{app:bootstrap-proof}
We control the loadings and residuals used by the bootstrap generator in
Section~\ref{sec:bootstrap}, establish a uniform oracle limit, and
prove Theorem~\ref{thm:column-bootstrap} after refitting.

Write $X_n=\Opl(b_n)$ for $X_n=O_{\mathbb P}(\Lnp^C b_n)$ with fixed $C<\infty$; $\Ops$ denotes the conditional version in probability. Put $u_n=n/\sfac^2$ and $v_n=p^{1/2+\delta}/n$. Assumption~\ref{ass:D} and Lemma~\ref{lem:rates} give
\[
\begin{gathered}
u_n+v_n=O(n^{-\eta}),\qquad \sqrt n\epsT\asymp\sqrt{u_n},\\
\rn^2p/n=v_n^2p^{-\delta},\qquad n/p+\sqrt p/n\longrightarrow0.
\end{gathered}
\]

The pilot-residual and uniform oracle arguments below use the common
loading condition and also support the PCA bootstrap. The final
refitting argument here concerns LAG; the PCA counterpart is proved
in Appendix~\ref{app:pca-bootstrap}.

\subsection{Fitted loadings and residuals}
For a true-rank loading estimator satisfying \eqref{eq:admission}, let
\[
\begin{gathered}
\widehat{\mat{F}}=(\widehat{\vct{F}}_1,\ldots,\widehat{\vct{F}}_n),\qquad
\widehat{\mat{G}}_c=\widehat{\mat{F}}\mat{M}_1=\widehat{\mat{A}}\T\Rmat,\\
\widehat{\mat{E}}=\Rmat-\widetilde{\mat{A}}\widehat{\mat{G}}_c,\qquad
\mat{S}_R=\widehat{\mat{E}}\widehat{\mat{E}}\T/n,\quad
\mat{D}_R=\diag(\mat{S}_R),\\
m_R=\tr(\mat{S}_R),\quad\tau_R=\tr(\mat{S}_R^2),\quad
v_R=n^{-1}\sum_t(\|\widehat{\vct{\varepsilon}}_t\|^2-m_R)^2.
\end{gathered}
\]
Let $\mat{M}_{c0}$ remove $[\one_n,\mat{Z},\Fm\T]$.

\begin{lemma}[Pilot residual bounds]\label{lem:column-pilot}
Under Assumptions~\ref{ass:A}--\ref{ass:D} and $H_0$, let the true-rank loading estimator satisfy \eqref{eq:admission}. Then
\[
\begin{gathered}
\widehat{\mat{E}}\one_n=0,\quad m_R\asymp_{\mathbb P}p,\quad
\|\widehat{\mat{E}}\|=\Op(\sqrt p),\quad
\max_{i,t}|\widehat\varepsilon_{it}|=\Opl(1),\\
\max_i|D_{R,ii}-(\Sg)_{ii}|=\Opl(n^{-1/2}+\epsT),\qquad
v_R=\Opl(p+p^2/n),\\
\max_i\|\vct{e}_i\T(\widehat{\mat{E}}-\Em\mat{M}_{c0})\|
=\Op(\sqrt n\epsT).
\end{gathered}
\]
There is a symmetric matrix $\mat{\Delta}_R$ such that
\begin{equation}\label{eq:residual-gram-column}
\begin{gathered}
\widehat{\mat{E}}\T\widehat{\mat{E}}
=\frac{m_R}{1-1/n}\mat{M}_1+\mat{\Delta}_R,\\
\mat{\Delta}_R\one_n=0,\quad\tr\mat{\Delta}_R=0,\qquad
\|\mat{\Delta}_R\|_F^2=\Op(p^2+n^2p).
\end{gathered}
\end{equation}
Consequently,
\begin{equation}\label{eq:column-trace-size}
\tau_R=\frac{m_R^2}{n}\{1+\Op(n^{-1}+n/p)\}
\asymp_{\mathbb P}p^2/n.
\end{equation}
For $\mat{A}_0=\widetilde{\mat{A}}(\widetilde{\mat{A}}\T\widetilde{\mat{A}})^{-1/2}$,
\begin{equation}\label{eq:fixed-loading-orthogonality}
\begin{gathered}
\|\widetilde{\mat{A}}-\widehat{\mat{A}}\|=\Op(\rn),\qquad
\max_i\|\vct{e}_i\T\mat{A}_0\|=\Opl(p^{-1/2}),\\
a_R:=\tr(\mat{A}_0\T\mat{S}_R\mat{A}_0)
=\Opl(p^{1+\delta}/n^2)=\op(1),\qquad
\|\bar{\widehat{\vct{F}}}\|/\sqrt p=\Op(1).
\end{gathered}
\end{equation}
\end{lemma}
\begin{proof}
Since $\Rmat\one_n=0$ and $\Rmat\mat{Z}=0$, substitution gives
\[
\begin{gathered}
\widehat{\mat{G}}_c\one_n=0,\quad \widehat{\mat{G}}_c\mat{Z}=0,\quad
\widehat{\mat{E}}\one_n=0,\quad \widehat{\mat{E}}\mat{Z}=0,\quad
\widehat{\mat{E}}\widehat{\mat{G}}_c\T=0,\\
\widehat{\mat{A}}\T\widetilde{\mat{A}}=\mat{I}_r,\qquad
\widehat{\mat{A}}\T\widehat{\mat{E}}=0.
\end{gathered}
\]
For the centered design, the gap is $\sqrt n\sfac$.
Lemma~\ref{lem:time} gives, for its fitted residual projector
$\widehat{\mat{M}}_c$,
\[
\|\widehat{\mat{M}}_c-\mat{M}_{c0}\|=O_{\mathbb P}(\epsT),\qquad
\|\Fm\widehat{\mat{M}}_c\|=O_{\mathbb P}(\sqrt n+\rn\sqrt p).
\]
Since $\widehat{\mat{E}}=(\mat{A}\Fm+\Em)\widehat{\mat{M}}_c$,
\[
\begin{aligned}
&\max_i\|\vct{e}_i\T(\widehat{\mat{E}}-\Em\mat{M}_{c0})\|\\
&\le Cp^{-1/2}\|\Fm\widehat{\mat{M}}_c\|
+\max_i\|\vct{e}_i\T\Em\|\,
\|\widehat{\mat{M}}_c-\mat{M}_{c0}\|\\
&=\Op(\sqrt{n/p}+\sqrt n\epsT)=\Op(\sqrt n\epsT).
\end{aligned}
\]
Here $\rn\sqrt p=o(\sqrt n)$ and $\epsT\ge p^{-1/2}$.
A sub-Gaussian union bound for $\Em\mat{M}_{c0}$ and
$\sqrt n\epsT=o(1)$ give the entry bound. Lemma~\ref{lem:noise}
and the row approximation give
\[
\begin{gathered}
\max_i\left|n^{-1}\|\vct{e}_i\T\Em\mat{M}_{c0}\|^2
-(\Sg)_{ii}\right|=\Opl(n^{-1/2}),\\
\max_i n^{-1}\left|
\|\vct{e}_i\T\widehat{\mat{E}}\|^2
-\|\vct{e}_i\T\Em\mat{M}_{c0}\|^2\right|=\Opl(\epsT).
\end{gathered}
\]
This proves the diagonal bound. For
$\mat{D}_E=\widehat{\mat{E}}-\Em$,
\[
\begin{gathered}
\rank(\mat{D}_E)\le m+2r+1,\qquad
\|\mat{D}_E\|\le\|\Fm\widehat{\mat{M}}_c\|+\|\Em\|
=\Op(\sqrt p),\\
\|\mat{D}_E\|_F^2=\Op(p),\qquad
\|\widehat{\mat{E}}\|=\Op(\sqrt p),\qquad
\max_t(\|\widehat{\vct{\varepsilon}}_t\|+
\|\vct{\varepsilon}_t\|)=\Opl(\sqrt p).
\end{gathered}
\]
The difference-of-squares identity and Lemma~\ref{lem:grammom} give
\[
\begin{aligned}
v_R&\le\frac2n\sum_t(\|\vct{\varepsilon}_t\|^2-\mnoise)^2
+\frac2n\sum_t(\|\widehat{\vct{\varepsilon}}_t\|^2
-\|\vct{\varepsilon}_t\|^2)^2\\
&\le\Op(p)+\Opl(p/n)\|\mat{D}_E\|_F^2
=\Opl(p+p^2/n).
\end{aligned}
\]

In the centered time coordinates, Lemma~\ref{lem:residual} gives the exact
Gram decomposition
\[
\widehat{\mat{E}}\T\widehat{\mat{E}}
=\mnoise\widehat{\mat{M}}_c+
\mat{M}_{c0}\mat{K}_\varepsilon\mat{M}_{c0}+\mat{B}_R,
\quad \rank(\mat{B}_R)\le5r,\quad\|\mat{B}_R\|=O_{\mathbb P}(n).
\]
The difference $\widehat{\mat{M}}_c-\mat{M}_1$ has fixed rank, whereas
$\|\mat{K}_\varepsilon\|_F=O_{\mathbb P}(n\sqrt p)$.
Taking traces shows
$|m_R-\mnoise|=O_{\mathbb P}(p/n+\sqrt{p/n}+1)$,
so $m_R\asymp_{\mathbb P}p$. Subtract $m_R\mat{M}_1/(1-1/n)$ to obtain
\eqref{eq:residual-gram-column}. Its two terms are Frobenius-orthogonal,
so
\[
\tau_R=\frac{m_R^2}{n-1}+\frac{\|\mat{\Delta}_R\|_F^2}{n^2},\qquad
\frac{\|\mat{\Delta}_R\|_F^2}{np^2}
=O_{\mathbb P}(n^{-1}+n/p)=o_{\mathbb P}(1).
\]

Put $\mat{V}=\widehat{\mat{A}}$, $\mat{D}=\widetilde{\mat{A}}-\mat{V}$ and $\mat{S}=\mat{V}\T\mat{A}$. The loading and time-space bounds give
\[
\begin{gathered}
\widehat{\mat{G}}_c=\mat{S}\Fm\mat{M}_c+\mat{V}\T\Em\mat{M}_c,\qquad
\lambda_{\min}(\widehat{\mat{G}}_c\widehat{\mat{G}}_c\T)\asymp_{\mathbb P}n\sfac^2,\\
\widetilde{\mat{A}}=\Rmat\widehat{\mat{G}}_c\T
(\widehat{\mat{G}}_c\widehat{\mat{G}}_c\T)^{-1},\qquad
\|\mat{D}\|=\Op\{\rn+\sqrt{p/n}/\sfac\}=\Op(\rn).
\end{gathered}
\]
For the row bound, conditional sub-Gaussian concentration and
$\|\Em\|=\Op(\sqrt p)$ yield
\[
\begin{gathered}
\max_i\|\vct{e}_i\T\Em\mat{M}_c\Fm\T\|=\Opl(\sqrt n\sfac),\\
\max_i\|\vct{e}_i\T\Em\mat{M}_c\Em\T\mat{V}\|
\le\max_i\|\vct{e}_i\T\Em\|\|\Em\|=\Opl(\sqrt{np}),\\
\max_i\|\vct{e}_i\T\widetilde{\mat{A}}\|
\le\Opl\left(p^{-1/2}+\frac1{\sqrt n\sfac}
+\frac{\sqrt p}{\sqrt n\sfac^2}\right)=\Opl(p^{-1/2}).
\end{gathered}
\]
The exact orthogonality relations give
\[
\begin{gathered}
\mat{V}\T\mat{D}=0,\quad\widetilde{\mat{A}}\T\widetilde{\mat{A}}
=\mat{I}_r+\mat{D}\T\mat{D},\\
\mat{A}_0\T\widehat{\mat{E}}
=(\mat{I}_r+\mat{D}\T\mat{D})^{-1/2}\mat{D}\T\widehat{\mat{E}},\\
a_R\le r\|\mat{D}\|^2\|\mat{S}_R\|
=\Opl\left(\frac{p^{1+\delta}}{n^2}\right)
=\Opl(v_n^2p^{-\delta})=\op(1).
\end{gathered}
\]
Finally, $\bar{\widehat{\vct{F}}}=\mat{V}\T\widehat{\vct{\mu}}$ and
$\widehat{\vct{\mu}}=\mat{A}(\vct{\lambda}+\vct{f}_w)+\vct{\varepsilon}_w$
under $H_0$, with $\|\vct{f}_w\|=\Op(\sfac/\sqrt n)$ and
$\|\vct{\varepsilon}_w\|=\Op(\sqrt{p/n})$. Thus
$\|\bar{\widehat{\vct{F}}}\|/\sqrt p=\Op(1)$.
\end{proof}

\subsection{A uniform oracle limit}
Let ${\vct{\theta}}=(\mat{C},\mat{H},\vct{b})$ range over
\[
\Theta_B=\left\{{\vct{\theta}}:\begin{array}{l}
\mat{C}=\mat{C}\T\succ0,\quad 10^{-3}\mat{I}\preceq\mat{C}^2\preceq10^3\mat{I},\\
\|\mat{H}\|\le0.95,\quad \|\vct{b}\|\le B
\end{array}\right\}.
\]
Use common independent Gaussian innovations and $\vct{w}_0\sim N(\vct{0},\mat{I}_r)$ to generate $\vct{w}_t(\mat{H})=\mat{H}\vct{w}_{t-1}(\mat{H})+\mat{J}(\mat{H})\vct{\zeta}_t$. Let $\mat{M}_{\vct{\theta}}$ remove the columns of
\[
\begin{gathered}
\mat{X}_{\vct{\theta}}=[\mat{Z},\mat{V}_{\vct{\theta}}],\qquad
\vct{e}_t\T\mat{V}_{\vct{\theta}}=(\vct{b}+\mat{C}\vct{w}_t(\mat{H}))\T,\\
\vct{\beta}_{\vct{\theta}}=\frac{\mat{M}_{\vct{\theta}}\one_n}{\sqrt{\one_n\T\mat{M}_{\vct{\theta}}\one_n}}.
\end{gathered}
\]
The Gaussian innovations and residual indices are independent of the original sample. Write $\mathbb P^*$ for their conditional law and set
\[
\begin{gathered}
Q_{R,{\vct{\theta}}}^*=\frac{\|\mat{E}^*\vct{\beta}_{\vct{\theta}}\|^2-m_R}{\sqrt{2\tau_R}},\\
M_{R,{\vct{\theta}}}^*=\max_i\frac{(\mat{E}^*\vct{\beta}_{\vct{\theta}})_i^2}{D_{R,ii}}-2\log p+\log\log p.
\end{gathered}
\]
\begin{lemma}[Uniform conditional calibration]\label{lem:var-oracle}
Under Assumptions~\ref{ass:A}--\ref{ass:D} and $H_0$, with the true-rank pilot residuals of Lemma~\ref{lem:column-pilot}, for each fixed $B<\infty$,
\[
\sup_{{\vct{\theta}}\in\Theta_B}\sup_{x,y}
|\mathbb P^*(Q_{R,{\vct{\theta}}}^*\le x,M_{R,{\vct{\theta}}}^*\le y)-\Phi(x)F_G(y)|
\longrightarrow_{\mathbb P}0.
\]
\end{lemma}
\begin{proof}
Gaussian concentration and $\|\mat{H}\|\le0.95$ give, uniformly in ${\vct{\theta}}$,
\begin{equation}\label{eq:var-weight-bounds}
\begin{gathered}
\one_n\T\mat{M}_{\vct{\theta}}\one_n\asymp_{\mathbb P^*}n,
\quad\|\vct{\beta}_{\vct{\theta}}\|=1,\quad
\max_t|\beta_{{\vct{\theta}},t}|=\Ops(n^{-1/2}),\\
\sup_{{\vct{\theta}}}\|\partial^\alpha\vct{\beta}_{\vct{\theta}}\|=\Ops(1),\qquad
\sup_{{\vct{\theta}},t}|\partial^\alpha\beta_{{\vct{\theta}},t}|=\Ops(n^{-1/2})
\quad(|\alpha|\le d_\Theta+2).
\end{gathered}
\end{equation}
Here $d_\Theta$ is fixed. Let $\widetilde{\mat{X}}_{\vct{\theta}}=[\one_n,\mat{X}_{\vct{\theta}}]$. On the original-factor regularity event,
\[
\begin{gathered}
\sum_{j\ge0}(j+1)^k0.95^j<\infty,\qquad
\sup_{\vct{\theta}}\|\partial^\alpha\widetilde{\mat{X}}_{\vct{\theta}}\|=\Ops(\sqrt n),\\
\sup_{\vct{\theta}}\left\|n^{-1}\widetilde{\mat{X}}_{\vct{\theta}}\T\widetilde{\mat{X}}_{\vct{\theta}}
-\mathbb E_w\{n^{-1}\widetilde{\mat{X}}_{\vct{\theta}}\T\widetilde{\mat{X}}_{\vct{\theta}}\}\right\|=\Ops(n^{-1/2}),\\
\inf_{\vct{\theta}}\lambda_{\min}\mathbb E_w
\{n^{-1}\widetilde{\mat{X}}_{\vct{\theta}}\T\widetilde{\mat{X}}_{\vct{\theta}}\}\ge c_B>0.
\end{gathered}
\]
Differentiating the inverse Gram matrix proves \eqref{eq:var-weight-bounds}. Here $\mathbb E_w$ integrates only the generated Gaussian path.

For $N_j=\sum_t\indc\{I_t=j\}$,
$\mathbb P_I(\max_jN_j\ge k)\le n(e/k)^k$. Hence
\[
\|\mat{E}^*\|=\Ops(\sqrt p),\qquad
\max_i\|\vct{e}_i\T\mat{E}^*\|=\Ops(\sqrt n).
\]
For fixed ${\vct{\theta}}$, $\vct{\beta}_{\vct{\theta}}$ is independent of the original error matrix conditional on the original factors. Put $b_p=2\log p-\log\log p$ and
\[
F_p(y)=\begin{cases}[2\Phi(\sqrt{b_p+y})-1]^p,&b_p+y\ge0,\\0,&b_p+y<0.\end{cases}
\]
Then $\sup_y|F_p(y)-F_G(y)|\to0$ and the densities of $F_p$ are uniformly bounded for large $p$.

For fixed ${\vct{\theta}}$, condition on its generated path, write $\vct{\beta}=\vct{\beta}_{\vct{\theta}}$, and define
\[
\begin{gathered}
C_n=2\sum_{s<t}\beta_s\beta_t(1\{I_s=I_t\}-1/n),\qquad
v_4=\sum_t\beta_t^4,\\
c_j=\sum_t\beta_t1\{I_t=j\},\quad
\vct{d}=\vct{c}-n^{-1}\Big(\sum_t\beta_t\Big)\one_n,\\
\E C_n=0,\quad\sigma_C^2=\E C_n^2=2(1-1/n)(1-v_4)/n,\\
\E\vct{d}\vct{d}\T=\mat{M}_1/n,\quad
\|\vct{d}\|^2=1-1/n+C_n.
\end{gathered}
\]
For $D_t=2\beta_t\sum_{s<t}\beta_s(1\{I_t=I_s\}-1/n)$ and
$\mathcal F_t=\sigma(I_1,\ldots,I_t)$,
\[
\begin{gathered}
\sum_t\E D_t^4\le C(v_4^2/n+v_4/n^2),\\
\sigma_C^{-4}\sum_t\E D_t^4\le C(nv_4^2+v_4)=O(\Lnp^2/n),\\
V_n:=\sum_t\E(D_t^2\mid\mathcal F_{t-1})
=\frac4n\sum_t\beta_t^2(1-1/n)\sum_{s<t}\beta_s^2\\
\hfill{}+\frac8n\sum_t\beta_t^2
\sum_{s<u<t}\beta_s\beta_u(1\{I_s=I_u\}-1/n),\\
\E V_n=\sigma_C^2,\qquad\operatorname{Var}(V_n)\le Cn^{-3}.
\end{gathered}
\]
The quantitative martingale bound of \citet{haeusler1988} gives
\[
\begin{aligned}
&\sup_x\left|\mathbb P_I(C_n/\sigma_C\le x)-\Phi(x)\right|\\
&\quad\le C\left\{\sigma_C^{-4}\sum_t\mathbb E_I D_t^4
+\mathbb E_I|V_n/\sigma_C^2-1|^2\right\}^{1/5}
\le C(\Lnp^C/n)^{1/5}.
\end{aligned}
\]
Thus $\sqrt{n/2}C_n\Rightarrow N(0,1)$ uniformly over these weights.
Independent draws also satisfy
\[
\E\langle\vct{d}^{(1)},\vct{d}^{(2)}\rangle^2=(n-1)/n^2.
\]

Conditioning on the original data,
$\mat{E}^*\vct{\beta}=\widehat{\mat{E}}\vct{d}$, and
Lemma~\ref{lem:column-pilot} gives
\[
\begin{gathered}
\|\widehat{\mat{E}}\vct{d}\|^2-m_R=\frac{m_R}{1-1/n}C_n+R_n,\\
R_n=\sum_t\beta_t^2\Delta_{R,I_tI_t}
+2\sum_{s<t}\beta_s\beta_t\Delta_{R,I_sI_t},\\
\E^*R_n^2=v_4v_R+2(1-v_4)\|\mat{\Delta}_R\|_F^2/n^2,\\
\E^*R_n^2/\tau_R=\Opl(p^{-1}+n^{-1}+n/p).
\end{gathered}
\]
Also $\E^*\|(\mat{I}_n-\mat{M}_{c0})\vct{d}\|^2=(m+r)/n$,
so
\[
\begin{aligned}
\|\widehat{\mat{E}}\vct{d}-\mat{E}\vct{d}\|_\infty
&\le\Opl(\sqrt n\epsT)\|\vct{d}\|
+\max_i\|\vct{e}_i\T\mat{E}(\mat{I}_n-\mat{M}_{c0})\|
\|(\mat{I}_n-\mat{M}_{c0})\vct{d}\|\\
&=\Ops(\sqrt n\epsT+n^{-1/2})=\ops(\Lnp^{-1/2}).
\end{aligned}
\]
Define
\[
Q_{R,	heta}^*=\frac{\|\mat{E}^*\vct{\beta}_{\vct{\theta}}\|^2-m_R}{\sqrt{2\tau_R}},
\qquad M_{R,	heta}^*=\max_i\frac{(\mat{E}^*\vct{\beta}_{\vct{\theta}})_i^2}{D_{R,ii}}
-2\log p+\log\log p.
\]
By the preceding residual and diagonal bounds,
\[
\begin{gathered}
Q_{R,	heta}^*=\sqrt{n/2}C_n+\Ops(\sqrt{u_n}+n^{-1/2}),\\
M_{R,	heta}^*=\max_i\frac{(\mat{E}\vct{d})_i^2}{\Sigma_{\varepsilon,ii}}
-2\log p+\log\log p+\Ops(\sqrt{u_n}+n^{-1/2}).
\end{gathered}
\]

For one or two independent resampling draws, the occupancy bound and
\eqref{eq:var-weight-bounds} give
\[
\begin{gathered}
\max_j|d_j^{(b)}|=O_{\mathbb P}(\Lnp^{3/2}/\sqrt n),\qquad
\|\vct{d}^{(b)}\|^2-1=\Op(n^{-1/2}),\\
\langle\vct{d}^{(1)},\vct{d}^{(2)}\rangle=\Op(n^{-1/2}).
\end{gathered}
\]
Condition on the original factors and the indices. For $b=1,2$, define
\[
\vct{V}_j=\sqrt n\big(d_j^{(1)}\vct{\varepsilon}_j\T,
d_j^{(2)}\vct{\varepsilon}_j\T\big)\T.
\]
These vectors are independent and centered. Their coordinate variances
average to quantities bounded away from zero, and, for $q=1,2$,
\[
\begin{gathered}
\max_{j,i}\mathbb E\exp\{|V_{j,i}|/(C\Lnp^{3/2})\}\le2,\qquad
\max_i n^{-1}\sum_j\mathbb E|V_{j,i}|^{2+q}
\le C\Lnp^{3q/2},\\
\Cov\!\begin{pmatrix}\Em\vct{d}^{(1)}\\\Em\vct{d}^{(2)}\end{pmatrix}
=\begin{pmatrix}
\|\vct{d}^{(1)}\|^2\Sg&
\langle\vct{d}^{(1)},\vct{d}^{(2)}\rangle\Sg\\
\langle\vct{d}^{(1)},\vct{d}^{(2)}\rangle\Sg&
\|\vct{d}^{(2)}\|^2\Sg
\end{pmatrix}.
\end{gathered}
\]
Proposition~2.1 of \citet{cck2017}, with $B_n=C\Lnp^{3/2}$,
approximates the probabilities of rectangles by those of the Gaussian
vector with this covariance, with error
\[
C(B_n^2\log^7(2pn)/n)^{1/6}
\le C(\Lnp^{10}/n)^{1/6}\longrightarrow0.
\]
Let $\mat{R}_\varepsilon=\diag(\Sg)^{-1/2}\Sg\diag(\Sg)^{-1/2}$.
Assumption~\ref{ass:C} implies
\[
\max_{i\ne j}|R_{\varepsilon,ij}|\le1-c_0,\quad
\sum_{i,j}R_{\varepsilon,ij}^2\le Cp,\quad
\sum_{i,j}|R_{\varepsilon,ij}|\le Cp^{3/2}.
\]
The Gaussian comparison in Lemma~\ref{lem:gaussian}, with
$u_p^2=2\log p-\log\log p+t$, gives
\[
\sum_{i<j}|R_{\varepsilon,ij}|e^{-u_p^2/(1+|R_{\varepsilon,ij}|)}
\le C\Lnp^C\{p^{-1/2}+p^{1-2/(2-c_0)}\}\longrightarrow0.
\]
The cross-block comparison is
$O_{\mathbb P}(n^{-1/2}p^{-1/2}\Lnp^C)$, and each variance rescaling
changes the squared maximum by $O_{\mathbb P}(n^{-1/2}\Lnp)$.
Thus, for $b_0=1,2$, uniformly on index events of probability tending to one,
\[
\begin{aligned}
&\mathbb P\left\{
\max_i\frac{(\Em\vct{d}^{(b)})_i^2}{\Sigma_{\varepsilon,ii}}
-2\log p+\log\log p\le t_b,\ b\le b_0
\,\middle|\,\sigma(\vct{z}_t,\vct{f}_t:t\in\mathbb Z),(I_t^{(b)})\right\}\\
&\hspace{4em}=\prod_{b=1}^{b_0}F_p(t_b)+O\{\Lnp^C(n^{-1/12}+p^{-\kappa_0})\}.
\end{aligned}
\]

Write $\widetilde Q_{\vct{\theta}}=\sqrt{n/2}C_n({\vct{\theta}})$ and
$\widetilde M_{\vct{\theta}}=\max_i(\Em\vct{d}_{\vct{\theta}})_i^2/\Sigma_{\varepsilon,ii}-b_p$.
For fixed ${\vct{\theta}}$, the one- and two-draw calculations give, conditional on the original factors,
\begin{equation}\label{eq:var-pointwise-cdf}
\mathbb E_\varepsilon\left|
\mathbb P^*(\widetilde Q_{\vct{\theta}}\le x,\widetilde M_{\vct{\theta}}\le y)
-\Phi(x)F_p(y)\right|^2
\le C\Lnp^C(n^{-1/12}+p^{-\kappa_0})
\end{equation}
for some $\kappa_0>0$. The collision bound controls the normal component; the Gaussian comparison controls the maximum and its independence from the collision statistic.

To pass to data-dependent parameters, differentiate under the common path construction. Degeneracy of the collision kernel gives
\[
\begin{gathered}
\mathbb E_I|\partial^\alpha\widetilde Q_{\vct{\theta}}|^2
=(1-1/n)\sum_{s\ne t}
\{\partial^\alpha(\beta_{{\vct{\theta}},s}\beta_{{\vct{\theta}},t})\}^2=\Ops(1),\\
\mathbb E_I\|\partial^\alpha\vct{d}_{\vct{\theta}}\|^2
=(1-1/n)\|\partial^\alpha\vct{\beta}_{\vct{\theta}}\|^2,\qquad
\max_j|\partial^\alpha d_{{\vct{\theta}},j}|=\Ops(n^{-1/2}).
\end{gathered}
\]
To establish uniformity over $\Theta_B$, enlarge the parameter
set slightly while keeping $\mat{C}$ positive definite, $\|\mat{H}\|<1$
and $\|\vct{b}\|$ bounded, and cover $\Theta_B$ by finitely many fixed
coordinate boxes. On each box $K\subset\mathbb R^{d_\Theta}$, repeated
one-dimensional integration gives
\[
\sup_K|g|\le C_K\sum_{\alpha\in\{0,1\}^{d_\Theta}}
\int_K|\partial^\alpha g(\vct{\theta})|\,d\vct{\theta}.
\]
Apply this inequality to each first derivative of $\widetilde Q$ and of
$a_i(\vct{\theta})=(\Em\vct{d}_{\vct{\theta}})_i/
\sqrt{\Sigma_{\varepsilon,ii}}$. The displayed collision moment bounds
control the integrals for $\widetilde Q$. Conditional sub-Gaussianity and
a union bound over $i\le p$ control the integrals of
$\max_i|\partial^\alpha a_i|$ by a fixed power of $\Lnp$ in probability.
Thus the suprema of $|a_i|$ and their first derivatives, and of the first
derivatives of $\widetilde Q$, have the same logarithmic bounds. The
derivatives required have order at most $d_\Theta+1$ and are covered by
\eqref{eq:var-weight-bounds}. Since
$|\max_i a_i^2-\max_i b_i^2|\le
\max_i|a_i-b_i|\max_i(|a_i|+|b_i|)$, these bounds imply
\begin{equation}\label{eq:var-modulus}
\sup_{\|{\vct{\theta}}-{\vct{\theta}}'\|\le h}
\{ |\widetilde Q_{\vct{\theta}}-\widetilde Q_{{\vct{\theta}}'}|
+|\widetilde M_{\vct{\theta}}-\widetilde M_{{\vct{\theta}}'}|\}
=O_{\mathbb P}(h\Lnp^C).
\end{equation}
The same derivative calculation for the residual replacement above gives
\[
\sup_{{\vct{\theta}}\in\Theta_B}
\{|Q_{R,{\vct{\theta}}}^*-\widetilde Q_{\vct{\theta}}|
+|M_{R,{\vct{\theta}}}^*-\widetilde M_{\vct{\theta}}|\}
=\Ops(\sqrt{u_n}+n^{-1/2}).
\]
For the quadratic remainder, the centered-kernel calculation gives
\[
\mathbb E_I|\partial^\alpha R_n({\vct{\theta}})|^2/\tau_R
\le C\Lnp^C(p^{-1}+n^{-1}+n/p),
\]
and the vector remainder uses the row bound in Lemma~\ref{lem:column-pilot}. Thus the derivative argument applies to both terms.

Choose an $h_n=\Lnp^{-K}$ net $\mathcal N_n$ of $\Theta_B$, with
$|\mathcal N_n|\le C\Lnp^{Kd_\Theta}$. Equation~\eqref{eq:var-pointwise-cdf} and a union bound give, for each fixed $x,y,\epsilon>0$,
\[
\begin{aligned}
&\mathbb P\left\{\max_{{\vct{\theta}}\in\mathcal N_n}
|\mathbb P^*(\widetilde Q_{\vct{\theta}}\le x,\widetilde M_{\vct{\theta}}\le y)
-\Phi(x)F_p(y)|>\epsilon\right\}\\
&\quad\le C\epsilon^{-2}\Lnp^{C+Kd_\Theta}(n^{-1/12}+p^{-\kappa_0})\to0.
\end{aligned}
\]
Take $K$ larger than the exponent in \eqref{eq:var-modulus}. Let $\omega_n$ be its left side at $h=h_n$, and write $\widetilde H_{\vct{\theta}}$ for the conditional joint CDF. For fixed $a>0$,
\[
\begin{aligned}
&\mathbb P^*(\omega_n>a)\to_{\mathbb P}0,\\
&\sup_{\vct{\theta}}|\widetilde H_{\vct{\theta}}(x,y)-\Phi(x)F_p(y)|\\
&\quad\le \max_{\substack{\vct{\nu}\in\mathcal N_n\\e\in\{-1,1\}}}
|\widetilde H_{\vct{\nu}}(x+ea,y+ea)-\Phi(x+ea)F_p(y+ea)|\\
&\qquad+Ca+\mathbb P^*(\omega_n>a).
\end{aligned}
\]
Let $a\downarrow0$. A finite threshold grid and monotonicity give uniformity in $(x,y)$. The residual replacement and $F_p\to F_G$ finish the proof.
\end{proof}

\subsection{Validity after refitting}
\begin{proof}[Proof of Theorem~\ref{thm:column-bootstrap}]
\noindent\textit{Step 1: the generated oracle statistics.}
First fix the rank at $r$, using the zero-extension convention on undefined fits. Put
\[
\begin{gathered}
\mat{K}_{\widetilde A}=(\widetilde{\mat{A}}\T\widetilde{\mat{A}})^{1/2},\qquad
\mat{A}_0=\widetilde{\mat{A}}\mat{K}_{\widetilde A}^{-1},\\
\mat{F}^*=(\vct{F}_1^*,\ldots,\vct{F}_n^*),\qquad \mat{F}_R=\mat{K}_{\widetilde A}\mat{F}^*.
\end{gathered}
\]
Then $\vct{Y}_t^*=\widehat{\mat{B}}\vct{z}_t+\mat{A}_0\vct{F}_{R,t}+\vct{\varepsilon}_t^*$, with $\mat{A}_0\T\mat{A}_0=\mat{I}_r$. The two factor designs have the same time span. Let $\mat{M}_0^*$ remove $[\mat{Z},\mat{F}^{*\mathsf T}]$ and set
\[
\vct{\beta}^*=\mat{M}_0^*\one_n/\sqrt{\one_n\T\mat{M}_0^*\one_n},
\qquad \mat{Y}^*\vct{\beta}^*=\mat{E}^*\vct{\beta}^*.
\]
Since $\widehat{\vct{\theta}}=(\widehat{\mat{C}}_f,\widehat{\mat{H}}_f,
\bar{\widehat{\vct{F}}}/\sqrt p)$ lies in $\Theta_B$ with probability tending to one as $B\to\infty$, Lemma~\ref{lem:var-oracle} gives
\begin{equation}\label{eq:oraclelimit}
\begin{gathered}
(Q_R^*,M_R^*)=(Q_{R,\widehat{\vct{\theta}}}^*,M_{R,\widehat{\vct{\theta}}}^*),\\
\sup_{x,y}|\mathbb P^*(Q_R^*\le x,M_R^*\le y)-\Phi(x)F_G(y)|\to_{\mathbb P}0.
\end{gathered}
\end{equation}
Condition on the generated factor path when taking moments over residual indices. By Lemma~\ref{lem:column-pilot}, for a unit time vector $\vct{b}$ independent of these indices,
\begin{equation}\label{eq:noise}
\begin{gathered}
\mat{K}^*=\mat{E}^{*\mathsf T}\mat{E}^*-m_R\mat{I}_n,\qquad
\mathbb E_I\|\mat{K}^*\vct{b}\|^2=(n-1)\tau_R+v_R=\Opl(p^2),\\
\mathbb E_I\|\mat{A}_0\T\mat{E}^*\|_F^2=na_R=o_{\mathbb P}(n),\qquad
\mathbb E_I\|\mat{A}_0\T\mat{E}^*\vct{b}\|^2=a_R=o_{\mathbb P}(1).
\end{gathered}
\end{equation}
\noindent\textit{Step 2: loading and time-space estimation.}
The generator has covariance and lag-one singular values of order $p$, uniformly over its fitted parameters. Conditional Gaussian concentration gives
\[
\begin{gathered}
\sigma_{\min}(\mat{F}_R\mat{M}_c)\asymp_{\mathbb P^*}\sqrt{np},
\qquad\|\mat{F}_R\|=\Ops(\sqrt{np}),\\
\lambda_r\left\{\sum_{h=1}^{k_0}\mat{\Gamma}_{R,h}\mat{\Gamma}_{R,h}\T\right\}
\asymp_{\mathbb P^*}p^2.
\end{gathered}
\]
Here $\mat{\Gamma}_{R,h}=p\mat{K}_{\widetilde A}\widehat{\mat{C}}_f\widehat{\mat{H}}_f^h
\widehat{\mat{C}}_f\mat{K}_{\widetilde A}$ is the generated centered factor lag covariance. For fixed $h>0$,
\[
\begin{gathered}
\mathbb E_I\left\|n^{-1}\sum_t\vct{\varepsilon}_{t+h}^*\vct{\varepsilon}_t^{*\mathsf T}\right\|_F^2\le Cm_R^2/n,\\
\mathbb E_I\|\mat{E}^*\mat{F}_R\T/n\|_F^2=m_R\|\mat{F}_R\|_F^2/n^2=\Ops(p^2/n),\\
\mathbb E_I\|\mat{E}^*(\mat{I}_n-\mat{M}_c)\|_F^2=(m+1)m_R.
\end{gathered}
\]
The lag-product perturbation is $\Ops(p^2/\sqrt n+p^2/n)$. Dividing by the lag eigenvalue gap of order $p^2$ in the eigenvector identity of Lemma~\ref{lem:loading} gives
\[
\|\widehat{\mat{P}}^*-\mat{A}_0\mat{A}_0\T\|=\Ops(n^{-1/2}).
\]
Put $\rho_n^*=n^{-1/2}$, $a_n^*=p^{-1/2}+n^{-1}$ and
$h_n^*=(a_n^*)^2+(np)^{-1/2}+n^{-1}$. Then
\[
\begin{gathered}
a_n^*\asymp p^{-1/2},\qquad h_n^*=O(n^{-1}),\\
\sqrt n a_n^*+n^{3/2}(h_n^*)^2+
\frac{n^{3/2}}{\sqrt p}(a_n^*)^2=O(\sqrt{n/p}+n^{-1/2}).
\end{gathered}
\]
Let $\mat{O}_z$ span the complement of $\mat{Z}$, and put
\[
\begin{gathered}
\vct{u}_z=\frac{\mat{O}_z\T\one_n}{\|\mat{O}_z\T\one_n\|},\quad
\mat{G}_R=\mat{F}_R\mat{O}_z,
\quad\mat{W}^*=\mat{E}^*\mat{O}_z,\\
\mat{\Pi}_0=\mat{G}_R\T(\mat{G}_R\mat{G}_R\T)^{-1}\mat{G}_R,\quad
\mat{J}_0=\mat{I}-\mat{\Pi}_0,\\
\vct{v}=\mat{J}_0\vct{u}_z,\quad
\mat{G}_R\vct{v}=0,\quad\|\vct{v}\|^2\asymp_{\Prob^*}1.
\end{gathered}
\]
For the orthonormal refitted loading directions $\widehat{\mat{A}}^*$,
\[
\begin{gathered}
\mat{T}_A^*=(\widehat{\mat{A}}^{*\mathsf T}\mat{A}_0)^{-1}
\widehat{\mat{A}}^{*\mathsf T}-\mat{A}_0\T,\qquad
\|\mat{T}_A^*\|=\Ops(\rho_n^*),\\
\mat{N}_A^*=(\mat{A}_0\T+\mat{T}_A^*)\mat{W}^*,\\
\|\mat{N}_A^*\|=\Ops(\sqrt n+\rho_n^*\sqrt p),\qquad
\|\mat{N}_A^*\vct{v}\|=\Ops(1+\rho_n^*\sqrt p).
\end{gathered}
\]
Let $\widehat{\mat{\Pi}}^*$ project onto the row space of
$\mat{G}_R+\mat{N}_A^*$ and
$\widehat{\mat{J}}^*=\mat{I}-\widehat{\mat{\Pi}}^*$.
The gap $\sigma_{\min}(\mat{G}_R)\asymp\sqrt n\sqrt p$ gives
\begin{equation}\label{eq:time}
\begin{aligned}
\|\widehat{\mat{\Pi}}^*-\mat{\Pi}_0\|
&\le C\|\mat{N}_A^*\|/(\sqrt n\sqrt p)=\Ops(a_n^*),\\
\|\mat{\Pi}_0\widehat{\mat{J}}^*\vct{u}_z\|
&\le\|\widehat{\mat{\Pi}}^*-\mat{\Pi}_0\|^2
+C\|\mat{N}_A^*\vct{v}\|/(\sqrt n\sqrt p)=\Ops(h_n^*).
\end{aligned}
\end{equation}
Put $\mat{X}^*=\mat{A}_0\mat{G}_R+\mat{W}^*$,
$\mat{Q}_F=\mat{G}_R\T\mat{G}_R$, and
\[
\begin{gathered}
\mat{D}_0=\mat{G}_R\T\mat{A}_0\T\mat{W}^*
+\mat{W}^{*\mathsf T}\mat{A}_0\mat{G}_R
+\mat{O}_z\T\mat{K}^*\mat{O}_z,
\qquad \mat{X}^{*\mathsf T}\mat{X}^*-m_R\mat{I}=\mat{Q}_F+\mat{D}_0,\\
\|\mat{D}_0\|=\Ops(n\sqrt p+p)=\Ops(n\sqrt p),\\
\|\mat{D}_0\vct{v}\|=\Ops(\sqrt n\sqrt p+p)=\Ops(p),\qquad
\|\mat{Q}_F\|=\Ops(np).
\end{gathered}
\]
Since $(\mat{G}_R+\mat{N}_A^*)\widehat{\mat{J}}^*=0$,
\begin{equation}\label{eq:finite}
\begin{gathered}
\widehat{\mat{J}}^*(\mat{Q}_F+\mat{D}_0)\widehat{\mat{J}}^*
=\mat{W}_0+\mat{B}_R^*,\qquad
\mat{W}_0=\mat{J}_0\mat{O}_z\T\mat{K}^*\mat{O}_z\mat{J}_0,\\
\rank(\mat{B}_R^*)\le5r,\qquad
\|\mat{B}_R^*\|\le\|\mat{N}_A^*\|^2+
2\|\mat{D}_0\|\|\widehat{\mat{J}}^*-\mat{J}_0\|\\
=\Ops(n+(\rho_n^*)^2p+n\sqrt pa_n^*)=\Ops(n).
\end{gathered}
\end{equation}
\noindent\textit{Step 3: the fitted covariance scales.}
Conditional on the generated path, for a fixed symmetric time matrix $\mat{T}_0$,
\[
\operatorname{Var}_I\{\tr(\mat{T}_0\mat{K}^*)\}
=2\tau_R\tr(\mat{T}_0^2)+(v_R-2\tau_R)\sum_tT_{0,tt}^2.
\]
With $k=n-m-r$, this and \eqref{eq:finite} imply
\begin{equation}\label{eq:center}
\begin{gathered}
\tr\mat{W}_0=\Ops(p),\qquad
 t_0:=k(\tr(\widehat{\mat{\Sigma}}_\varepsilon^*)-m_R)
=\tr(\mat{W}_0+\mat{B}_R^*)=\Ops(p+n),\\
(\tr(\widehat{\mat{\Sigma}}_\varepsilon^*)-m_R)/\sqrt{\tau_R}
=\Ops(n^{-1/2}+\sqrt n/p).
\end{gathered}
\end{equation}
Define
$U^*=[n(n-1)]^{-1}\sum_{s\ne t}
(\vct{\varepsilon}_s^{*\mathsf T}\vct{\varepsilon}_t^*)^2$.
The empirical moment bounds give
\[
\begin{gathered}
\mathbb E_IU^*=\tau_R,\qquad
\mathbb E_I(\vct{\varepsilon}_1^{*\mathsf T}\mat{S}_R\vct{\varepsilon}_1^*)^2
\le\|\mat{S}_R\|^2\mathbb E_I\|\vct{\varepsilon}_1^*\|^4=\Opl(p^4/n^2),\\
\mathbb E_I(\vct{\varepsilon}_1^{*\mathsf T}\vct{\varepsilon}_2^*)^4
\le n^{-2}\max_{i,j}|\widehat{\vct{\varepsilon}}_i\T\widehat{\vct{\varepsilon}}_j|^2
\|\widehat{\mat{E}}\T\widehat{\mat{E}}\|_F^2=\Opl(p^4/n),\\
\|\mat{K}^*\|_F^2=n(n-1)U^*+
\sum_t(\|\vct{\varepsilon}_t^*\|^2-m_R)^2,\\
v_R/(n\tau_R)=\Opl(p^{-1}+n^{-1}),\\
\operatorname{Var}_I(U^*)/\tau_R^2=\Opl(n^{-1}),\qquad
\frac{\|\mat{K}^*\|_F^2}{n^2\tau_R}-1
=\Ops(n^{-1/2}+p^{-1}).
\end{gathered}
\]
Let $\mat{P}_0=\mat{I}_n-\mat{O}_z\mat{J}_0\mat{O}_z\T$.
By \eqref{eq:noise},
\[
0\le\|\mat{K}^*\|_F^2-\|\mat{W}_0\|_F^2
\le2\|\mat{P}_0\mat{K}^*\|_F^2=\Ops(p^2).
\]
As $\|\mat{W}_0\|=\Ops(p)$ and
$\widehat v_\varepsilon^*=
\{\|\mat{W}_0+\mat{B}_R^*\|_F^2-t_0^2/k\}/\{k(k+1)\}$,
\begin{equation}\label{eq:scale}
\begin{gathered}
\left|\|\mat{W}_0+\mat{B}_R^*\|_F^2-t_0^2/k-\|\mat{W}_0\|_F^2\right|\\
\le2\|\mat{W}_0\|\|\mat{B}_R^*\|_*+\|\mat{B}_R^*\|_F^2+t_0^2/k
=\Ops(np+n^2+p^2/n),\\
\widehat v_\varepsilon^*/\tau_R-1
=\Ops(n^{-1/2}+p^{-1}+n/p^2).
\end{gathered}
\end{equation}
Entry concentration and the fixed-rank projection bound yield
\[
\begin{gathered}
\max_i\left|n^{-1}\|\vct{e}_i\T\mat{E}^*\|^2-D_{R,ii}\right|
=\Ops(n^{-1/2}),\quad
\max_i\|\vct{e}_i\T\mat{E}^*\mat{P}_0\|^2=\Ops(1),\\
\max_i\|\vct{e}_i\T(\mat{X}^*\widehat{\mat{J}}^*-\mat{W}^*\mat{J}_0)\|\\
\le\Ops(p^{-1/2})\|\mat{N}_A^*\|
+\max_i\|\vct{e}_i\T\mat{W}^*\|\|\widehat{\mat{J}}^*-\mat{J}_0\|
=\Ops(\sqrt na_n^*).
\end{gathered}
\]
The difference-of-squares identity therefore gives
\begin{equation}\label{eq:diag}
\max_i|(\widehat{\mat{\Sigma}}_\varepsilon^*)_{ii}-D_{R,ii}|=\Ops(n^{-1/2}+a_n^*).
\end{equation}

\noindent\textit{Step 4: replacement by the oracle statistics.}
Set $\widehat\eta=\vct{u}_z\T\widehat{\mat{J}}^*\vct{u}_z$,
$\eta_0=\|\vct{v}\|^2$ and
$\vct{d}_u=\widehat{\mat{J}}^*\vct{u}_z-\vct{v}$.
By \eqref{eq:time},
$|\widehat\eta-\eta_0|=\Ops(a_n^*)$ and
\[
\begin{gathered}
\|\sqrt{\widehat{\mathcal I}^*}\,\widehat{\alp}^*\|^2-m_R
=\frac{\vct{u}_z\T\widehat{\mat{J}}^*(\mat{Q}_F+\mat{D}_0)
\widehat{\mat{J}}^*\vct{u}_z}{\widehat\eta},\qquad
\|\mat{E}^*\vct{\beta}^*\|^2-m_R
=\vct{v}\T\mat{D}_0\vct{v}/\eta_0,\\
\left|\vct{u}_z\T\widehat{\mat{J}}^*(\mat{Q}_F+\mat{D}_0)
\widehat{\mat{J}}^*\vct{u}_z-\vct{v}\T\mat{D}_0\vct{v}\right|\\
\le2\|\mat{D}_0\vct{v}\|\|\vct{d}_u\|
+\|\mat{Q}_F\|\|\mat{\Pi}_0\vct{d}_u\|^2
+\|\mat{D}_0\|\|\vct{d}_u\|^2.
\end{gathered}
\]
Using $Q_R^*=O_{\Prob^*}(1)$ from \eqref{eq:oraclelimit},
\[
\begin{aligned}
&\frac{|\|\sqrt{\widehat{\mathcal I}^*}\,\widehat{\alp}^*\|^2-\|\mat{E}^*\vct{\beta}^*\|^2|}
{\sqrt{\tau_R}}\\
&\quad=\Ops\left(\sqrt na_n^*+n^{3/2}(h_n^*)^2+
\frac{n^{3/2}\sqrt p}{p}(a_n^*)^2+a_n^*\right)\\
&\quad=\Ops(\sqrt{n/p}+n^{-1/2}),\\
&\|\sqrt{\widehat{\mathcal I}^*}\,\widehat{\alp}^*-\mat{E}^*\vct{\beta}^*\|_\infty\\
&\quad=\Ops(\sqrt nh_n^*+\sqrt na_n^*+a_n^*)
=\Ops(\sqrt{n/p}).
\end{aligned}
\]
Assumption~\ref{ass:D} absorbs the fixed logarithmic factors.
Equations \eqref{eq:center}--\eqref{eq:diag}
then give
\[
T_{\mathrm{sum}}^*=Q_R^*+\ops(1),\qquad
T_{\mathrm{max}}^*=M_R^*+\ops(1),
\]
and \eqref{eq:oraclelimit} proves the joint limit.
The continuous map
\[
(s,t)\longmapsto\tfrac12\cot\{\pi[1-\Phi(s)]\}
+\tfrac12\cot\{\pi[1-\FG(t)]\}
\]
gives the standard Cauchy limit. If $F_j$ denotes the corresponding
continuous null limit, then
\[
\begin{aligned}
&\sup_x|\Prob^*(T_j^*\le x)-\Prob_0(T_j\le x)|\\
&\le\sup_x|\Prob^*(T_j^*\le x)-F_j(x)|\\
&\quad+\sup_x|\Prob_0(T_j\le x)-F_j(x)|\to_{\Prob}0.
\end{aligned}
\]
The selected-rank transfer costs at most $1\{\widehat r\ne r\}=\op(1)$.
\end{proof}

\clearpage
\section{Proof of the mean-square comparison}
\label{app:mse}
This appendix proves Proposition~\ref{prop:mse}. We define the population weights, establish $L^2$ expansions of the two estimation errors, and compare their leading risks by orthogonality. Moment bounds then control the contribution of factor-number selection errors to expected risk.

For the selected-rank comparison, also impose \eqref{eq:lag-rank-separation} and,
for some fixed $\kappa>0$,
\begin{equation}\label{eq:mse-rank-condition}
\frac{p^{\delta}}{n a_{{\rm sel},n}}=O(n^{-\kappa}),\qquad
n^2p^{-2\delta}\mathbb P\left\{
\min_{r<j\le R_*}q_{j,n}\le a_{{\rm sel},n}
\right\}^{1/2}\to0.
\end{equation}
The strengthened condition controls the contribution of incorrect factor-number selections to mean-square risk.

\subsection{Notation and population weights}

Within this proof, $\widehat{\vct{\alpha}}_{\rm OLS}^{[a]}$ denotes the rank-$a$ version of the Step~3 estimator $\widehat{\vct{\alpha}}$. The comparison index is $j\in\{\mathrm{dir},\mathrm{OLS}\}$. Use the decomposition
\[
\vct{Y}_t=\vct{\alpha}+\mat{B}\vct{z}_t
 +\mat{A}(\vct{\lambda}+\vct{f}_t)+\vct{\varepsilon}_t,
\qquad \mat{A}^\top\mat{A}=\mat{I}_r,
\qquad \mat{A}^\top\vct{\alpha}=\vct{0}.
\]
Set $s=\sfac$, $\vct{\tau}_n=\vct{\lambda}/s$,
$v_p=\tr(\mat{Q}_A\Sg\mat{Q}_A)$, $\vct{x}_t=\vct{f}_t/s$,
$\vct{z}_t^0=\vct{z}_t-\vct{m}_z$, and
\begin{gather*}
\mat{\Gamma}_z=\E(\vct{z}_t^0\vct{z}_t^{0\top}),\qquad
\widetilde{\mat{\Gamma}}_f(0)=\mat{\Gamma}_f(0)/s^2,\qquad
\widetilde{\mat{\Gamma}}_f(h)=\mat{\Gamma}_f(h)/s^2,\\
\mat{K}=\sum_{h=1}^{k_0}\widetilde{\mat{\Gamma}}_f(h)\widetilde{\mat{\Gamma}}_f(h)^\top,\\
\vct{q}_t=\mat{K}^{-1}\sum_{h=1}^{k_0}\widetilde{\mat{\Gamma}}_f(h)\vct{x}_{t-h},\qquad
\mat{C}_{qz}=\E(\vct{q}_t\vct{z}_t^{0\top}),\\
\vct{q}_t^\perp=\vct{q}_t-\mat{C}_{qz}\mat{\Gamma}_z^{-1}\vct{z}_t^0,\\
\vct{d}_t=\vct{q}_t^\perp-\widetilde{\mat{\Gamma}}_f(0)^{-1}\vct{x}_t,\qquad
\mat{\Delta}=\E(\vct{d}_t\vct{d}_t^\top).
\end{gather*}
Let $\mat{Z}_c=\mat{Z}-\vct{1}_n\bar{\vct{z}}^{\top}$ and
\begin{gather*}
\vct{h}=\vct{1}_n/n
 -\mat{Z}_c(\mat{Z}_c^\top\mat{Z}_c)^{-1}\bar{\vct{z}},\qquad
c_{z,n}=n\norm{\vct{h}}_2^2,\\
\mat{M}_c=\mat{I}_n-\vct{1}_n\vct{1}_n^\top/n
 -\mat{Z}_c(\mat{Z}_c^\top\mat{Z}_c)^{-1}\mat{Z}_c^\top,\qquad
\mat{O}\mat{O}^\top=\mat{M}_c,\quad
\mat{O}^\top\mat{O}=\mat{I}_{n-m-1},\\
\mat{E}=(\vct{\varepsilon}_1,\ldots,\vct{\varepsilon}_n),\qquad
\nu_n=\lambda_{\min}\!\left(p^{-1}\mat{O}^\top\mat{E}^\top
                  \mat{Q}_A\mat{E}\mat{O}\right).
\end{gather*}

Put
\[
\eta_{n,p}=p^{\delta/2}/\sqrt n=\rn,\qquad
b_{{\rm sel},n}=n^2p^{-2\delta}\mathbb P(\widehat r\ne r)^{1/2}.
\]

\subsection{Risk comparison}
\begin{proof}[Proof of Proposition~\ref{prop:mse}]
\noindent\textit{Step 1: inverse moments and global risk bounds.}
Under the assumptions of Proposition~\ref{prop:mse}, Gaussian conditional covariance and fixed-dimensional
Gram bounds give, for each fixed $q$ and all sufficiently large $n$,
\[
\E\lambda_{\min}(\mat{Z}_c\T\mat{Z}_c/n)^{-q}\le C_q,
\qquad \E c_{z,n}^q\le C_q.
\]
Indeed, condition on the other columns of $\mat{Z}_c$. The squared
distance of a remaining column from their span has a Gaussian covariance
bounded below on a subspace of dimension at least $n-m$; its negative
moments are bounded by those of $c\chi^2_{n-m}$.
Use $\tr\{(\mat{Z}_c\T\mat{Z}_c)^{-1}\}$ as the sum of the inverse
squared distances, and apply H\"older's inequality to the sample mean.
Conditionally on $\mat{Z}$, diagonalizing
$\Sg^{1/2}\mat{Q}_A\Sg^{1/2}$ gives
\[
\nu_n\ge \frac c p\lambda_{\min}(\mat{W}\T\mat{W}),\qquad
W_{ij}\overset{\rm iid}{\sim}N(0,1),\quad
\mat{W}\in\mathbb R^{(p-r)\times(n-m-1)}.
\]
To verify the inverse-moment bound, put $a=p-r$, $b=n-m-1$ and
$\mat{V}=\mat{W}\T\mat{W}$. A fixed-radius net of the unit sphere in
$\mathbb R^b$, together with Gaussian quadratic-form concentration, gives
$\mathbb P\{\lambda_{\min}(\mat{V})<a/2\}\le C e^{-ca}$ for large $n$,
because $b/a\to0$. Moreover,
$(\mat{V}^{-1})_{jj}$ is the inverse squared distance of column $j$ of
$\mat{W}$ from the span of its other columns; that squared distance has
the $\chi^2_{a-b+1}$ law. Its negative moments and
$(\sum_{j=1}^b x_j)^{2q}\le b^{2q-1}\sum_{j=1}^b x_j^{2q}$ yield
$\E\{p\,\tr(\mat{V}^{-1})\}^{2q}\le C_q b^{2q}$ for fixed $q\ge1$.
Splitting according to the preceding event and using Cauchy--Schwarz gives
\[
\E\{p/\lambda_{\min}(\mat{V})\}^q
\le (2p/a)^q+C_q b^q e^{-ca/2}\le C_q.
\]
Consequently $\operatorname*{ess\,sup}_{\mat{Z}}
\E(\nu_n^{-q}\mid\mat{Z})\le C_q$; smaller positive orders follow
by H\"older's inequality.
All sample inverses below consequently exist almost surely.

Put $\eta=\eta_{n,p}$,
$\Fm=(\vct{f}_1,\ldots,\vct{f}_n)$,
$\mat{G}=\Fm\mat{M}_c$, and $\vct{\ell}=\vct{\lambda}+\Fm\vct{h}$. Then
\[
\mat{M}_c\vct{h}=\vct{0},\qquad
\widehat{\vct{\mu}}=\vct{\alpha}+\mat{A}\vct{\ell}+\mat{E}\vct{h},
\qquad
\mat{R}=\mat{A}\mat{G}+\mat{E}\mat{M}_c,
\qquad \widehat{\mat{C}}_\eta=\mat{R}\mat{R}^\top/n.
\]
Let $\mat{J}_h$ be the lag shift matrix and set
\[
\widehat{\mat{\Gamma}}_\eta(h)=\mat{R}\mat{J}_h\mat{R}^\top/(n-h),\qquad
\widehat{\mat{M}}_A=\sum_{h=1}^{k_0}
             \widehat{\mat{\Gamma}}_\eta(h)\widehat{\mat{\Gamma}}_\eta(h)^\top.
\]
For a candidate rank $a$, $\widehat{\mat{A}}^{[a]}$ contains the leading $a$
orthonormal eigenvectors of $\widehat{\mat{M}}_A$. Write
\[
\widetilde{\mat{A}}^{[a]}=\widehat{\mat{C}}_\eta\widehat{\mat{A}}^{[a]}
 (\widehat{\mat{A}}^{[a]\top}\widehat{\mat{C}}_\eta
  \widehat{\mat{A}}^{[a]})^{-1},\qquad
\widehat{\vct{\ell}}^{[a]}=\widehat{\mat{A}}^{[a]\top}\widehat{\vct{\mu}}.
\]
With $\mat{T}=\mat{R}\mat{O}$ and $\mat{L}_h=\mat{O}^\top\mat{J}_h\mat{O}$,
\begin{gather*}
\mat{T}^\top\mat{T}\succeq
 \mat{O}^\top\mat{E}^\top\mat{Q}_A\mat{E}\mat{O}
 \succeq p\nu_n\mat{I}_{n-m-1},\\
\widehat{\mat{M}}_A
=\mat{T}\left\{\sum_{h=1}^{k_0}
 \frac{\mat{L}_h(\mat{T}^\top\mat{T})\mat{L}_h^\top}{(n-h)^2}\right\}\mat{T}^\top,
\qquad
\operatorname{rank}(\mat{L}_1)\geq n-2m-3>R_*.
\end{gather*}
Hence $\operatorname{col}(\widehat{\mat{A}}^{[a]})\subseteq\operatorname{col}(\mat{T})$ for
$1\leq a\leq R_*$. For $\mat{B}_a=\mat{T}^\top\widehat{\mat{A}}^{[a]}$,
\[
\widetilde{\mat{A}}^{[a]}=\mat{T}\mat{B}_a(\mat{B}_a^\top\mat{B}_a)^{-1},\qquad
s_{\min}(\mat{B}_a)\geq s_{\min}(\mat{T}),\qquad
\norm{\widetilde{\mat{A}}^{[a]}}
\leq\frac{\norm{\Fm}+\norm{\mat{E}}}{\sqrt{p\nu_n}}.
\]
Let $d_F=\norm{\Fm}/(s\sqrt n)$ and $e_E=\norm{\mat{E}}/\sqrt p$.
For each fixed $q$, $\E d_F^q+\E e_E^q\leq C_q$, and
\begin{gather*}
\norm{\widehat{\vct{\alpha}}^{[a]}_j-\vct{\alpha}}
\leq C\frac s\eta(1+d_F+\eta e_E)^2c_{z,n}^{1/2}
                       (1+\nu_n^{-1/2}),\\
\E\{(1+d_F)^8c_{z,n}^2\}
\leq\{\E(1+d_F)^{16}\E c_{z,n}^4\}^{1/2}\leq C,\\
\E(e_E^8\nu_n^{-2}\mid\mat{Z})
\leq\{\E(e_E^{16}\mid\mat{Z})
             \E(\nu_n^{-4}\mid\mat{Z})\}^{1/2}\leq C.
\end{gather*}
Conditioning first on the factors gives
\[
\max_{0\leq a\leq R_*}
\norm{\widehat{\vct{\alpha}}^{[a]}_j-\vct{\alpha}}_{L^4}
\leq Cs/\eta,
\qquad j\in\{\mathrm{dir},\mathrm{OLS}\},
\]
where $\widehat{\vct{\alpha}}^{[0]}_j=\widehat{\vct{\mu}}$ and
$\norm{\vct{V}}_{L^q}=(\E\norm{\vct{V}}^q)^{1/q}$.

\noindent\textit{Step 2: expansions of the two loading estimators.}
For the centered first- and second-order factor products required below, the
mixing moment bound gives, for fixed $q\geq2$,
\[
\norm{n^{-1}\sum_{t=1}^n W_t}_{L^q}\leq C_q n^{-1/2};
\]
see \citet[Theorem 1 and Remark 1]{mpr2011}.
Taking $q=12$, choose a factor-measurable event $\mathcal G_F$ on which the
sample means and covariance matrices lie in fixed small neighborhoods of their
population values, and put
$\mathcal G_E=\{\norm{\mat{E}}\leq C\sqrt p\}$,
$\mathcal G=\mathcal G_F\cap\mathcal G_E$. Then
\[
\Prb(\mathcal G_F^c)\leq Cn^{-6},\qquad
\Prb(\mathcal G_E^c)\leq Ce^{-cp}.
\]
Define
\begin{gather*}
\mat{S}_h=\mat{G}\mat{J}_h\mat{G}^\top/(n-h),\qquad
\mat{V}_G=\mat{G}\mat{G}^\top/n,\qquad
\mat{K}_G=\sum_{h=1}^{k_0}\mat{S}_h\mat{S}_h^\top,\\
\mat{U}_{\mathrm{dir}}
=\mat{M}_c\left\{\sum_{h=1}^{k_0}
 \frac{\mat{J}_h\mat{G}^\top\mat{S}_h^\top}{n-h}\right\}\mat{K}_G^{-1},
\qquad
\mat{U}_{\mathrm{OLS}}=\mat{G}^\top(\mat{G}\mat{G}^\top)^{-1}.
\end{gather*}
On $\mathcal G_F$,
\[
\begin{gathered}
\norm{\vct{h}}\leq Cn^{-1/2},\quad
\norm{\mat{G}}\leq Cs\sqrt n,\quad
\mat{V}_G\succeq cs^2\mat{I}_r,\\
\mat{K}_G\succeq cs^4\mat{I}_r,\quad
\norm{\vct{\ell}}\leq Cs,\quad
\norm{\mat{U}_j}\leq C/(s\sqrt n).
\end{gathered}
\]
Expanding the actual lag matrix in powers of $\mat{E}$ yields
\begin{gather*}
\widehat{\mat{M}}_A=\mat{A}\mat{K}_G\mat{A}^\top
                       +\mat{M}^{(1)}+\mat{M}_{\geq2},\\
\norm{\mat{M}^{(1)}}\leq Cs^4\eta,\qquad
\norm{\mat{M}_{\geq2}}\leq Cs^4\eta^2,\\
\mat{Q}_A\mat{M}^{(1)}\mat{A}\mat{K}_G^{-1}
=\mat{Q}_A\mat{E}\mat{U}_{\mathrm{dir}}
\quad\text{on }\mathcal G.
\end{gather*}
After an orthogonal alignment, write
$\widehat{\mat{A}}=(\mat{A}+\mat{V})(\mat{I}_r+\mat{V}^\top\mat{V})^{-1/2}$,
$\mat{A}^\top\mat{V}=\mat{0}$, and $\norm{\mat{V}}\leq C\eta$.
For $\mat{H}=\widehat{\mat{M}}_A-\mat{A}\mat{K}_G\mat{A}^\top$, the eigenvector
equation becomes
\[
\mat{V}\mat{K}_G
=\mat{Q}_A\mat{H}\mat{A}+\mat{Q}_A\mat{H}\mat{V}
 -\mat{V}\mat{A}^\top\mat{H}\mat{A}
 -\mat{V}\mat{A}^\top\mat{H}\mat{V}.
\]
Consequently,
\[
\widehat{\mat{A}}=\mat{A}+\mat{Q}_A\mat{E}\mat{U}_{\mathrm{dir}}+\mat{R}_A,
\qquad \norm{\mat{R}_A}\leq C\eta^2\quad\text{on }\mathcal G.
\]
Similarly,
\begin{gather*}
\widehat{\mat{C}}_\eta=\mat{A}\mat{V}_G\mat{A}^\top+\mat{C}_1+\mat{C}_2,\\
\mat{C}_1=\frac{\mat{A}\mat{G}\mat{E}^\top+\mat{E}\mat{G}^\top\mat{A}^\top}{n},
\qquad \mat{C}_2=\mat{E}\mat{M}_c\mat{E}^\top/n,\\
\norm{\mat{C}_1}\leq Cs^2\eta,\qquad
\norm{\mat{C}_2}\leq Cs^2\eta^2,\\
\widehat{\mat{A}}^{\top}\widehat{\mat{C}}_\eta\widehat{\mat{A}}
=\mat{V}_G+\mat{A}^\top\mat{C}_1\mat{A}+O(s^2\eta^2).
\end{gather*}
Using
\[
(\mat{V}_G+\mat{H}_0)^{-1}
=\mat{V}_G^{-1}-\mat{V}_G^{-1}\mat{H}_0\mat{V}_G^{-1}
 +\mat{V}_G^{-1}\mat{H}_0\mat{V}_G^{-1}\mat{H}_0
  (\mat{V}_G+\mat{H}_0)^{-1},
\]
we obtain
\[
\widetilde{\mat{A}}
=\mat{A}+\mat{Q}_A\mat{C}_1\mat{A}\mat{V}_G^{-1}+\mat{R}_D
=\mat{A}+\mat{Q}_A\mat{E}\mat{U}_{\mathrm{OLS}}+\mat{R}_D,
\qquad \norm{\mat{R}_D}\leq C\eta^2.
\]
Substitution into the two estimators, with $s\eta=\sqrt{p/n}\to\infty$, gives
\[
\norm{\widehat{\vct{\alpha}}^{[r]}_j-\vct{\alpha}
       -\mat{Q}_A\mat{E}(\vct{h}-\mat{U}_j\vct{\ell})}
\leq C\{s\eta^2+\eta\norm{\vct{\alpha}}\}
\leq Cs\eta^2\quad\text{on }\mathcal G.
\]
Put
$\vct{L}_j^*=\indc\{\mathcal G_F\}\mat{Q}_A\mat{E}
 (\vct{h}-\mat{U}_j\vct{\ell})$, extended by zero on $\mathcal G_F^c$.
Factor--error independence implies
$\norm{\vct{L}_j^*}_{L^4}\leq C\sqrt{p/n}$.
The global fourth-moment bound and H\"older's inequality now give
\begin{align*}
\E\norm{\widehat{\vct{\alpha}}^{[r]}_j-\vct{\alpha}-\vct{L}_j^*}^2
&\leq Cs^2\eta^4
 +C(s^2\eta^{-2}+p/n)(n^{-6}+e^{-cp})^{1/2}\\
&\leq Cs^2\eta^4.
\end{align*}

\noindent\textit{Step 3: population weights and the leading risk difference.}
Let $h_t^0=(1-\vct{z}_t^{0\top}\mat{\Gamma}_z^{-1}\vct{m}_z)/n$.
The factor sample-moment bounds yield
\begin{gather*}
\norm{\indc\{\mathcal G_F\}(\mat{S}_h/s^2-\widetilde{\mat{\Gamma}}_f(h))}_{L^4}
+\norm{\indc\{\mathcal G_F\}(\mat{V}_G/s^2-\widetilde{\mat{\Gamma}}_f(0))}_{L^4}
\leq Cn^{-1/2},\\
\norm{\indc\{\mathcal G_F\}(\mat{K}_G/s^4-\mat{K})}_{L^4}
\leq Cn^{-1/2},\\
\norm{\indc\{\mathcal G_F\}(\vct{h}-\vct{h}^0)}_{L^4}\leq C/n,
\qquad
\norm{\indc\{\mathcal G_F\}(\vct{\ell}/s-\vct{\tau}_n)}_{L^4}
\leq Cn^{-1/2}.
\end{gather*}
For the matrix $\mat{T}_q$ whose $t$th row is $\vct{q}_t^\top$,
\[
\frac1n\mat{M}_c\mat{T}_q
=\frac1n\left\{\mat{T}_q-\vct{1}_n\bar{\vct{q}}^{\top}
 -\mat{Z}_c(\mat{Z}_c^\top\mat{Z}_c/n)^{-1}
              (\mat{Z}_c^\top\mat{T}_q/n)\right\}.
\]
Writing $\mat{V}_{\mathrm{dir}}$ and $\mat{V}_{\mathrm{OLS}}$ for the matrices
with rows $(\vct{q}_t^\perp)^\top$ and $(\widetilde{\mat{\Gamma}}_f(0)^{-1}\vct{x}_t)^\top$,
respectively, the preceding identities, including the fixed lag-boundary terms,
give
\[
\norm{\indc\{\mathcal G_F\}(s\mat{U}_j-\mat{V}_j/n)}_{L^4}\leq C/n.
\]
Define
\begin{gather*}
\omega_{\mathrm{dir},t}
=1-\vct{z}_t^{0\top}\mat{\Gamma}_z^{-1}\vct{m}_z
        -\vct{\tau}_n^\top\vct{q}_t^\perp,\qquad
\omega_{\mathrm{OLS},t}
=1-\vct{z}_t^{0\top}\mat{\Gamma}_z^{-1}\vct{m}_z
        -\vct{\tau}_n^\top\widetilde{\mat{\Gamma}}_f(0)^{-1}\vct{x}_t,\\
\vct{\omega}_j=(\omega_{j,1},\ldots,\omega_{j,n})^\top,
\qquad \vct{L}_j=\mat{Q}_A\mat{E}\vct{\omega}_j/n.
\end{gather*}
Since $\norm{\vct{\omega}_j/n}_{L^4}\leq Cn^{-1/2}$,
\begin{gather*}
\norm{\indc\{\mathcal G_F\}(\vct{h}-\mat{U}_j\vct{\ell})
          -\vct{\omega}_j/n}_{L^2}\leq C/n,\\
\norm{\vct{L}_j^*-\vct{L}_j}_{L^2}\leq C\sqrt p/n
\leq Cs\eta^2,\\
\norm{\widehat{\vct{\alpha}}^{[r]}_j-\vct{\alpha}-\vct{L}_j}_{L^2}
\leq Cs\eta^2= C\sqrt{p/n}\,\eta.
\end{gather*}
Here the first bound also uses
$\norm{\indc\{\mathcal G_F^c\}\vct{\omega}_j/n}_{L^2}
\leq Cn^{-1/2}\Prb(\mathcal G_F^c)^{1/4}\leq Cn^{-2}$.

Temporal independence of the errors gives
\[
\E\norm{\vct{L}_j}^2=\frac{v_p}{n}\E\omega_{j,t}^2,
\qquad \norm{\vct{L}_j}_{L^2}\leq C\sqrt{p/n}.
\]
Thus, for $\mathcal R_j^{[r]}
=\E\norm{\widehat{\vct{\alpha}}^{[r]}_j-\vct{\alpha}}^2$,
\[
\begin{aligned}
\left|\mathcal R_j^{[r]}-\frac{v_p}{n}\E\omega_{j,t}^2\right|
&\leq2\norm{\vct{L}_j}_{L^2}
 \norm{\widehat{\vct{\alpha}}^{[r]}_j-\vct{\alpha}-\vct{L}_j}_{L^2}\\
 &\quad+\norm{\widehat{\vct{\alpha}}^{[r]}_j-\vct{\alpha}-\vct{L}_j}_{L^2}^2
\leq C\frac pn\eta.
\end{aligned}
\]
Moreover,
\begin{gather*}
\E(\vct{x}_t\vct{q}_t^\top)
=\sum_{h=1}^{k_0}\widetilde{\mat{\Gamma}}_f(h)\widetilde{\mat{\Gamma}}_f(h)^\top\mat{K}^{-1}
=\mat{I}_r,\qquad
\E(\vct{z}_t^0\vct{q}_t^{\perp\top})=\mat{0},\\
\E\vct{d}_t=\vct{0},\qquad
\E(\vct{x}_t\vct{d}_t^\top)=\mat{0},\qquad
\E(\vct{z}_t^0\vct{d}_t^\top)=\mat{0},\\
\E\omega_{\mathrm{dir},t}^2-\E\omega_{\mathrm{OLS},t}^2
=\E(\vct{\tau}_n^\top\vct{d}_t)^2
=\vct{\tau}_n^\top\mat{\Delta}\vct{\tau}_n,\qquad
\mat{\Delta}=\E(\vct{d}_t\vct{d}_t^\top)\succeq\mat{0}.
\end{gather*}
\noindent\textit{Step 4: factor-number selection.}
On $\mathcal G$, singular-value perturbation of the stacked lag matrix gives
\[
q_{r,n}\le C\eta^2<a_{{\rm sel},n},\qquad
\min_{j<r}q_{j,n}>c_0>a_{{\rm sel},n}
\]
eventually. Therefore \eqref{eq:mse-rank-condition} and
$\mathbb P(\mathcal G^c)\le Cn^{-6}+Ce^{-cp}$ give
$b_{{\rm sel},n}\to0$.
Finally, the selected-rank and true-rank estimators agree on
$\{\widehat r=r\}$. Therefore,
\begin{align*}
|\mathcal R_j-\mathcal R_j^{[r]}|
&\leq\E\!\left[\left\{
 \norm{\widehat{\vct{\alpha}}^{[\widehat r]}_j-\vct{\alpha}}^2
 +\norm{\widehat{\vct{\alpha}}^{[r]}_j-\vct{\alpha}}^2
 \right\}\indc\{\widehat r\ne r\}\right]\\
&\leq Cs^2\eta^{-2}\Prb(\widehat r\ne r)^{1/2}
=C\frac pn b_{\mathrm{sel},n}.
\end{align*}
Combining the bounds gives
\[
\mathcal R_{\rm dir}-\mathcal R_{\rm OLS}
=\frac{v_p}{n}\vct{\tau}_n\T\mat{\Delta}\vct{\tau}_n
+O\!\left\{\frac pn(\eta_{n,p}+b_{{\rm sel},n})\right\}.
\]
For the true-rank comparison set $b_{{\rm sel},n}=0$.
Since $v_p\asymp p$ and $\eta_{n,p}+b_{{\rm sel},n}\to0$, the
proposition and its strict inequality follow.
\end{proof}

\clearpage
\section{Residual PCA: construction, theory and proofs}\label{app:pca}
The residual PCA comparison uses the pricing target and two OLS regressions
from the main text, with PCA for loading extraction. Its results use
Assumptions~\ref{ass:A}--\ref{ass:D} and, where stated, the signal condition
in Assumption~\ref{ass:signal}. We give the construction and sample identities,
verify loading accuracy and rank consistency, and then apply the common
bounds from Appendix~\ref{app:technical}. Each result is followed by its proof.

\subsection{Construction and exact sample identities}\label{app:pca-method}
Use the first-stage residual matrix $\Rmat$ from Section~\ref{sec:selectors}
and let $\widehat\lambda_1^{\rm pc}\ge\cdots\ge\widehat\lambda_p^{\rm pc}$
be the eigenvalues of $\widehat{\mat{C}}_\eta=\Rmat\Rmat\T/n$.
For fixed $R_*>r$ with $R_*<\min(p,n-m-1)$, define
\begin{equation}\label{eq:pca-selector}
\begin{aligned}
V_n(j)&=p^{-1}\sum_{\ell>j}\widehat\lambda_\ell^{\rm pc},
&g_{n,p}&=(n^{-1}+p^{-1})\log\{\min(n,p)\},\\
\mathrm{IC}_{p2,n}(j)&=\log V_n(j)+jg_{n,p},
&\widehat r_{\rm pc}&=\min\mathop{\arg\min}_{0\le j\le R_*}
\mathrm{IC}_{p2,n}(j).
\end{aligned}
\end{equation}
This is the residual $\mathrm{IC}_{p2}$ rule of \citet{baiNg2002}.
Let $\widehat{\mat{A}}_{\rm pc}$ contain the leading
$\widehat r_{\rm pc}$ orthonormal eigenvectors. Construct
\[
\widehat{\vct{F}}_{t,\rm pc}
=\widehat{\mat{A}}_{\rm pc}\T
(\vct{Y}_t-\widehat{\mat{B}}\vct{z}_t),\qquad
\bar{\widehat{\vct{F}}}_{\rm pc}
=\widehat{\mat{A}}_{\rm pc}\T\widehat{\vct{\mu}},
\]
and perform Step 3 of Section~\ref{sec:selectors}. The statistics, information
normalization and analytic component $p$-values are exactly those in
Section~\ref{sec:tests}, evaluated using these factors. If
$\widehat r_{\rm pc}=0$, the latent block is empty and the final fit is
the observed-factor regression. Undefined fits or nonpositive test scales
can be assigned zero statistics; the regularity result below shows that
this convention is asymptotically immaterial.

Write $\widehat{\mat{P}}_{\rm pc}
=\widehat{\mat{A}}_{\rm pc}\widehat{\mat{A}}_{\rm pc}\T$ and
$\widehat{\mat{Q}}_{\rm pc}=\mat{I}_p-\widehat{\mat{P}}_{\rm pc}$.
The population matrix used by PCA is
\[
\mat{C}_\eta=\mat{A}\mat{\Gamma}_f(0)\mat{A}\T+\Sg.
\]
It contains both the factor covariance component and the contemporaneous error covariance. Residual PCA estimation follows
\citet{bai2003,giglioLiaoXiu2021}; its interaction with the pricing signal
is described by the following exact identities.

\begin{proposition}[PCA post-regression and projection]\label{prop:pca-identity}
Whenever the fitted inverses exist,
\begin{equation}\label{eq:pca-post}
\widetilde{\mat{A}}_{\rm pc}=\widehat{\mat{A}}_{\rm pc},\qquad
\widehat{\alp}_{\rm pc}=\widehat{\mat{Q}}_{\rm pc}\widehat{\vct{\mu}}.
\end{equation}
For
\[
w_t=\frac1n\{1-(\vct{z}_t-\bar{\vct{z}})\T
\mat{S}_z^{-1}\bar{\vct{z}}\},\qquad
\vct{f}_w=\sum_t w_t\vct{f}_t,\quad
\vct{\varepsilon}_w=\sum_t w_t\vct{\varepsilon}_t,
\]
the exact decomposition is
\begin{equation}\label{eq:apr-pca-decomposition}
\widehat{\alp}_{\rm pc}
=\alp-\widehat{\mat{P}}_{\rm pc}\alp
+\widehat{\mat{Q}}_{\rm pc}\mat{A}(\vct{\lambda}+\vct{f}_w)
+\widehat{\mat{Q}}_{\rm pc}\vct{\varepsilon}_w.
\end{equation}
Moreover,
\begin{equation}\label{eq:pca-projection-norms}
\begin{aligned}
\|\widehat{\mat{Q}}_{\rm pc}\alp\|^2
&=\|\alp\|^2-\|\widehat{\mat{P}}_{\rm pc}\alp\|^2,\\
\|\widehat{\mat{Q}}_{\rm pc}\vct{\varepsilon}_w\|^2
&=\|\vct{\varepsilon}_w\|^2
-\|\widehat{\mat{P}}_{\rm pc}\vct{\varepsilon}_w\|^2.
\end{aligned}
\end{equation}
\end{proposition}

\begin{proof}[Proof of Proposition~\ref{prop:pca-identity}]
Let $\widehat{\mat{\Lambda}}_{\rm pc}$ be the diagonal matrix of the
selected positive eigenvalues. The eigenvector equation gives
\[
\widehat{\mat{C}}_\eta\widehat{\mat{A}}_{\rm pc}
=\widehat{\mat{A}}_{\rm pc}\widehat{\mat{\Lambda}}_{\rm pc},\qquad
\widehat{\mat{A}}_{\rm pc}\T\widehat{\mat{C}}_\eta
\widehat{\mat{A}}_{\rm pc}=\widehat{\mat{\Lambda}}_{\rm pc}.
\]
Substitution into \eqref{eq:post} proves \eqref{eq:pca-post}. Also,
\[
\sum_t w_t=1,\qquad \sum_t w_t\vct{z}_t=0,\qquad
\widehat{\vct{\mu}}=\sum_t w_t\vct{Y}_t
=\alp+\mat{A}(\vct{\lambda}+\vct{f}_w)+\vct{\varepsilon}_w.
\]
Multiplication by $\widehat{\mat{Q}}_{\rm pc}$ gives
\eqref{eq:apr-pca-decomposition}. Since
$\widehat{\mat{P}}_{\rm pc}\widehat{\mat{Q}}_{\rm pc}=0$,
Pythagoras gives \eqref{eq:pca-projection-norms}.
\end{proof}

Thus PCA post-regression and direct projection have exactly the same
intercept. Proposition~\ref{prop:mse} concerns the distinct LAG estimators.
For PCA, selected noise directions remove both realized noise and any
pricing signal along those directions. These norm identities describe
signal retention and noise removal, which contribute to the finite-sample behavior of the tests. The additional leakage term in
\eqref{eq:apr-pca-decomposition} also remains relevant in finite samples.

For comparison with the cross-sectional construction associated with
\citet{giglioLiaoXiu2021}, regress the same $\widehat{\vct{\mu}}$ on
$[\one_p,\widehat{\mat{A}}_{\rm pc}]$ and denote its coefficients by
$(\widehat\kappa,\widehat{\vct{\gamma}})$. If this design has full column
rank, its normal equations give
\[
\widehat{\vct{\gamma}}
=\widehat{\mat{A}}_{\rm pc}\T
(\widehat{\vct{\mu}}-\widehat\kappa\one_p).
\]
Consequently the alpha that retains the fitted common intercept satisfies
\begin{equation}\label{eq:apr-glx}
\begin{aligned}
\widehat{\alp}_G
&:=\widehat{\vct{\mu}}-\widehat{\mat{A}}_{\rm pc}\widehat{\vct{\gamma}}
=\widehat{\alp}_{\rm pc}+\widehat\kappa\widehat{\mat{P}}_{\rm pc}\one_p,\\
\widehat{\mat{Q}}_{\rm pc}\widehat{\alp}_G&=\widehat{\alp}_{\rm pc}.
\end{aligned}
\end{equation}
This is an equality after projection for common inputs. The individual
multiple-testing procedure and the global tests here have different
testing objectives and calibrations.

\subsection{Loading accuracy and factor-number consistency}\label{app:pca-technical}\label{app:pca-rank}
In the next loading lemma, $\widehat{\mat{A}}_{\rm pc}^{[r]}$ and
$\widehat{\mat{P}}_{\rm pc}^{[r]}$ use the true rank. All spectral
tie-breaking rules are chosen measurably from $\Rmat$.

\begin{lemma}[Refined PCA loading rate]\label{lem:pca}
Under Assumptions~\ref{ass:A}--\ref{ass:D},
\[
\|\widehat{\mat{P}}_{\rm pc}^{[r]}-\mat{P}_A\|
=\Op(p^{\delta/2}/\sqrt n)=\Op(\rn).
\]
\end{lemma}
\begin{proof}
Write $\widehat{\mat{C}}_\eta
=\mat{A}\mat{\Gamma}_f(0)\mat{A}\T+\mat{E}_{C,n}$.
Lemma~\ref{lem:refined-blocks}, at $h=0$, gives
\[
\|\mat{E}_{C,n}\|
=\Op\{\sfac^2n^{-1/2}+\sfac\sqrt{p/n}+p/n\},\qquad
g_C=\lambda_{\min}\{\mat{\Gamma}_f(0)\}\ge c\sfac^2,
\]
\[
\|\mat{E}_{C,n}\|/g_C
=\Op(n^{-1/2}+\rn+\rn^2)=\Op(\rn).
\]
Let $\widehat{\mat{\Lambda}}_{\rm pc}$ contain the leading $r$
eigenvalues. On $\|\mat{E}_{C,n}\|<g_C/2$,
\[
\lambda_r(\widehat{\mat{C}}_\eta)\ge g_C-\|\mat{E}_{C,n}\|,\qquad
\mat{Q}_A\widehat{\mat{A}}_{\rm pc}^{[r]}
\widehat{\mat{\Lambda}}_{\rm pc}
=\mat{Q}_A\mat{E}_{C,n}\widehat{\mat{A}}_{\rm pc}^{[r]}.
\]
Equation~\eqref{eq:projection-geometry} yields
\[
\|\widehat{\mat{P}}_{\rm pc}^{[r]}-\mat{P}_A\|
\le\frac{\|\mat{E}_{C,n}\|}{g_C-\|\mat{E}_{C,n}\|}
=\Op(\rn).
\]
The spectral rule is centered-residual measurable and unchanged under
$\Ymat\mapsto\Ymat+\vct{a}\one_n\T$. Moreover,
$\sqrt p\,\rn^2=p^{1/2+\delta}/n\to0$, so
\eqref{eq:admission} holds for PCA.
\end{proof}

\begin{lemma}[Residual spectral bounds for information-criterion selection]\label{lem:ic-spectral}
For fixed $R_*>r$, under Assumptions~\ref{ass:A}--\ref{ass:D},
\[
\widehat\lambda^{\rm pc}_{r+1}=O_\mathbb P(1+p/n),\qquad
\mathbb P\{\widehat\lambda^{\rm pc}_r\ge c_1\sfac^2\}\to1,
\]
and there are constants $0<c_V<C_V<\infty$ such that
\[
\mathbb P\{c_V\le V_n(R_*)\le V_n(0)\le C_V\}\to1.
\]
\end{lemma}
\begin{proof}
Write $\mat{F}_c=\Fm\mat{M}_c$, $\mat{W}_c=\Em\mat{M}_c$ and
$d_c=n-m-1$. Since $\Rmat=\mat{A}\mat{F}_c+\mat{W}_c$,
Lemma~\ref{lem:noise} and rank-$r$ approximation give
\[
\|\mat{W}_c\|^2=\Op(n+p),\qquad
\widehat\lambda_{r+1}^{\rm pc}\le\|\mat{W}_c\|^2/n.
\]
The signal lower bound is in Lemma~\ref{lem:pca}. With the innovations
stacked, the quadratic-form matrix for $\|\mat{W}_c\|_F^2$ is
$\mat{M}_c\otimes\Sg$. Equation~\eqref{eq:quadmoment} gives
\[
\Ex\|\mat{W}_c\|_F^2=d_c\mnoise,\qquad
\Var_\xi(\|\mat{W}_c\|_F^2)\le Cd_c\taunoise,
\]
\[
\frac{\|\mat{W}_c\|_F^2}{np}
=\frac{d_c}{n}\frac{\mnoise}{p}+\Op((np)^{-1/2}).
\]
Subtracting the rank-$r$ signal from a rank-$R_*$ approximation and using
the preceding spectral inequalities yields
\[
\begin{aligned}
npV_n(R_*)&\ge\|\mat{W}_c\|_F^2-(R_*+r)\|\mat{W}_c\|^2,\\
npV_n(0)&\le2\|\Fm\|_F^2+2\|\mat{W}_c\|_F^2.
\end{aligned}
\]
Now $c\le\mnoise/p\le C$,
$\|\mat{W}_c\|^2/(np)=\Op(n^{-1}+p^{-1})$, and
$\|\Fm\|_F^2/(np)=\tr(\mat{\Gamma}_f(0))/p+\Op(p^{-\delta}n^{-1/2})$
by Lemma~\ref{lem:weak-moments}. These prove the two-sided bounds.
\end{proof}

\begin{lemma}[Consistency of the PCA information criterion]\label{lem:pc-rank-original}
Under Assumptions~\ref{ass:A}--\ref{ass:D}, $\mathbb P(\widehat r_{\rm pc}\ne r)\to0$.
\end{lemma}
\begin{proof}
Put $\nu_j=\widehat\lambda_j^{\rm pc}$ and $g=g_{n,p}$.
Assumption~\ref{ass:D} gives $g/(n^{-1}+p^{-1})\to\infty$ and
$p^\delta g\le2p^\delta\log n/n\to0$.
On the event in Lemma~\ref{lem:ic-spectral}, all $V_n(j)$ lie in
$[c_V,C_V]$. For $j<r$, $\log(1+x)\ge x/(1+x)$ gives
\[
\begin{aligned}
V_n(j)-V_n(r)&\ge(r-j)c_1p^{-\delta},\\
\mathrm{IC}_{p2,n}(j)-\mathrm{IC}_{p2,n}(r)
&\ge(r-j)\{c_1p^{-\delta}/C_V-g\}>0.
\end{aligned}
\]
For $j>r$, use $\log(1+x)\le x$ to obtain
\[
\begin{aligned}
V_n(r)-V_n(j)&\le(j-r)\nu_{r+1}/p,\\
\mathrm{IC}_{p2,n}(j)-\mathrm{IC}_{p2,n}(r)
&\ge(j-r)\{g-\nu_{r+1}/(pc_V)\}>0
\end{aligned}
\]
with probability tending to one, because
$\nu_{r+1}/(pg)=\Op\{(n^{-1}+p^{-1})/g\}\to_{\mathbb P}0$.
A union bound over the fixed candidate set proves consistency.
\end{proof}

\subsection{Regularity and first-order approximation}\label{app:pca-inference}
The preceding PCA loading and rank results
give the following first-order approximation. The rate terms are defined
in Appendix~\ref{app:proof-notation}.

\begin{proposition}[PCA regularity and approximation]\label{prop:pca-approximation}
Under Assumptions~\ref{ass:A}--\ref{ass:D} and \ref{ass:signal}, use
residual PCA at the true rank or at the rank in \eqref{eq:pca-selector}.
The fits and test scales are defined with probability tending to one, and
\[
\begin{gathered}
\widehat{\mathcal I}_{\rm pc}/\Ieff-1=\Op(\epsT),\qquad
n/\widehat{\mathcal I}_{\rm pc}-\cstar=\Op(\epsT+n^{-1/2}),\\
T_{\mathrm{sum},\rm pc}-T_{\mathrm{sum}}^\circ=\Op(\bsum).
\end{gathered}
\]
Under $H_0$,
\[
T_{\mathrm{max},\rm pc}-T_{\mathrm{max}}^\circ
=\Op(\bmax+\bmax^2).
\]
\end{proposition}

\begin{proof}[Proof of Proposition~\ref{prop:pca-approximation}]
At rank $r$, Lemma~\ref{lem:pca} gives \eqref{eq:admission}.
The PCA map is centered-residual measurable and shift invariant.
Lemmas~\ref{lem:geometry} and~\ref{lem:time} therefore give
\[
\|\widehat{\mat{\Pi}}_{\rm pc}-\mat{\Pi}_0\|=\Op(\epsT),\quad
|\widehat\eta_{\rm pc}-\eta_0|=\Op(\epsT),\quad
\eta_0^{-1}=\Op(1).
\]
Using $\widehat{\mathcal I}_{\rm pc}=\omega_z^2\widehat\eta_{\rm pc}$
and $\Ieff=\omega_z^2\eta_0$ yields the information bounds. Also,
\[
\mat{S}_z\succeq c\mat{I}_m,\quad
\widehat{\mat{K}}_{g,\rm pc}\succeq c\sfac^2\mat{I}_r,\quad
\widehat{\mathcal I}_{\rm pc}\ge cn
\]
with probability tending to one. Lemma~\ref{lem:scales} gives
\[
\begin{gathered}
\widehat v_{\varepsilon,\rm pc}/\taunoise=1+\Op(n^{-1}),\\
\max_i\left|(\widehat{\mat{\Sigma}}_{\varepsilon,\rm pc})_{ii}
-(\Sg)_{ii}-\Ieff\alpha_i^2/k\right|=\Op(\vdiag).
\end{gathered}
\]
Since $\taunoise\asymp p$, $(\Sg)_{ii}\ge c$ and $\vdiag\to0$,
the fits and scales are regular. Lemma~\ref{lem:replacement} gives
\[
T_{\mathrm{sum},\rm pc}=Q_n^\circ+\Delta_n+\Op(\bsum),\qquad
T_{\mathrm{max},\rm pc}=M_n^\circ+\Op(\bmax+\bmax^2)\quad(H_0).
\]
For the oracle intercept,
\[
T_{\mathrm{sum}}^\circ-Q_n^\circ-\Delta_n
=\frac{(\Ieff-n/\cstar)\|\alp\|^2
+2\sqrt{\Ieff}\,\alp\T\vct{\xi}_n^\circ}{\snoise}
=\Op(\sqrt{n/p}).
\]
The bound $\sqrt{n/p}\le\sqrt n/\sfac\le\bvec\le\bsum$ proves
the Sum assertion, and $T_{\mathrm{max}}^\circ=M_n^\circ$ under $H_0$.
Finally, Lemma~\ref{lem:pc-rank-original} and
\eqref{eq:number-rate} transfer each bound to the selected rank.
\end{proof}

\subsection{Null distributions, power and comparison with LAG}\label{app:pca-limits}
\begin{theorem}[PCA null calibration and power]\label{thm:pca-inference}
Under Assumptions~\ref{ass:A}--\ref{ass:D}, use the true-rank PCA
implementation or the selector \eqref{eq:pca-selector}. Under $H_0$,
\[
(T_{\mathrm{max},\rm pc},T_{\mathrm{sum},\rm pc})
\Rightarrow(G_0,Z_0),\qquad G_0\ind Z_0,\qquad
T_{\mathrm{cauchy},\rm pc}\Rightarrow\operatorname{Cauchy}(0,1).
\]
Under Assumption~\ref{ass:signal}, if $\Delta_n\to\Delta<\infty$,
\[
T_{\mathrm{sum},\rm pc}\Rightarrow N(\Delta,1),\qquad
\mathbb P\{T_{\mathrm{sum},\rm pc}>\Phi^{-1}(1-\gamma)\}
\to1-\Phi\{\Phi^{-1}(1-\gamma)-\Delta\}.
\]
If $\Delta_n\to\infty$, then
$T_{\mathrm{sum},\rm pc}/\Delta_n\to_{\mathbb P}1$.
If $\sqrt n\|\alp\|_\infty/\sqrt{\log(2p)}\to\infty$, then
$T_{\mathrm{max},\rm pc}\to_{\mathbb P}\infty$.
In either consistent regime,
$T_{\mathrm{cauchy},\rm pc}\to_{\mathbb P}\infty$.
\end{theorem}

\begin{proof}[Proof of Theorem~\ref{thm:pca-inference}]
Proposition~\ref{prop:pca-approximation} and Lemma~\ref{lem:oracle}
give the joint null limit, since $\bsum,\bmax\to0$. The same continuous-mapping argument as in
Appendix~\ref{app:null-proof} gives the standard Cauchy limit. For bounded signals,
\[
T_{\mathrm{sum},\rm pc}=Q_n^\circ+\Delta_n+\Op(\bsum),\qquad
Q_n^\circ\Rightarrow N(0,1),\qquad \bsum\to0.
\]
These yield the dense limits and the local rejection probability.

Choose $i_0$ with $|\alpha_{i_0}|=\|\alp\|_\infty$.
Equation~\eqref{eq:vectorreplacement}, applicable by Lemma~\ref{lem:pca},
gives
\[
\sqrt{\widehat{\mathcal I}_{\rm pc}}\widehat\alpha_{i_0,\rm pc}
=\sqrt{\Ieff}\alpha_{i_0}+\xi_{n,i_0}^\circ+\Op(\bvec),\qquad
\xi_{n,i_0}^\circ=\Op(1).
\]
The coordinate variance is bounded above and away from zero in probability,
so
\[
\frac{|\sqrt{\widehat{\mathcal I}_{\rm pc}}\widehat\alpha_{i_0,\rm pc}|}
{\sqrt{(\widehat{\mat{\Sigma}}_{\varepsilon,\rm pc})_{i_0i_0}}
\sqrt{\log(2p)}}
\ge c\frac{\sqrt n\|\alp\|_\infty}{\sqrt{\log(2p)}}
-\frac{\Op(1)+\Op(\bvec)}{\sqrt{\log(2p)}}\to_{\mathbb P}\infty.
\]
This proves sparse consistency.

For the combination, the negative part of $T_{\mathrm{sum},\rm pc}$
is tight because $\Delta_n\ge0$. Put
$\mathcal J_n=\{i:|\alpha_i|\le(n\Lnp)^{-1/2}\}$.
Then $p-|\mathcal J_n|\le C_\alpha^2n\Lnp=o(p)$, and the same
score and variance bounds give
\[
\max_{i\in\mathcal J_n}
\left|\frac{\widehat{\mathcal I}_{\rm pc}\widehat\alpha_{i,\rm pc}^2}
{(\widehat{\mat{\Sigma}}_{\varepsilon,\rm pc})_{ii}}
-\frac{(\xi_{n,i}^\circ)^2}{(\Sg)_{ii}}\right|
=\Op\{1+\bvec\sqrt{\Lnp}+\vdiag\Lnp\}=\Op(1).
\]
Lemmas~\ref{lem:coupling} and~\ref{lem:gaussian} on $\mathcal J_n$
give a Gumbel limit with the same centering, because
$|\mathcal J_n|/p\to1$. Thus the negative part of
$T_{\mathrm{max},\rm pc}$ is tight. For a consistent component $j$
and the other component $j'$, writing $x^-=\max(-x,0)$,
\[
\cot(\pi p_{j,\rm pc})\to_{\mathbb P}\infty,\qquad
\{\cot(\pi p_{j',\rm pc})\}^-=\Op(1),
\]
\[
2T_{\mathrm{cauchy},\rm pc}
\ge\cot(\pi p_{j,\rm pc})-\{\cot(\pi p_{j',\rm pc})\}^-
\to_{\mathbb P}\infty.
\]
Rank consistency transfers all assertions to $\widehat r_{\rm pc}$.
\end{proof}

\begin{proposition}[First-order comparison with LAG]\label{prop:pca-lag-equivalence}
Under Assumptions~\ref{ass:A}--\ref{ass:rank}, use LAG with its
consistent selector and PCA with \eqref{eq:pca-selector}, on the same
sample. Under $H_0$,
\[
T_{j,\rm pc}-T_{j,\rm lag}\to_{\mathbb P}0,
\qquad j\in\{\mathrm{sum},\mathrm{max},\mathrm{cauchy}\}.
\]
Under Assumption~\ref{ass:signal}, their Sum statistics also differ by
$o_{\mathbb P}(1)$ and have the same dense local-power limit.
\end{proposition}

\begin{proof}[Proof of Proposition~\ref{prop:pca-lag-equivalence}]
The two implementations share the oracle quantities. Their approximation
bounds give
\[
\begin{aligned}
|T_{\mathrm{sum},\rm pc}-T_{\mathrm{sum},\rm lag}|&=\Op(\bsum),\\
|T_{\mathrm{max},\rm pc}-T_{\mathrm{max},\rm lag}|
&=\Op(\bmax+\bmax^2)\quad(H_0).
\end{aligned}
\]
For the Cauchy map, restrict both pairs of component $p$-values to
$[\epsilon,1-\epsilon]^2$. The map is Lipschitz on this compact set,
so the difference converges to zero there. The joint uniform null limits
bound the probability of its complement by $8\epsilon+o(1)$.
Let $n\to\infty$ and then $\epsilon\downarrow0$. The dense conclusion
follows from the common $\Delta_n$.
\end{proof}
Finite-sample power also depends on factor-number selection and the projection effects in
\eqref{eq:apr-pca-decomposition}, as illustrated in Section~\ref{sec:sim-power}.

\subsection{Bootstrap with a fixed factor number for PCA}\label{app:pca-bootstrap}
Use the generator in Section~\ref{sec:bootstrap} with the PCA pilot fit.
In particular,
\[
\widetilde{\mat{A}}_{\rm pc}=\widehat{\mat{A}}_{\rm pc},\qquad
\widehat{\mat{E}}_{\rm pc}=\widehat{\mat{Q}}_{\rm pc}\Rmat,
\]
and, in each pseudo-sample, replace LAG extraction by the leading
$\widehat r_{\rm pc}$ eigenvectors of the centered residual covariance.
All other operations follow Section~\ref{sec:bootstrap}. If the outer rank
is zero, omit the latent generator; this event has probability tending
to zero.

\begin{theorem}[PCA bootstrap validity]\label{thm:pca-bootstrap}
Under Assumptions~\ref{ass:A}--\ref{ass:D} and $H_0$, the preceding
PCA bootstrap, at the true rank or using \eqref{eq:pca-selector}, satisfies
\[
\sup_{x,y}\left|
\mathbb P^*(T_{\mathrm{sum},\rm pc}^*\le x,
T_{\mathrm{max},\rm pc}^*\le y)-\Phi(x)F_G(y)
\right|\to_{\mathbb P}0.
\]
Also $T_{\mathrm{cauchy},\rm pc}^*\Rightarrow_{\mathbb P^*}
\operatorname{Cauchy}(0,1)$ in probability, and, for
$j\in\{\mathrm{sum},\mathrm{max},\mathrm{cauchy}\}$,
\[
\sup_x\left|\mathbb P^*(T_{j,\rm pc}^*\le x)
-\mathbb P_0(T_{j,\rm pc}\le x)\right|\to_{\mathbb P}0.
\]
\end{theorem}

\begin{proof}[Proof of Theorem~\ref{thm:pca-bootstrap}]
First fix the rank at $r$. Put $\mat{A}_0=\widehat{\mat{A}}_{\rm pc}$
and suppress the subscript $\mathrm{pc}$ on pilot residual quantities.
By Proposition~\ref{prop:pca-identity},
\[
\widehat{\mat{E}}\one_n=0,\quad
\mat{A}_0\T\widehat{\mat{E}}=0,\quad
\mat{A}_0\T\mat{A}_0=\mat{I}_r,\quad
a_R=\tr(\mat{A}_0\T\mat{S}_R\mat{A}_0)=0.
\]
Lemma~\ref{lem:pca} verifies \eqref{eq:admission}, so
Lemma~\ref{lem:column-pilot} gives
\[
\begin{gathered}
\|\widehat{\mat{E}}\|=\Op(\sqrt p),\quad
\|\mat{S}_R\|=\Op(p/n),\quad m_R\asymp_{\mathbb P}p,\quad
\tau_R\asymp_{\mathbb P}p^2/n,\\
\max_i\|\vct{e}_i\T\mat{A}_0\|=\Opl(p^{-1/2}),\quad
\max_i|D_{R,ii}-(\Sg)_{ii}|=\Opl(n^{-1/2}+\epsT),\\
v_R=\Opl(p+p^2/n),\qquad
\|\bar{\widehat{\vct{F}}}\|/\sqrt p=\Op(1).
\end{gathered}
\]
Here $\Opl$ and $\Ops$ have the logarithmic-rate meanings specified
in Appendix~\ref{app:bootstrap-proof}. Define $\mat{M}_0^*$ to remove
$[\mat{Z},\mat{F}^{*\mathsf T}]$, and let
$\vct{\beta}^*=\mat{M}_0^*\one_n/
\sqrt{\one_n\T\mat{M}_0^*\one_n}$. Lemma~\ref{lem:var-oracle} yields
\begin{equation}\label{eq:pca-boot-oracle}
\begin{gathered}
Q_R^*=\frac{\|\mat{E}^*\vct{\beta}^*\|^2-m_R}{\sqrt{2\tau_R}},\qquad
M_R^*=\max_i\frac{(\mat{E}^*\vct{\beta}^*)_i^2}{D_{R,ii}}
-2\log p+\log\log p,\\
\sup_{x,y}\left|\mathbb P^*(Q_R^*\le x,M_R^*\le y)
-\Phi(x)F_G(y)\right|\to_{\mathbb P}0.
\end{gathered}
\end{equation}
The oracle uses the empirical residual covariance $\mat{S}_R$, with $\tr(\mat{S}_R^2)$ of order $p^2/n$. The residual-resampling argument establishes conditional calibration at this quadratic scale.

The PCA-specific refitting bound follows from its covariance eigenproblem.
Set $\mat{F}_c^*=\mat{F}^*\mat{M}_c$ and
$\mat{W}_c^*=\mat{E}^*\mat{M}_c$. Exactly,
\[
\Rmat^*=\mat{A}_0\mat{F}_c^*+\mat{W}_c^*,\qquad
\widehat{\mat{C}}_\eta^*
=\mat{A}_0\frac{\mat{F}_c^*\mat{F}_c^{*\mathsf T}}n\mat{A}_0\T
+\mat{E}_{C,n}^*.
\]
The standardized Gaussian generator and the observed-design bounds give
\[
\lambda_{\min}(\mat{F}_c^*\mat{F}_c^{*\mathsf T}/n)
\asymp_{\mathbb P^*}p,\qquad
\|\mat{F}_c^*\|_F^2=\Ops(np).
\]
Conditional on the factor path, independence of the residual draws gives
\[
\mathbb E_I\|\mat{E}^*\mat{F}_c^{*\mathsf T}/n\|_F^2
=m_R\|\mat{F}_c^*\|_F^2/n^2=\Ops(p^2/n).
\]
For the contemporaneous noise covariance and the removed fixed-dimensional
time space,
\[
\begin{gathered}
\mathbb E_I\|n^{-1}\mat{E}^*\mat{E}^{*\mathsf T}-\mat{S}_R\|_F^2
\le n^{-1}\mathbb E_I\|\vct{\varepsilon}_1^*\|^4
=\Opl(p^2/n),\\
\mathbb E_I\|\mat{E}^*(\mat{I}_n-\mat{M}_c)\|_F^2
=(m+1)m_R=\Op(p).
\end{gathered}
\]
Therefore
\begin{equation}\label{eq:pca-boot-loading}
\begin{gathered}
\|\mat{E}_{C,n}^*\|=\Ops(p/\sqrt n+p/n),\\
\|\widehat{\mat{P}}_{\rm pc}^*-\mat{A}_0\mat{A}_0\T\|
\le\frac{\|\mat{E}_{C,n}^*\|}
{\lambda_{\min}(\mat{F}_c^*\mat{F}_c^{*\mathsf T}/n)
-\|\mat{E}_{C,n}^*\|}
=\Ops(n^{-1/2}).
\end{gathered}
\end{equation}
The inequality holds on an event with conditional probability tending to
one, by the eigenvector calculation in Lemma~\ref{lem:pca}.

The time-space and scale estimates follow from the preceding loading bound. Put $\rho_n^*=n^{-1/2}$,
$a_n^*=p^{-1/2}+n^{-1}$ and
$h_n^*=(a_n^*)^2+(np)^{-1/2}+n^{-1}$. Each resampled column is
orthogonal to $\mat{A}_0$, and the maximum column multiplicity is
$O_{\mathbb P^*}(\log n)$. Hence
\[
\mat{A}_0\T\mat{E}^*=0,\qquad
\|\mat{E}^*\|=\Ops(\sqrt p).
\]
With the aligned inverse
\[
\mat{T}_A^*=(\widehat{\mat{A}}_{\rm pc}^{*\mathsf T}\mat{A}_0)^{-1}
\widehat{\mat{A}}_{\rm pc}^{*\mathsf T}-\mat{A}_0\T,
\qquad \|\mat{T}_A^*\|=\Ops(\rho_n^*),
\]
the perturbation in the extracted time scores is
$\mat{N}_A^*=(\mat{A}_0\T+\mat{T}_A^*)\mat{E}^*\mat{O}_z$.
It obeys the bounds used in \eqref{eq:time}:
\[
\begin{gathered}
\|\mat{N}_A^*\|=\Ops(\sqrt n+\rho_n^*\sqrt p),\qquad
\|\mat{N}_A^*\vct{v}\|=\Ops(1+\rho_n^*\sqrt p),\\
\|\widehat{\mat{\Pi}}_{\rm pc}^*-\mat{\Pi}_0\|=\Ops(a_n^*),\qquad
\|\mat{\Pi}_0\widehat{\mat{J}}_{\rm pc}^*\vct{u}_z\|=\Ops(h_n^*).
\end{gathered}
\]
The gap is $\sigma_{\min}(\mat{F}^*\mat{O}_z)\asymp_{\mathbb P^*}
\sqrt{np}$; the projection inequalities proving \eqref{eq:time}
therefore apply. The finite-rank expansion \eqref{eq:finite} and the
residual-index moment identities \eqref{eq:noise}--\eqref{eq:diag},
which depend on the preceding time-space and residual moment bounds,
give
\[
\begin{gathered}
\frac{\tr(\widehat{\mat{\Sigma}}_{\varepsilon,\rm pc}^*)-m_R}
{\sqrt{\tau_R}}=\Ops(n^{-1/2}+\sqrt n/p),\\
\widehat v_{\varepsilon,\rm pc}^*/\tau_R-1
=\Ops(n^{-1/2}+p^{-1}+n/p^2),\\
\max_i|(\widehat{\mat{\Sigma}}_{\varepsilon,\rm pc}^*)_{ii}-D_{R,ii}|
=\Ops(n^{-1/2}+a_n^*),\\
\frac{\left|\|\sqrt{\widehat{\mathcal I}_{\rm pc}^*}
\widehat{\alp}_{\rm pc}^*\|^2-\|\mat{E}^*\vct{\beta}^*\|^2\right|}
{\sqrt{\tau_R}}=\Ops(\sqrt{n/p}+n^{-1/2}),\\
\|\sqrt{\widehat{\mathcal I}_{\rm pc}^*}\widehat{\alp}_{\rm pc}^*
-\mat{E}^*\vct{\beta}^*\|_\infty=\Ops(\sqrt{n/p}).
\end{gathered}
\]
All fixed logarithmic factors are absorbed by Assumption~\ref{ass:D}.
The empirical diagonals are bounded away from zero in probability, and
the oracle coordinate maximum is $\Ops(1)$. Substitution into the
statistics yields
\[
T_{\mathrm{sum},\rm pc}^*=Q_R^*+\ops(1),\qquad
T_{\mathrm{max},\rm pc}^*=M_R^*+\ops(1).
\]
Equation~\eqref{eq:pca-boot-oracle} proves the joint limit. Continuity
of the cotangent combination at interior component $p$-values gives
the Cauchy limit. For the corresponding continuous limit $F_j$,
\[
\begin{aligned}
&\sup_x|\mathbb P^*(T_{j,\rm pc}^*\le x)-\mathbb P_0(T_{j,\rm pc}\le x)|\\
&\le\sup_x|\mathbb P^*(T_{j,\rm pc}^*\le x)-F_j(x)|
+\sup_x|\mathbb P_0(T_{j,\rm pc}\le x)-F_j(x)|\to_{\mathbb P}0,
\end{aligned}
\]
where the original-sample limit follows from Theorem~\ref{thm:pca-inference}.
Finally, the selected-rank conditional procedure agrees with the
true-rank one on $\{\widehat r_{\rm pc}=r\}$. Their conditional distribution
distance is bounded by $\indc\{\widehat r_{\rm pc}\ne r\}=o_{\mathbb P}(1)$
by Lemma~\ref{lem:pc-rank-original}. Each bootstrap refit uses this same outer factor number.
\end{proof}

\end{document}